# Simple techniques for likelihood analysis of univariate and multivariate stable distributions:
## with extensions to multivariate stochastic volatility and dynamic factor models

Efthymios G. Tsionas
Department of Economics,
Athens University of Economics and Business,
76 Patission Street, 104 34 Athens, Greece
Tel.: (++30210) 8203 338, Fax: (++30210) 8203 301, email: tsionas@aueb.gr

## Abstract

In this paper we consider a variety of procedures for numerical statistical inference in the family of univariate and multivariate stable distributions. In connection with univariate distributions (i) we provide approximations by finite location-scale mixtures and (ii) versions of approximate Bayesian computation (ABC) using the characteristic function and the asymptotic form of the likelihood function. In the context of multivariate stable distributions we propose several ways to perform statistical inference and obtain the spectral measure associated with the distributions, a quantity that has been a major impediment in using them in applied work. We extend the techniques to handle univariate and multivariate stochastic volatility models, static and dynamic factor models with disturbances and factors from general stable distributions, a novel way to model multivariate stochastic volatility through time-varying spectral measures and a novel way to multivariate stable distributions through copulae. The new techniques are applied to artificial as well as real data (ten major currencies, SP100 and individual returns). In connection with ABC special attention is paid to crafting well-performing proposal distributions for MCMC and extensive numerical experiments are conducted to provide critical values of the "closeness" parameter that can be useful for further applied econometric work.

**Acknowledgements**: The author wishes to thank seminar participants at the Department of Economics, University of Leicester, and seminar participants of CRETE 2012 at Milos, Greece.

# 1. Introduction

Univariate stable distributions have been thoroughly studied in econometrics, statistics and finance over the past few decades (Samorodnitsky and Taqqu, 1994). Their empirical application is still hampered by the fact that their density is not available in closed form, despite advances in Bayesian computation using MCMC. Buckle (1995) and Tsionas (1999) provided Gibbs sampling schemes for general and symmetric stable distributions, respectively. The problem is that the conditional posterior distributions of certain latent variables are cumbersome to deal with and require careful tuning.

The analogous problem in the multivariate case is exceedingly difficult although a few attempts have been made to solve it. The impediment is that multivariate stable distributions, unlike the univariate case, are defined through their spectral measure which, in practice, is unknown. Ravishanker and Qiou (1999) for example, proposed an EM algorithm based on Buckle (1995) in the case of symmetric isotropic stable distributions but this class is too narrow to be of empirical importance. It is defined by the transformation $\mathbf{X} = \boldsymbol{\mu} + \boldsymbol{\Sigma}^{1/2}\boldsymbol{\xi}$, where $\boldsymbol{\xi}$ is a vector of independent random variables each one distributed as standard symmetric stable, $\boldsymbol{\mu}$ is a vector of location parameters, $\boldsymbol{\Sigma} = \mathbf{C}'\mathbf{C}$ is a scale matrix, and $\mathbf{C}$ denotes its Cholesky decomposition.

The idea of the paper is that Approximate Bayesian Computation (ABC) can be implemented easily in connection with univariate or multivariate stable distributions since it can be tailored to use the characteristic function, which summarizes fully all sampling empirical evidence which is available through the likelihood function. Although ABC has not have found many applications in econometrics, there are many in the statistical literature (for example Wilkinson, 2008). ABC is a technique that can be used when the density of observations is not available in closed form or when the model is implicitly defined (such models are known as "computer code models"). It is related to calibration in the sense that in its simplest version, a draw of parameters is made from the prior, a set of moments are selected and simulated moments are compared to empirical counterparts. If a measure of "closeness" is "small" then the draw is accepted. Finding such moments can be difficult, especially in stable distributions, where the moments that should be matched can be non-intuitive. Measures of "closeness" are also hard to formulate and given such a measure, an appropriate definition of "close" can be hard. In the context of ABC it is clear that whenever possible the characteristic function can be used since it is equivalent to the density function. For stable distributions, at least in the univariate case, it is well known that the characteristic function has a very simple closed form.

In univariate stable Paretian distributions, implementation of ABC is conceptually straightforward although certain problems remain open and must be addressed: First, the choice of a proposal distribution for the parameters and, second, the choice of a measure of "closeness" between the theoretical and empirical characteristic functions. The third problem is the number and configuration of grid points for the computation of the characteristic functions which remains open and unsettled in the statistics and econometrics literature for many decades. In connection with multivariate stable Paretian distributions, even the computation of the characteristic functions becomes complicated because they are only defined through their spectral measure, an object that is needed to retain the equivalence between the density and the characteristic function. The estimation of the spectral measure itself has proved itself to be quite cumbersome even for bivariate distributions. We examine several existing approximations to the spectral measure for use with ABC for Bayesian inference, and propose a new one based on a multivariate normal approximation which, along with another technique known as method of principal directions (Meerschaert and Scheffler, 1999), are found to perform well. Specifically, the method of principal directions is found to perform extremely well in discovering and approximating the "important" linear combinations of stable variates that can be used to configure the grid points for evaluation of the multivariate characteristic function. We consider this feature very important as it can facilitate considerably joint likelihood analysis in multivariate stable distributions for all parameters, including the grid configuration and the placement of grid points for the characteristic function.

Joint inferences for the spectral measure and the configuration of the grid for the evaluation of the characteristic functions are provided using well-crafted proposal distributions for use with ABC in the context of multivariate stable Paretian distributions. Moreover, we show how another smooth approximation to the spectral measure, the so called spherical harmonic analysis (Pivato, 2001, Pivato and Seco, 2003) can be implemented in ABC along with the asymptotic normal form of the likelihood function. Since Bayesian inference for multivariate distributions is found quite simple to implement and perform well, we generalize the stable Paretian distributions in two important directions. First, in the context of univariate and multivariate stochastic volatility and second, in connection with static and dynamic factor models, including a new factor model that uses a Markov model for the fundamental parameters of the model. In addition, we propose a new stochastic volatility model, that is more appropriate for multivariate stable variates. Clearly, the techniques developed can be used to obtain statistical inferences for multivariate volatility models based on the normal distribution, with or without jumps and leverage effects. Their application is, naturally, simpler when compared to their stable Paretian counterparts. However, the techniques remain simple and efficient even in high-dimensional multivariate stable distributions.

The approximation of univariate stable distributions by finite mixtures of normal distributions should not be discounted given the main emphasis of the paper on the multivariate case which, undoubtedly, imposed so many difficulties and obstacles so far. For univariate stable distributions we apply ABC inference in detail, showing how well-crafted proposal distributions can be constructed and used in artificial and real data, based on the characteristic function. Bayesian inferences organized around the usage of finite mixtures of normal distributions are found to be quite close to those provided either by ABC or the exact approach based on the fast Fourier transform to compute the stable Paretian densities. Since the finite mixtures of normals approximation depend on results that can be easily be tabulated (as in this paper) the routine application of Bayesian inference for linear regression with stable disturbaces or factor analysis based on stable distributions, becomes possible almost effortlessly.

This work falls squarely within recent advances in the econometrics of stable distributions. Dominicy and Veredas (2012) propose a method of quantiles to fit symmetric stable distributions. Since the quantiles are not available in closed form they are obtained using simulation resulting in the method of simulated quantiles or MSQ. Hallin, Swan, Verdebout and Veredas (2012) propose an easy-to-implement R-estimation procedure which remains $\sqrt{n}$-consistent contrary to least squares with stable disturbances. Broda, Haas, Krause, Paolella and Steude (2012) propose a new stable mixture GARCH model that encompasses several alternatives and can be extended easily to the multivariate asset returns case using independent components analysis. Ogata (2012) uses a discrete approximation to the spectral measure of multivariate stable distributions and proposes estimating the parameters by equating the theoretical and empirical characteristic function in a generalized empirical likelihood / GMM framework. Relative to this work, (*i*) we show how to implement Bayesian inference for multivariate stable distributions by treating the grid points of the characteristic function and the support of the spectral measure as parameters. Moreover, (*ii*) we generalize the model to multivariate stochastic volatility and factor models, and (*iii*) we provide alternatives to the discretization of the spectral measure that are easier to compute and perform better in artificial as well as real data. Regarding (Bayesian) indirect inference for the parameters of univariate stable distributions, we provide useful results that can be used to implement MCMC for any data set when draws from the posterior distribution of normal mixtures are available. Classical indirect inference has been examined by Garcia, Renault and Veredas (2011) among others, using the skewed Student-*t* as auxiliary model. As the authors mention this "appears as a good candidate since it has the same number of parameters as the $\alpha$-stable distribution, with each parameter playing a similar role". Lombardi and Veredas (2007) used a multivariate Student-t to perform indirect estimation for for elliptical stable distributions based on the same argument.

To summarize, in this paper, we break new ground along the following directions. In connection with univariate stable distributions we propose, *first*, a mixture of normals approximation with few components. The approximation is obtained through minimizing the Kullback – Leibler distance between stable and mixture-of-normals distributions and can be used to perform efficient Bayesian inference using MCMC. *Second*, we examine Approximate Bayesian Computation (ABC) which relies only on the theoretical and empirical characteristic functions. *Third*, we provide several computational approaches to statistical inference in multivariate, general stable distributions using ABC. *Fourth*, we provide a new copula – based approach for multivariate, general stable distributions. *Fifth*, we consider extensions to multivariate stochastic volatility and static and dynamic factor models whose factors and disturbances are members of the multivariate stable family. All techniques are applied to exchange rate and stock return data and are supplemented by Monte Carlo simulations.

## 2. Univariate General Stable Distributions

A random variable $X$ is called (strictly) stable if for all $n$, $\sum_{i=1}^{n} X_i \stackrel{D}{=} c_n X$, for some constant $c_n$, where $X_1,...,X_n$ are independently distributed with the same distribution as $X$. It is known that the only possible choice is to have $c_n = n^{1/\alpha}$, for some $\alpha \in (0,2]$. General non-symmetric stable distributions are defined via the log characteristic function which is given by the following expression (Samorodnitsky and Taqqu, 1994, and Zolotarev, 1986):

$$\log \varphi(\tau) \triangleq \log \mathbb{E} \exp(\iota\tau X) = \begin{cases} \iota\mu\tau - \sigma^\alpha |\tau|^\alpha \left[1 - \iota\beta \operatorname{sgn}(\tau) \tan \frac{\pi\alpha}{2}\right], & \alpha \neq 1, \\ \iota\mu\tau - \sigma|\tau|\left[1 + \iota\beta \operatorname{sgn}(\tau) \frac{2}{\pi} \log |\tau|\right], & \alpha = 1, \end{cases} \quad (1)$$

for any $\tau \in \mathbb{R}$, where $\mu$ and $\sigma$ are the location and scale parameters of the distribution, respectively, and $\iota = \sqrt{-1}$. We denote a general stable random variable by: $X \sim \mathscr{S}_{\alpha,\beta}(\mu,\sigma)$. The density is given by

$$f(x) = \frac{1}{2\pi} \int_{-\infty}^{\infty} \exp(-\iota\tau x) \varphi(\tau) d\tau. \quad (2)$$

The density is not available in closed form making it difficult to implement maximum likelihood or Bayesian MCMC procedures. Buckle (1995) and Tsionas (1999) considered scale mixture representation of stable distributions for Bayesian analysis in the general and symmetric class respectively. Tsionas (1999) exploited the fact that for symmetric laws, that is $X \sim S_{\alpha,0}(\mu,\sigma)$ we have: $X = W^{1/2} Z$, where $Z \sim N(0,1)$, and independently $W \sim S_{\frac{\alpha}{2},1}(0,1)$. Since this is a scale mixture of normal distributions, Bayesian numerical procedures are greatly facilitated. Buckle (1995) used another representation due to Zolotarev (1986, pp. 65-66):

$$X = \frac{\sin \alpha U}{(\cos \alpha U)^{1/\alpha}} \left[\frac{\cos(\alpha-1)U}{E}\right]^{(1-\alpha)/\alpha}, \text{ for } \alpha \neq 1,$$

where $U$ is uniformly distributed in $\left[-\frac{\pi}{2}, \frac{\pi}{2}\right]$, and $E$ is standard exponential[1]. Approximate computation of the general stable densities is facilitated by the Fast Fourier Transform (FFT), see Mittnik, Doganoglu, and Chenyao (1999) and Mittnik, Rachev, Doganoglu, and Chenyao (1999)[2]. The integral representation in (2) is computed at N equally spaced points with distance h, that is $x_k = \left(k - 1 - \frac{N}{2}\right)h$, $k = 1,...,N$. If $\tau = 2\pi\omega$, the integral becomes $f\left(\left(k - 1 - \frac{N}{2}\right)h\right) = \int_{-\infty}^{\infty} \varphi(2\pi\omega) \exp\left(-\iota 2\pi\omega\left(k - 1 - \frac{N}{2}\right)h\right) d\omega$ which can be approximated using the rectangle rule as:

$$f\left(\left(k - 1 - \frac{N}{2}\right)h\right) \approx \frac{1}{hN}(-1)^{k-1-\frac{N}{2}} \sum_{n=1}^{N} (-1)^{n-1} \varphi\left(2\pi\left(n - 1 - \frac{N}{2}\right)h / N\right) \exp\left(-\iota 2\pi(n-1)(k-1)/N\right). \quad (3)$$

---
[1] Where not needed, we present results for the "focal" case $\alpha \neq 1$.
[2] See also Matsui and Takemura (2006).

In turn, this is equivalent to performing a FFT to the sequence:

$$(-1)^{n-1} \varphi\left(2\pi\left(n - 1 - \frac{N}{2}\right)h / N\right), \; n = 1,...,N.$$

A fairly accurate procedure results when $N = 2^{16}$, and $h = 10^{-4}$. Accuracy of the FFT has been examined in detail by Tsionas (2012a) in a different context. In the case of symmetric stable distributions, $\mathscr{S}_{\alpha,0}(\mu,\sigma)$, McCulloch (1998) developed a more efficient procedure without sacrificing accuracy.

## 3. Approximate representation by mixtures of normal

The stable distribution with density $f(u)$ is amenable to approximation by families of distributions for which Bayesian inference is tractable. One such family is the finite mixture of normal[3]:

$$p(u;\boldsymbol{\theta}) = \sum_{m=1}^{M} \pi_m f_N(u; \mu_m, \sigma_m^2), \quad (4)$$

where $f_N(u;\mu,\sigma^2)$ denotes the density of a normal distribution with mean $\mu$ and variance $\sigma^2$, and $\boldsymbol{\theta} = [\boldsymbol{\pi}', \boldsymbol{\mu}', \boldsymbol{\sigma}']'$ in obvious notation. To determine the parameters $\boldsymbol{\theta}^*$ we use the Kullback-Leibler distance: $KL(f \| p) = \int_{\mathbb{R}} f(u) \log \frac{f(u)}{p(u;\boldsymbol{\theta})} du$. The range of integration is truncated to $E = [-12, 12]$. Part of the problem is to determine the optimal number of components, $M$. Initial experimentations indicated that the number of components is quite small (3 or 4) by fitting general mixtures of up to M=15 components: Several probabilities were practically zero so we considered approximations by mixtures of lower dimensions. These approximations are shown in Figure 1 for M=2, 3, and 4. The approximation for M=4 is indistinguishable from the true density while M=3 performs well[4]. The comparison of the density with the mixture of normals is presented in Figure 1 and the Kullback-Leibler distance as a function of the number of components (M) is presented in Figure 2.

Suppose a scale mixture of normals with M components is fitted to the data using MCMC. Denote the draws by $A^{(s)} = \left[\sigma^{(s)\prime}, \pi^{(s)\prime}\right]'$, $s = 1,...,S$. To transform to draws from a symmetric stable distribution $\theta^{(s)} = \left[\mu^{(s)}, \delta^{(s)}, \alpha^{(s)}\right]$ whose characteristic function is $\varphi(\tau) = \exp\left(\iota\mu\tau - \delta|\tau|^\alpha\right)$, consider the characteristic function of the approximating normal mixture: $\varphi_N(\tau) = \sum_{m=1}^{M} \pi_m \exp\left(-\frac{1}{2}\sigma_m^2 \tau^2\right)$. Given the parameters of the mixture in $A^{(s)}$ we consider the optimization problem: $\min_{\theta^{(s)}} : \sum_{i=1}^{I} |\varphi(\tau_i) - \varphi_N(\tau_i)|^2$ ,

where $(\tau_i, i = 1,...,I)$ is a grid of points. The optimization problem was found to be very easy to solve and 10,000 draws were obtained in less than a minute. In Figure 1 we present the marginal posterior densities of α from an "exact" Metropolis MCMC and two mixture approximations with M=3 and M=5 components. We have constructed an artificial data set with n=1,500 observations, μ=0, δ=1, and α=1.40. With M=5 components the mixture approximation and the "exact" posterior are indistinguishable[5].

---

[3] As far as we know the only other relevant work is Georgiadis and Mulgrew (2001) who used a mixture of Cauchy and normal.
[4] Plot of the KL criterion at the optimum indicate that the fit with two or more components is approximately constant. We considered finite scale mixtures with M=2 up to M=15 components.
[5] We have tried to fit mixtures with M>5 components but the optimization failed because many σ's and π's are actually zero in the optimal approximation using the KL criterion.

Turning attention to the more general non-symmetric stable distributions, a family which is a simple reparametrization of the above and is continuous with respect to the parameters is the following (Nolan, 1997):

$$\log \varphi(t) = \begin{cases} \iota\mu\tau - \sigma^{\alpha}|\tau|^{\alpha}\left[1 - \iota\beta \operatorname{sgn}(\tau) \tan\frac{\pi\alpha}{2}\left(\sigma|\tau|^{\alpha-1} - 1\right)\right], & \alpha \neq 1, \\ \iota\mu\tau - \sigma|\tau|\left[1 + \iota\beta \operatorname{sgn}(\tau) \frac{2}{\pi} \log|\tau|\right], & \alpha = 1. \end{cases} \quad (5)$$

Critical values (90%) of the $D$ statistic in the case of non-symmetric stable distributions are provided in Table 1b. The critical values vary little depending on sample size, $n$, as well as the important parameters α and β. This fact will be of considerable interest in the next section where we take up Bayesian inference using Approximate Bayesian Computation (ABC).

## 4. Approximate Bayesian Computation

ABC is a way to perform Bayesian inference in complex models whose likelihood is not available in closed form. Developed by Marjoram, Molitor, Plagnol, and Tavare (2003) it gained wide acceptance in the statistical community (Toni et al., 2009). Suppose the observed data, $\mathbf{X} \in \mathbb{R}^d$ $\mathbf{X} = \{\mathbf{X}_t, t = 1, ..., T\}$ has been generated from a distribution $F(\cdot; \boldsymbol{\theta})$, where $\boldsymbol{\theta}$ is a parameter vector. When the likelihood function is not available, ABC in its original form generates $\boldsymbol{\theta}$ from the prior, and artificial data $\tilde{\mathbf{X}}_{\boldsymbol{\theta}} = \{\tilde{\mathbf{X}}_{\boldsymbol{\theta},t}, t = 1, ..., T\} \sim F(\cdot; \boldsymbol{\theta})$. If $\tilde{\mathbf{X}}_{\boldsymbol{\theta}} = \mathbf{X}$ then the parameter is accepted. Since the equality $\tilde{\mathbf{X}}_{\boldsymbol{\theta}} = \mathbf{X}$ has measure zero, the ABC method has been modified as follows. Suppose $\mathbf{S}(\mathbf{X})$ is a vector of summary statistics, $\mathbf{S}: \mathbb{R}^d \to \mathbb{R}^s$ and $\mathbf{S}(\tilde{\mathbf{X}}_{\boldsymbol{\theta}})$ is the same set computed from the artificial data. Again, a parameter vector $\boldsymbol{\theta}$ is generated from the prior, and it is accepted if $D(\mathbf{S}(\mathbf{X}), \mathbf{S}(\tilde{\mathbf{X}}_{\boldsymbol{\theta}})) \leq \varepsilon$, where $D$ is a certain distance function and $\varepsilon > 0$ is a constant. To avoid drawing from the prior the following Metropolized version of ABC is often used (Plagnol and Tavare, 2004): Suppose $\boldsymbol{\theta} \sim Q(\cdot; \boldsymbol{\theta}^o)$, where $\boldsymbol{\theta}^o$ denotes the currently available draw. If $D(\mathbf{S}(\mathbf{X}), \mathbf{S}(\tilde{\mathbf{X}}_{\boldsymbol{\theta}})) \leq \varepsilon$ then accept the draw with probability $A(\boldsymbol{\theta}, \boldsymbol{\theta}^o) = \min\left\{1, \frac{p(\boldsymbol{\theta}) q(\boldsymbol{\theta}; \boldsymbol{\theta}^o)}{p(\boldsymbol{\theta}^o) q(\boldsymbol{\theta}^o; \boldsymbol{\theta})}\right\}$, where $q$ is the density associated with the measure $Q$.

The problem in ABC is to select the summary statistics and the constant $\varepsilon$. In complex models the choice of summary statistics is ad hoc to a certain extent and it is based on whatever is known about the model. As Wilkinson (2008) argues: "*It cannot be known whether these summaries are sufficient for the data, and so in most cases the use of summaries means that there is another layer of approximation.*" However, Wilkinson (2008) also shows that when the summaries are sufficient statistics then ABC provides exact results. See also Wegmann, Leuenberger, and Excoffier (2009).

*In the context of stable distributions the natural set of sufficient statistics is the characteristic function*, defined by $\varphi(\tau) = \int \exp(\iota\tau y) f(y) dy$, where $\iota = \sqrt{-1}$, $\tau \in \mathbb{R}$. The empirical characteristic function is defined as $\hat{\varphi}(\tau) = n^{-1} \sum_{t=1}^{n} \exp(\iota\tau y_t)$. For any simulated data set $\tilde{Y}(\theta) = (\tilde{y}_t(\theta), t = 1, ..., n)$, the empirical characteristic function can be computed as $\tilde{\varphi}(\tau) = n^{-1} \sum_{t=1}^{n} \exp(\iota\tau \tilde{y}_t(\theta))$. In the simplest case the characteristic function of the symmetric stable distributions is $\varphi(\tau) = \exp(-|\tau|^{\alpha})$, where $\alpha \in (0, 2]$ is the characteristic exponent.

Clearly we accept a parameter draw $\boldsymbol{\theta}$ from the prior if a measure of distance $d(\hat{\varphi}, \overline{\varphi}) \leq \varepsilon$. Various measures are available, for example the $L_{\infty}$-distance between the log characteristic functions, $d(\hat{\varphi}, \overline{\varphi}) = \max_{(\tau_i, i=1,...,I)} \left|\log \hat{\varphi}(\tau_i) - \log \overline{\varphi}(\tau_i)\right|$. Since the function $\varphi$ completely characterizes the distribution we have in fact a complete set of sufficient statistics. In the more general case with a location ($\mu$) and scale parameter ($\sigma$) the log characteristic function of symmetric stable distributions is:

$$\log \varphi(\tau) = \iota\mu\tau - \sigma|\tau|^{\alpha}, \text{ for all } \tau \in \mathbb{R}.$$

Suppose we propose a draw for $\boldsymbol{\theta} = [\alpha, \beta, \mu, \sigma]'$. In standard MCMC, the draw should be accepted with probability: $A(\boldsymbol{\theta}, \boldsymbol{\theta}^o) = \min\left\{1, \frac{p(\boldsymbol{\theta}|\mathbf{Y})q(\boldsymbol{\theta};\boldsymbol{\theta}^o)}{p(\boldsymbol{\theta}^o|\mathbf{Y})q(\boldsymbol{\theta}^o;\boldsymbol{\theta})}\right\}$, where $p(\boldsymbol{\theta}|\mathbf{Y}) \propto p(\boldsymbol{\theta})\prod_{t=1}^{T}f(y_t;\alpha,\beta,\mu,\sigma)$, $f(y_t;\alpha,\mu,\sigma)$ denotes the density of stable laws, $\mathscr{S}_{\alpha,\beta}(\mu,\sigma)$, and $\boldsymbol{\theta} \sim Q(\cdot;\boldsymbol{\theta}^o)$. Clearly, the obstacle is that the density is not available in closed form. However, the characteristic function is available in closed form and it can be computed easily, while simulating random variables $Y \sim S_{\alpha,\beta}(\mu,\sigma)$ is straightforward (Chambers, Mallows, and Stuck, 1976, but see also Modarres and Nolan, 1994, and Weron, 1996).

## 5. Multivariate stable distributions

Suppose $\mathbf{X} \in \mathbb{R}^d$ is a vector of multivariate $\alpha$-stable random variables, with characteristic exponent $0 < \alpha \leq 2$. Its characteristic function is $\varphi_{\mathbf{X}}(\boldsymbol{\tau}) = \mathbb{E}\exp(\iota\langle\mathbf{X},\boldsymbol{\tau}\rangle) = \exp(-I_{\mathbf{X}}(\boldsymbol{\tau}) + \iota\langle\boldsymbol{\mu},\boldsymbol{\tau}\rangle)$, where $\langle\mathbf{a},\mathbf{b}\rangle = \mathbf{a}'\mathbf{b}$ denotes the inner product, and

$$I_{\mathbf{X}}(\boldsymbol{\tau}) = \int_{\mathbb{S}^{d-1}} \psi_\alpha(\langle\boldsymbol{\tau},\mathbf{s}\rangle)\Gamma(d\mathbf{s}), \tag{6}$$

where $\mathbb{S}^d$ is the boundary of the unit sphere in $\mathbb{R}^d$, $\mathbb{S}^d = \{\mathbf{u} \in \mathbb{R}^d : \|\mathbf{u}\| = 1\}$, $\Gamma$ is a finite Borel measure of the vector $\mathbf{X}$, called the spectral measure, $\boldsymbol{\mu}$ is a parameter vector, and the function $\psi$ is defined as follows:

$$\psi_\alpha(u) = \begin{cases} |u|^\alpha(1 - \iota\,\mathrm{sgn}(u)\tan\frac{\pi\alpha}{2}), & \alpha \neq 1, \\ |u|(1 + \iota\frac{2}{\pi}\mathrm{sgn}(u)\log u), & \alpha = 1. \end{cases} \tag{7}$$

Press (1972) attempted to define a multivariate $\alpha$-stable distribution without using the spectral measure $\Gamma$. Later on Paulauskas (1976) provided some corrections as not all $\alpha$-stable distribution can be represented using Press' (1972) characteristic function. Chen and Rachev (1995) in an interesting paper provided estimates of $\alpha$ and the spectral measure as well as applications to stable portfolia. It is notable that the projection of $\mathbf{X}$ on $\boldsymbol{\tau}$, viz. $\langle\mathbf{X},\boldsymbol{\tau}\rangle$ has a univariate stable distribution whose characteristic function is $\mathbb{E}\exp(\iota\langle\mathbf{X},\boldsymbol{\tau}\rangle u) = \exp(-I_{\mathbf{X}}(u\boldsymbol{\tau}))$. Suppose now that the spectral measure is approximated by a discrete measure,

$$\Gamma(d\mathbf{s}) = \sum_{j=1}^{J}\gamma_j\delta_{\{\mathbf{s}_j\}}(\mathbf{s}), \tag{8}$$

where $\gamma_j > 0$, $\mathbf{s}_j \in \mathbb{S}^{d-1}$, $j = 1,...,J$, and $\delta$ denotes Dirac's *delta*. Since $\varphi_{\mathbf{X}}(\boldsymbol{\tau}) = \exp\left(-\int_{\mathbb{S}^{d-1}}\psi_\alpha(\langle\boldsymbol{\tau},\mathbf{s}\rangle)\Gamma(d\mathbf{s})\right)$ we obtain

$$\varphi_{\mathbf{X}}(\boldsymbol{\tau}) = \exp\left(-\sum_{j=1}^{J}\gamma_j\psi_\alpha(\langle\boldsymbol{\tau},\mathbf{s}_j\rangle)\right). \tag{9}$$

In this case, for $\alpha \neq 1$, it can be shown that $\mathbf{X} \stackrel{d}{=} \sum_{j=1}^{J}\gamma_j^{1/\alpha}\mathscr{Z}_j\mathbf{s}_j$, where $\mathscr{Z}_j \sim iid\mathscr{S}_{\alpha,1}(0,1)$, $j = 1,...,J$, see Modarres and Nolan (1994). The interpretation is that a multivariate $\alpha$-stable random vector can be represented as a finite mixture of univariate $\alpha$-stable variates which are totally skewed to the right (that is, they have skewness coefficients $\beta = 1$). For $\alpha = 1$ we have $\mathbf{X} \stackrel{d}{=} \sum_{j=1}^{J}\gamma_j^{1/\alpha}\left(\mathscr{Z}_j + \frac{2}{\pi}\log\gamma_j\right)\mathbf{s}_j$. To proceed, it is clear that if the spectral measure is discrete, we have: $\varphi_{\mathbf{X}}(\boldsymbol{\tau}) = \exp\left(-\sum_{j=1}^{J}\gamma_j\psi_\alpha(\langle\boldsymbol{\tau},\mathbf{s}_j\rangle)\right)$ and therefore:

$$-\log\varphi_{\mathbf{X}}(\boldsymbol{\tau}_j) = \sum_{j=1}^{J}\gamma_j\psi_\alpha(\langle\boldsymbol{\tau},\mathbf{s}_j\rangle), \quad j = 1,...,J,$$

from which we obtain the following system of linear equations:

$\mathbf{I} = \boldsymbol{\Phi}\boldsymbol{\gamma}$, where $\mathbf{I} = [-\log\varphi_{\mathbf{X}}(\boldsymbol{\tau}_1),...,-\log\varphi_{\mathbf{X}}(\boldsymbol{\tau}_J)]' \triangleq [I_{\mathbf{X}}(\boldsymbol{\tau}_1),...,I_{\mathbf{X}}(\boldsymbol{\tau}_J)]'$, $\boldsymbol{\Phi} = [\Phi_{ij}]$, $\Phi_{ij} = \psi_\alpha(\boldsymbol{\tau}'_i\mathbf{s}_j)$, $i,j = 1,...,J$, and $\boldsymbol{\gamma} = [\gamma_1,...,\gamma_J]$. One can then obtain $\boldsymbol{\gamma} = \boldsymbol{\Phi}^{-1}\mathbf{I}$. In practice, the system of equations suffers from singularities and the estimates of **γ** are not always nonnegative. McCulloch has proposed the use of quadratic programming imposing the nonnegativity and reports that, at least in small dimensions, the procedure works well.

## 6. Asymptotic Normal Form Quasi - Likelihoods

### 6.1 Introduction

Feuerverger and Mureika (1977) and Feuerverger and McDunnough (1981a,b) pioneered the so called Asymptotically Normal Form (ANF) for stable laws[6]. Given the characteristic function $\varphi(\tau;\theta) = \exp(\iota\mu\tau - \sigma|\tau|^\alpha)$, where the parameter vector is $\boldsymbol{\theta} = [\mu,\sigma,\alpha]'$, and its empirical counterpart, $\hat{\varphi}(\tau;\boldsymbol{\theta}) = n^{-1}\sum_{t=1}^{n}\exp(\iota\tau u_t)$, with $u_t = (Y_t - \mu)/\sigma$, define $z_\theta(\tau) = \begin{bmatrix} \Re(\tau;\theta) \\ \Im\varphi(\tau;\theta) \end{bmatrix}$, where $\Re$ denotes the real part, $\Im$ denotes the imaginary part of a complex number, also $\hat{z}_\theta(\tau) = \begin{bmatrix} \Re\hat{\varphi}(\tau;\theta) \\ \Im\hat{\varphi}(\tau;\theta) \end{bmatrix}$. Since $\Psi(\tau) = n^{1/2}(\varphi(\tau;\boldsymbol{\theta}) - \hat{\varphi}(\tau;\boldsymbol{\theta}))$ is asymptotically normal at a finite number of points[7] $\boldsymbol{\tau} = (\tau_i, i=1,...,I)$ its covariance matrix is $\boldsymbol{\Sigma} = \mathrm{cov}(\Psi(\boldsymbol{\tau})) = \varphi(\boldsymbol{\tau} - \boldsymbol{\tau}') - \varphi(\boldsymbol{\tau})\varphi(\boldsymbol{\tau})'$. This expression can be written in a simpler form. If $\boldsymbol{\Sigma} = \begin{bmatrix} \mathbf{A} & \mathbf{B} \\ \mathbf{B} & \mathbf{C} \end{bmatrix}$, then we have:

$$\mathbf{A} = \tfrac{1}{2}[\Re\varphi(\tau-\tau') + \Re\varphi(\tau+\tau')] - \Re\varphi(\tau)\Re\varphi(\tau)',$$
$$\mathbf{B} = \tfrac{1}{2}[\Im\varphi(\tau-\tau') + \Im\varphi(\tau+\tau')] - \Re\varphi(\tau)\Im\varphi(\tau)', \qquad (10)$$
$$\mathbf{C} = \tfrac{1}{2}[\Im\varphi(\tau-\tau') - \Im\varphi(\tau+\tau')] - \Im\varphi(\tau)\Im\varphi(\tau)'.$$

Apparently, matrix $\boldsymbol{\Sigma} = \boldsymbol{\Sigma}_\theta$ depends on the parameter vector $\boldsymbol{\theta}$. The ANF of the log likelihood is:

$$\ln L = -\tfrac{1}{2}\ln|\boldsymbol{\Sigma}_\theta| - \tfrac{n}{2}(z_\theta(\boldsymbol{\tau}) - \hat{z}_\theta(\boldsymbol{\tau}))' \boldsymbol{\Sigma}_\theta^{-1} (z_\theta(\boldsymbol{\tau}) - \hat{z}_\theta(\boldsymbol{\tau})). \qquad (11)$$

Concentration with respect to $\boldsymbol{\Sigma}_\theta$ is not possible. The substantive issue in maximization of the ANF (or just the second term, a procedure known as min-Q) is the selection of the grid $\boldsymbol{\tau} = (\tau_i, i=1,...,I)$. Koutrouvelis (1980) and subsequently Pourahmadi (1987) favor a grid of the form $\tau_k = \pi k / 25$ for $k = 0, \pm 1, ..., \pm K$. Based on simulations by Koutrouvelis (1980) it seems that the optimal value of K is 10 to 15 as α decreases from 1.9 to 1.1. Pourahmadi (1987) showed that for a distribution whose support is $(-\Lambda, \Lambda)$ the rule $\tau_k = \pi k / \Lambda$, $k = 0, \pm 1, ..., \pm K$, is optimal in the sense that all other values of the characteristic function can be reconstructed (Pourahmadi, 1987, Theorem 4.3, p. 355). In this sense Koutrouvelis (1980) assumes that $\Lambda = 25$ for all stable laws[8].

When the distribution is not bounded the problem reduces to finding the period of the function $\hat{\varphi}(\tau;\theta)$, $-\infty < \tau < \infty$. The period is shown to be $p \approx 2\pi n^{-1}\sum_{t=1}^{n}|Y_t|^{-1} \triangleq 2\pi / H$, when $\{Y_t, t=1,...,n\}$ is a random sample and $H$ is the harmonic mean of the observations. Then, for a given one may set $K = [2\Lambda / H] \geq 2$, and $[\ ]$ denotes the integer part. Another approach is to set $K = [\Lambda A / \pi]$, where $A$ is the first positive zero of the characteristic function: $A = \inf\{\tau : \hat{\varphi}(\tau) = 0\}$, and the empirical characteristic function $\hat{\varphi}(\tau)$ does not depend on other parameters.

---

[6] For a more recent survey, see Yu (2004).
[7] See also Knight and Yu (2002) and Xu and Knight (2010, proposition 2, p. 28).
[8] Madan and Seneta (1987) also favour values of **τ** concentrated around the origin which, in the case of characteristic functions, is crucial for determining tail behavior.

The problem of choosing a grid can be bypassed if one uses a continuum of moments. Leitch and Paulson (1975) were the first to propose the following estimation procedure: $\min_{\boldsymbol{\theta}} : \int_{-\infty}^{\infty} |\varphi(\tau;\boldsymbol{\theta}) - \hat{\varphi}(\tau)|^2 \exp(-\tau^2/2) d\tau$, which can be solved using Hermitian quadrature. Tran (1998) used the second term in the ANF to estimate a mixture of normal distributions. Given the endpoint of the grid for $\tau$, Tran (1998) chose the uniform stepsize (say $\Delta$) which minimizes the determinant of the asymptotic covariance matrix. Many authors, and more recently Carrasco and Florens (2002) noted that the covariance matrix $\boldsymbol{\Sigma}_{\boldsymbol{\theta}}$ becomes singular as the grid becomes fine and extended, a condition which is necessary in order for the ANF to get arbitrarily close to the Cramer-Rao bound. The idea of Carrasco and Florens (2000, 2002) was to consider continuous values of $\tau$ as in Leitch and Paulson (1975) and solve $\min_{\boldsymbol{\theta}} : \int_{-\infty}^{\infty} |\varphi(\tau;\boldsymbol{\theta}) - \hat{\varphi}(\tau)|^2 g(\tau) d\tau$, where $g(\tau)$ is a continuous weighting function or alternatively follow Feuerverger and Mureika (1977) and Feuerverger and McDunnough (1981a,b) and solve $\int_{-\infty}^{\infty} w(\tau)(\varphi(\tau;\boldsymbol{\theta}) - \hat{\varphi}(\tau)) d\tau = 0$, where $w(\tau)$ is also a weighting function. Feuerverger and McDunnough (1981a) show that the optimal weight function is $w^*(\tau) = \frac{1}{2\pi} \int_{-\infty}^{\infty} \exp(-\iota\tau Y) \frac{\partial \log f(Y;\theta)}{\partial \theta} dY$. Since the density, $f$, is unknown, Carrasco and Florens (2002) propose the use of GMM with a continuum of moments whose kernel is $g(\tau,s) = \varphi(\tau,s) - \varphi(\tau)\varphi(s)$, as one would expect from the asymptotic covariance of the ANF procedure[9].

From the ANF, consider the likelihood function:

$$L(\boldsymbol{\theta}; \mathbf{Y}, \boldsymbol{\tau}) = |\boldsymbol{\Sigma}_{\boldsymbol{\theta},\boldsymbol{\tau}}|^{-1/2} \exp\left[-\frac{n}{2}(z_{\boldsymbol{\theta}}(\boldsymbol{\tau}) - \hat{z}_{\boldsymbol{\theta}}(\boldsymbol{\tau}))' \boldsymbol{\Sigma}_{\boldsymbol{\theta},\boldsymbol{\tau}}^{-1} (z_{\boldsymbol{\theta}}(\boldsymbol{\tau}) - \hat{z}_{\boldsymbol{\theta}}(\boldsymbol{\tau}))\right], \quad (12)$$

where $\mathbf{Y} = [Y_1,...,Y_n]'$ denotes the data and $\boldsymbol{\Sigma}_{\boldsymbol{\theta},\boldsymbol{\tau}}$ makes explicit the dependence of the covariance matrix on the grid. For a fixed grid one can consider various MCMC methods to derive dependent draws that converge in distribution to the posterior, whose kernel is given by $p(\boldsymbol{\theta} | \mathbf{Y}) \propto L(\boldsymbol{\theta}; \mathbf{Y}) p(\boldsymbol{\theta})$, where $p(\boldsymbol{\theta})$ denotes the prior. This posterior does not overcome the problems associated with the choice of grid. Suppose, in fact, we have a prior over the grid, so that the joint prior is $p(\boldsymbol{\theta}, \boldsymbol{\tau}) = p(\boldsymbol{\theta}) p(\boldsymbol{\tau})$. Then the new posterior becomes:

$$p(\boldsymbol{\theta}, \boldsymbol{\tau} | \mathbf{Y}) \propto L(\boldsymbol{\theta}; \mathbf{Y}, \boldsymbol{\tau}) p(\boldsymbol{\theta}) p(\boldsymbol{\tau}).$$

Clearly, if we think of the prior $p(\boldsymbol{\tau})$ as a weight function then $p(\boldsymbol{\theta} | \mathbf{Y}) = \int p(\boldsymbol{\theta}, \boldsymbol{\tau} | \mathbf{Y}) d\boldsymbol{\tau} = p(\boldsymbol{\theta}) \int L(\boldsymbol{\theta}; \mathbf{Y}, \boldsymbol{\tau}) p(\boldsymbol{\tau}) d\boldsymbol{\tau}$. Assuming that $p(\boldsymbol{\theta}) = \mathbb{I}(\boldsymbol{\theta} \in \Theta)$, where $\Theta$ is the parameter space, then $p(\boldsymbol{\theta} | \mathbf{Y}) = \int L(\boldsymbol{\theta}; \mathbf{Y}, \boldsymbol{\tau}) p(\boldsymbol{\tau}) d\boldsymbol{\tau}$, that is the integral of the ANF likelihood with respect to the "weight function", $p(\boldsymbol{\tau})$. Of course the grid has to satisfy some a priori reasonable properties. One of them is that 0 is a point in the grid, that the grid is symmetric and therefore we can set $\boldsymbol{\tau} = \{0, \pm\tau_1, ..., \pm\tau_K\}$, with $\tau_1 < ... < \tau_K$.

In this study it is desirable to remove the assumption that the points of the grid are equispaced and of course we do not wish to impose a priori the assumption of a fixed number of grid points, $K$. Treating $\tau_1 > 0$ as parameter, we set $\tau_j = \tau_{j-1} + h_j^2$, $j = 2,...,K$, where $\mathbf{h}^{(K)} = [h_2,...,h_K]'$ is a vector of K-1 *free* parameters. Of course we can also define $\tau_j = \tau_{j-1} + h_j^2$, $j = 1,2,...,K$, with $\tau_0 = 0$, and $\tilde{\mathbf{h}}^{(K)} = [h_1,...,h_K]'$, with $h_1 = 0$. Then we can place a prior on the free parameters $\left[\boldsymbol{\theta}, \tau_1, \mathbf{h}^{(K)\prime}\right]'$.

---

[9] Xu and Knight (2010) apply the continuous version of Leitch and Paulson (1975) using a weighting function which has the form $\exp(-b^2/2)$, see p. 28, under the name CECF. The authors apply the procedure to estimation of finite mixture of normals and find that the optimal b ranges from 1.22 to 2.15, see Table 2b.

Treating the problem in this manner, it then becomes clear that the grid points $\boldsymbol{\tau}$ can be treated as latent variables and in that sense the posterior distribution is augmented using these grid points, in the sense of Tanner and Wong (1987)[10]. The prior is specified as follows:

$$p\left(\boldsymbol{\theta},\tau_1,\mathbf{h}^{(K)},K\right) \propto p(\boldsymbol{\theta})\,p(\tau_1)\,p\left(\mathbf{h}^{(K)}\mid K\right)p(K), \qquad (13)$$

where $p(\boldsymbol{\theta}) = p(\mu,\sigma,\alpha) \propto \sigma^{-1}\mathbb{I}(0 < \alpha \leq 2,\ -1 \leq \beta \leq 1)$, $p(\tau_1) \propto \tau_1^{-1}$, $\mathbf{h}^{(K)} \sim \mathcal{N}_{K-1}\left(\mathbf{0}_{K-1},\ \omega^2 \mathbf{I}_{K-1}\right)$,

and $p(K)$ denotes the prior on the number of grid points (assuming the first one is always zero). Following Koutrouvelis (1980) it seems *a priori* reasonable to center the grid points around $\tau_k = \pi k / \Lambda$, for $k = 1,\ldots,K$. The points of the grid are equispaced with length $\Delta = \pi / \Lambda$. With $\Lambda = 25$ (typical for stable laws with α great than 1.1) this suggests uniform length about $\Delta = 0.12$. Therefore, a prior with ω=0.24 would be relatively uninformative relative to the likelihood. Regarding the number of grid points, $K$, a uniform prior over the set of integers $\{1,\ldots,50\}$ covers well the optimal values reported by Koutrouvelis (1980) and suggested by Pourahmadi (1987). An alternative is a Poisson distribution with parameter $\lambda = 15$, based again on the results of Koutrouvelis (1980, 1981) and Feuerverger and McDunnough (1981a,b).

As reported by other authors the main problem with the ANF is the bad conditioning of the matrix $\boldsymbol{\Sigma}_{\boldsymbol{\theta},\boldsymbol{\tau}}$. To overcome the problem we first regularize the matrix using $\tilde{\boldsymbol{\Sigma}}_{\boldsymbol{\theta},\boldsymbol{\tau}} = \boldsymbol{\Sigma}_{\boldsymbol{\theta},\boldsymbol{\tau}} + \varepsilon \mathbf{I}_{2K+1}$, where ε is a small constant. Second, we consider the singular value decomposition $\tilde{\boldsymbol{\Sigma}}_{\boldsymbol{\theta},\boldsymbol{\tau}} = \mathbf{USV}'$, where $\mathbf{S}$ is a diagonal matrix with the singular values, $[s_i,\ i=1,\ldots,2K+1]$, along its diagonal. If any element is zero it is replaced by $10^{-6}$. Then, the inverse matrix is $\boldsymbol{\Sigma}_{\boldsymbol{\theta},\boldsymbol{\tau}}^{-1} = \mathbf{VS}^{-1}\mathbf{U}$, and $\mathbf{S}^{-1} = diag[1/s_i,\ i=1,\ldots,2K+1]\mathbf{I}_{2K+1}$. Moreover, $\ln\left|\boldsymbol{\Sigma}_{\boldsymbol{\theta},\boldsymbol{\tau}}\right| = \sum_{i=1}^{2K+1}\log s_i$. The singular value decomposition was proved extremely fast, reliable and efficient in computing values of the log posterior without numerical problems.

## 6.2 Refinements of the Asymptotic Normal Form

The ANF has an asymprotic justification and can be used profitably in "large samples". Of course the notion of "large samples" is related both to the sample size as well as the information from particular samples. Consider the ANF of the likelihood in (12). Following the literature (Kohn, Li, and Villani, 2010) it is possible to use refinements based on mixtures of multivariate Student – $t$ distributions:

$$\tilde{L}_{St}(\boldsymbol{\theta};\mathbf{Y},\boldsymbol{\tau}) = \sum_{g=1}^{G} \pi_g \frac{\Gamma\left(\frac{\nu_g+p}{2}\right)}{\Gamma\left(\frac{\nu_g}{2}\right)(\pi\nu_g)^{p/2}} h_g^{-p/2}\left|\boldsymbol{\Sigma}_{\boldsymbol{\theta},\boldsymbol{\tau}}\right|^{-1/2}\left[1 + \frac{\nu_g}{2h_g}\left(z_{\boldsymbol{\theta}}(\boldsymbol{\tau}) - \hat{z}_{\boldsymbol{\theta}}(\boldsymbol{\tau})\right)'\boldsymbol{\Sigma}_{\boldsymbol{\theta},\boldsymbol{\tau}}^{-1}\left(z_{\boldsymbol{\theta}}(\boldsymbol{\tau}) - \hat{z}_{\boldsymbol{\theta}}(\boldsymbol{\tau})\right)\right]^{-\frac{\nu_g+p}{2}}, \quad (14)$$

where $\pi_g$ are mixing probabilities, $G$ denotes the number of groups, $p$ denotes the dimensionality of the parameter vector, $\nu_g$ are group-specific degrees of freedom, and $h_g$ are different scaling constants of the "basic" scale matrix $\boldsymbol{\Sigma}_{\boldsymbol{\theta},\boldsymbol{\tau}}$. We call this the ***Asymptotic Student-t Mixture Form*** (AtMF). Another possibility is, of course, an ***Asymptotic Normal Mixture Form*** (ANMF):

$$\tilde{L}_N(\boldsymbol{\theta};\mathbf{Y},\boldsymbol{\tau}) = \sum_{g=1}^{G} \pi_g h_g^{-p/2}\left|\boldsymbol{\Sigma}_{\boldsymbol{\theta},\boldsymbol{\tau}}\right|^{-1/2}\exp\left[-\frac{n}{2h_g}\left(z_{\boldsymbol{\theta}}(\boldsymbol{\tau}) - \hat{z}_{\boldsymbol{\theta}}(\boldsymbol{\tau})\right)'\boldsymbol{\Sigma}_{\boldsymbol{\theta},\boldsymbol{\tau}}^{-1}\left(z_{\boldsymbol{\theta}}(\boldsymbol{\tau}) - \hat{z}_{\boldsymbol{\theta}}(\boldsymbol{\tau})\right)\right], \qquad (15)$$

where $\tilde{L}_N(\boldsymbol{\theta};\mathbf{Y},\boldsymbol{\tau})$ denotes the "refined" normal – mixture likelihood and, again, the basic scale matrix of the ANF is multiplied by the constants $h_g$, $g=1,\ldots,G$.

---

[10] Consider $\varphi(\tau;\boldsymbol{\theta}) = \hat{\varphi}(\tau;\boldsymbol{\theta}) + \varepsilon(\tau;\boldsymbol{\theta})$, $\varepsilon(\tau;\boldsymbol{\theta}) \sim N\left(0,\boldsymbol{\Sigma}_{\boldsymbol{\theta}}(\tau)\right)$. This formulation, can be considered as a Gaussian process, $\varphi(\tau;\boldsymbol{\theta}) \sim \mathcal{G}\left(\hat{\varphi}(\tau;\boldsymbol{\theta}),\boldsymbol{\Sigma}_{\boldsymbol{\theta}}(\tau)\right)$, where $\hat{\varphi}(\tau;\boldsymbol{\theta})$ is the mean function, and $\boldsymbol{\Sigma}_{\boldsymbol{\theta}}(\tau)$ is the covariance matrix of the process. It is not clear how this idea can be used to facilitate Bayesian analysis unless the investigator is willing to assume that her data is "close" to a stable law but not exactly so.

To determine the usefulness of this approach we consider various sample sizes (n=50, 200, 500 and 2,000 and 5000) and various values of the characteristic exponents ($\alpha$=1.10, and 1.50). In the table below we report the modal values of $G$, the average values of scaling constants $h_g$ for fixed grids and optimal grids. We have examined 1,000 data sets for stable distributions with $\alpha=1.10$ and $\alpha=1.50$ and a case with $\beta=-0.20$ and $\alpha=1.50$ (which is empirically relevant in many applications). For each of the 1,000 data sets we have implemented the ANF procedure with various numbers of components ($G$) letting the configuration of the grid ($\tau$) and their number be parameters[11] (along with the degrees of freedom of the Student-$t$, viz. $\nu_g$) determined through MCMC. For each value of $G$ the marginal likelihood has been computed[12] and the optimal $G$ as well as the values of the scaling constants ($h_g$, $g=1,...,G$) were computed and saved. The optimal value of $G$ was selected to maximize the marginal likelihood. Finally, the modal values of $G$ and the medians of the scaling constants were computed and reported in Table 1.

Table 1. Optimal values of G, and scaling constants of the ANF for normal and Student-$t$ mixtures

|  | Modal G (normal mixture) | Values of $h_g$ | Modal G (Student-$t$ mixture) | Values of $h_g$ |
|---|---|---|---|---|
|  | Characteristic exponent, α=1.10, skewness β=0 | | | |
| n=50 | 3 | 0.73  1.08  1.78 | 3 | 0.35  1.14  2.12 |
| n=200 | 2 | 0.81  1.45 | 3 | 0.65  1.21  1.65 |
| n=500 | 1 | 0.87  1.21 | 2 | 0.63  1.45 |
| n=2000 | 1 | 1.12 | 1 | 1.43 |
| n=5000 | 1 | 1.04 | 1 | 1.12 |
|  | Characteristic exponent, α=1.50, skewness β=0 | | | |
| n=50 | 3 | 0.83  1.08  1.65 | 3 | 0.65  1.12  1.77 |
| n=200 | 2 | 0.85  1.35 | 3 | 0.88  1.09  1.32 |
| n=500 | 1 | 0.91  1.21 | 2 | 0.83  1.12 |
| n=2000 | 1 | 1.07 | 1 | 1.15 |
| n=5000 | 1 | 1.01 | 1 | 1.07 |
|  | Characteristic exponent, α=1.50, skewness β=-0.20 | | | |
| n=50 | 4 | 0.43  0.80  1.08  1.65 | 4 | 0.35  0.72  1.42  2.11 |
| n=200 | 3 | 0.45  1.35  2.22 | 3 | 0.38  1.21  1.85 |
| n=500 | 1 | 1.25 | 2 | 0.33  1.45 |
| n=2000 | 1 | 1.12 | 1 | 1.48 |
| n=5000 | 1 | 1.05 | 1 | 1.12 |

The message is that for "large" sample sizes (typically, $n \geq 500$) one component of the ANF in normal or Student-$t$ mixtures turns out to be optimal and the scaling constants ($h_g$) approach unity quickly especially in samples close to 2,000 observations.

## 6.3 The projection method

Given $\tau \in \mathbb{S}^d$, we can consider the projections $\tau'\mathbf{X} = \langle \tau, \mathbf{X} \rangle$ when $\mathbf{X}$ follows a multivariate stable distribution. Its characteristic function is $\mathbb{E}\exp(\iota u \langle \tau, \mathbf{X}\rangle) = \exp(-I_{\mathbf{X}}(u\tau))$. The linear projection will be univariate stable with the same characteristic exponent, α, and the scale, skewness and location parameters are given by the following (Zolotarev, 1986, p.20, Cambanis and Miller, 1981 or Nagaev, 2000):

$$\sigma^\alpha(\tau) = \Re I_{\mathbf{X}}(\tau) = \int_{\mathbb{S}^d} \langle \tau, \mathbf{s} \rangle^\alpha \Gamma(d\mathbf{s}), \qquad (16a)$$

$$\beta(\tau) = \sigma^{-\alpha}(\tau) \int_{\mathbb{S}^d} |\langle \tau, \mathbf{s}\rangle|^\alpha \operatorname{sgn}(\langle \tau, \mathbf{s}\rangle) \Gamma(d\mathbf{s}) = -\operatorname{Im} I_{\mathbf{X}}(\tau) / \left(\sigma^{-\alpha}(\tau)\tan\tfrac{\pi\alpha}{2}\right) \text{ (for } \alpha \neq 1\text{), (16b)}$$

and
$$\mu(\tau) = 0 \text{ for } \alpha \neq 1. \qquad (16c)$$

Suppose we have a random sample $\mathbf{X}_1,...,\mathbf{X}_n \in \mathbb{R}^d$, and $\underset{(n \times d)}{\mathbf{X}} = [\mathbf{X}_1,...,\mathbf{X}_n]$. McCulloch (1994) suggests to use a grid $\tau_1,...,\tau_n \in \mathbb{S}^d$ to define the one – dimensional data set $\langle \tau_i, \mathbf{X}_1\rangle, \langle \tau_i, \mathbf{X}_2\rangle,...,\langle \tau_i, \mathbf{X}_n\rangle$ for each $i=1,...,N$. Scale $\sigma(\tau_i)$ and $\beta(\tau_i)$ can be estimated using an estimate of $I_{\mathbf{X}}(\tau)$, as $\hat{I}_{\mathbf{X}}(\tau_i) = \hat{\sigma}^\alpha(\tau_i)\left(1 - \iota\hat{\beta}(\tau_i)\right)\tan\tfrac{\pi\alpha}{2}$, for $\alpha \neq 1$.

---
[11] To minimize computational costs the ANF is analyzed first and the optimal values of $\tau$, are determined. Then we proceed as if the configuration of the grid is known for the refined ANFs.
[12] The marginal likelihood is computed using the Laplace approximation based on the ANF (DiCiccio *et al*, 1997).

Since there is also available an estimate $\hat{\alpha}(\boldsymbol{\tau}_i)$ in each direction, McCulloch (1994) suggests the mean, $\hat{\alpha} = N^{-1}\sum_{i=1}^{N}\hat{\alpha}(\boldsymbol{\tau}_i)$ as an overall estimate of the characteristic exponent. As $N$ is likely to be small the properties of this estimator in finite samples are unclear.

Since the projections $\langle \boldsymbol{\tau}_i, \mathbf{X}_t \rangle$, are univariate stable and $\boldsymbol{\tau}_1,...,\boldsymbol{\tau}_N \in \mathbb{S}^d$, we have that $\mathbf{z}_i \triangleq \mathbf{X}\boldsymbol{\tau}_i$ is an $n \times 1$ vector (for each $i=1,...,N$) consisting of realizations of independent univariate stable variates. They have scale $\sigma(\boldsymbol{\tau}_i)$, skewness $\beta(\boldsymbol{\tau}_i)$ and location $\mu(\boldsymbol{\tau}_i) = \langle \boldsymbol{\tau}_i, \boldsymbol{\mu} \rangle$, where $\boldsymbol{\mu} \in \mathbb{R}^d$ is the vector of location parameters of each multivariate stable $\mathbf{X}_t$, $t=1,...,n$. There is a variety of methods to estimate location, skewness and scale parameters for each linear projection. Part of the problem is that the information that $\alpha$ is the *same* for *all* projections is *not* taken into account.

To implement an ABC approach, we first consider the $\boldsymbol{\tau}_i$s as fixed to form the characteristic function of the sample $\mathbf{z}_i$, $i=1,...,N$, given by $\varphi_i(u) = \exp(-I_\mathbf{X}(u\boldsymbol{\tau}_i))$, $u \in \mathbb{R}$. The empirical equivalent is

$$\hat{\varphi}_i(u) = n^{-1}\sum_{t=1}^{n}\exp(-I_{\mathbf{X}_t}(u\boldsymbol{\tau}_i)), \text{ for each } i=1,...,N. \qquad (17)$$

Since the $\mathbf{z}_i$s are independent it follows that the characteristic function of $\mathbf{z}_1,...,\mathbf{z}_N$ is $\varphi(u) = \sum_{i=1}^{N}\exp(-I_\mathbf{X}(u\boldsymbol{\tau}_i))$, and the empirical equivalent is

$$\hat{\varphi}(u) = n^{-1}\sum_{i=1}^{N}\sum_{t=1}^{n}\exp(-I_{\mathbf{X}_t}(u\boldsymbol{\tau}_i)) = \sum_{i=1}^{n}\hat{\varphi}_i(u). \qquad (18)$$

Then we can follow precisely the same ABC methods that we have developed in connection with univariate stable distributions. Next, the the $\boldsymbol{\tau}_i$s and their weights, can be treated as latent variables in ABC. The "prior" of the latent parameters is assumed to be uniform over the sphere $\mathbb{S}^d$. For each $\boldsymbol{\tau}_i$ there is an associated point mass since $\Gamma(d\mathbf{s}) \triangleq \sum_{t=1}^{N}\gamma_t \delta_{\boldsymbol{\tau}_t}(\bullet)$. Drawing the latent variables is facilitated by using a Hit-and-Run algorithm (Belisle, Romeijn, and Smith, 1993) which is (and also turned out to be) ideally suited for this type of problem. There is another useful approach that can be used in this problem. For each linear projection we clearly have:

$$\underset{(d \times 1)}{\mathbf{X}\,\boldsymbol{\tau}_i} = \mu(\boldsymbol{\tau}_i) + \underset{(n \times 1)}{\mathbf{u}_i}, \quad i=1,...,N, \qquad (19)$$

where each component of $\mathbf{u}_i$ follows a univariate stable distribution with zero location, and scale and skewness given by $\sigma(\boldsymbol{\tau}_i)$, and $\beta(\boldsymbol{\tau}_i)$. In this representation the advantage is that we can treat explicitly the $\boldsymbol{\tau}_i$s as random coefficients, with probability mass $\gamma_i$. This suggests that factor models can be profitably used in connection with multivariate stable distributions[13], an issue that we take up formally in section 10 for both static and dynamic factor models.

The likelihood and the posterior are easy to derive, although they depend on computing n different univariate general stable distributions with the same exponent $\alpha$ but different skewness parameters. Posterior inference is possible using Buckle's (1995) MCMC scheme which depends on the representation of stable distributions as mixtures (see also Tsionas, 1999). Buckle's (1995) approach has to be modified to accommodate the different skewness parameters, but accommodation of different location and scale parameters is trivial.

A certain approximation results if we assume that $\mathbf{u}_i$ can be approximated by a finite mixture of normals. Then Bayesian inference via MCMC is straightforward. The problem is that the scale and skewness coefficients of each element of $\mathbf{u}_i$ are different, and that the approximating finite mixture parameters will depend themselves on the grid points, $\boldsymbol{\tau}_i$. If we were, in addition, to assume that the measure $\Gamma$ has finite support with unit masses at $\boldsymbol{\tau}_i \in \mathbb{S}^d$ and weights $\gamma_i$, then the problem would admit an easy solution. The procedure would directly produce certain approximations $\hat{\boldsymbol{\tau}}_i$ and also $\hat{\gamma}_i$.

An alternative approach is to use the ANF in connection with the multivariate stable distributions. The approach has the advantage that it extends itself in a straightforward manner from univariate to multivariate stable distributions. One may proceed either directly from the joint characteristic function of the sample or explicitly consider the characteristic function of linear projections of multivariate stable random variables. Since there are explicit expressions for both representations, implementation of ABC along with the ANF procedure is quite feasible and no more difficult than ABC-ANF in the univariate case, where we have shown that the procedure performs very well.

---

[13] This idea has been taken up by Broda et al (2012) with the difference that they used Independent Components Analysis.

## 6.4 Gaussian approximation of the measure

Instead of discretizing the measure $\Gamma(d\mathbf{s})$, we can certainly use other approximations. The most prominent is a multivariate normal distribution in which case $I_\mathbf{X}(\boldsymbol{\tau}) = \int_{\mathbb{S}^{d-1}} \psi_\alpha(\langle\boldsymbol{\tau},\mathbf{s}\rangle)\Gamma(d\mathbf{s})$ can be interpreted as the expectation of $\psi_\alpha(\langle\boldsymbol{\tau},\mathbf{s}\rangle)$, when $\mathbf{s} \sim N_d(\mathbf{0}, \omega^2 \mathbf{I})\,|\,\mathbf{s} \in \mathbb{S}^d$, and $\omega$ is a parameter. Part of the attraction of the Gaussian approximation is that we can avoid sampling the relatively troublesome parameters $\mathbf{s}_i, \gamma_i$, $i = 1,...,N$, and of course $N$ itself.

Given the Gaussian approximation it is straightforward to compute $I_\mathbf{X}(\boldsymbol{\tau}) \approx \mathbb{E}_{\Gamma(d\mathbf{s})} \psi_\alpha(\langle\boldsymbol{\tau},\mathbf{s}\rangle)$, where the expectation $\mathbb{E}_{\Gamma(d\mathbf{s})}$ denotes an expectation taken with respect to $\mathbf{s} \sim N_d(\mathbf{0}, \omega^2 \mathbf{I})\,|\,\mathbf{s} \in \mathbb{S}^d$, given the parameter $\omega$. In Table 2 below we report the required number of draws (with $\omega=1$) so that we get the (absolute value of the) expectation (a complex valued function) to within $10^{-3}$ for random linear combinations of $\mathbf{s}$ and $\boldsymbol{\tau}$. We have examined 1,000 random directions $\boldsymbol{\tau}$ in the interval [-1, 1]. Since the draws for $\mathbf{s}$ are heavily concentrated in the unit hypersphere as the dimensionality (d) of the problem increases beyond a certain point we do not need as many draws as in lower dimensions. For the empirically relevant values of $\alpha$ (in excess of, say, 1.50) no more than 100 draws would be quite sufficient in most dimensions.

Table 2. Number of draws required to get the expectation to precision $10^{-3}$.

| A | d=2 | d=5 | d=20 | d=50 | d=200 | d=1,000 |
|---|---|---|---|---|---|---|
| 1.10 | 222 | 3,418 | 4,922 | 12,248 | 3,418 | 1,978 |
| 1.30 | 222 | 89 | 460 | 552 | 1,373 | 21 |
| 1.50 | 107 | 35 | 52 | 25 | 62 | 21 |
| 1.70 | 107 | 36 | 52 | 25 | 36 | 36 |
| 1.90 | 52 | 36 | 52 | 25 | 36 | 36 |

## 6.5 Spherical Harmonic Analysis and measure approximation

Pivato (2001) and Pivato and Seco (2003) proposed an approach that can be used to obtain a smooth estimate of the spectral measure. From the log characteristic function one can obtain the convolution:

$$\psi * \Gamma = \int_{\mathbb{S}^{d-1}} \psi_\alpha\left(\left\langle \boldsymbol{\tau}, \mathbf{s}^{-1} \right\rangle\right) \Gamma(d\mathbf{s}). \tag{20}$$

In dimensions d=2 or 4, $\mathbb{S}^{d-1}$ is a topological group and the convolution is well defined but unfortunately this is not so in other dimensions. First, the log characteristic function is expressed as a spherical Fourier series and second, to obtain the spectral measure one divides the spherical Fourier coefficients by certain constants $A_n$. Indeed, we have

$$\gamma = \sum_{n=1}^{\infty} \frac{1}{A_n} I_n \triangleq \sum_{n=0}^{\infty} \gamma_n,$$

$$A_n = \frac{(\psi * \zeta_n)(\mathbf{e}_1)}{\zeta_n(\mathbf{e}_1)}, \tag{21}$$

$$I_n \triangleq \mathcal{Z}_n * I, \ \mathcal{Z}_n : \mathbb{S}^{d-1} \to \mathbb{C}, \ \mathcal{Z}_n(\mathbf{s}, \sigma) = \zeta_n(e) \cdot \zeta_n(\langle \mathbf{s}, \sigma \rangle).$$

Moreover, $\zeta_n : \mathbb{S}^{d-1} \to \mathbb{C}$, are so called zonal harmonic polynomials. The eigenfunctions of the Laplacian operator are functions $\mathcal{E}_n(\mathbf{x}) = \exp(2\pi \cdot \boldsymbol{\iota} \cdot \langle \mathbf{n}, \mathbf{x} \rangle)$, where $\mathbf{x} \in [0,1)^d$, $\mathbf{n}$ is a d-dimensional vector of integers, and $\boldsymbol{\iota}$ is a vector in $\mathbb{C}^d$ whose elements are all equal to $\iota$. For the unit circle, $\mathbb{S}^1$, suppose polar coordinates are denoted by $(\sin\theta, \cos\theta)$, $\theta \in (0, 2\pi)$. For any complex valued function, $f$, we have[14] $\Delta_{\mathbb{S}^1} f = \frac{\partial^2 f}{\partial \theta^2}$.

In higher dimensions,

$$\Delta_{\mathbb{S}^d} f = \frac{\partial^2 f}{\partial \phi^2} + (d-1)\cot\phi \frac{\partial \phi}{\partial \theta} + \frac{1}{\sin^2\phi} \Delta_{\mathbb{S}^{d-1}} f,$$

given the diffeomorphism $(\mathbf{s}, \phi) \to (\cos\varphi, \sin\phi \cdot \mathbf{s})$. A complex number $\lambda$ is called an eigenvalue of the Laplacian if $\Delta f = -\lambda f$ for any complex, infinitely differentiable function $f$. The eigenfunctions of the Laplacian on $\mathbb{S}^{d-1}$ are called spherical harmonics. The zonal eigenfunctions $\zeta(\mathbf{x})$ can be defined through Gegenbeuaer polynomials of the form[15]

$$C_N^{(\nu)}(x) = \sum_{n=0}^{[N/2]} (-1)^n 2^{N-2n} c_{N,n}^{(\nu)} x^{N-2n},$$

where $c_{N,n}^{(\nu)} = \frac{\Gamma(\nu + n - N)}{\Gamma(\nu) n! (N - 2n)!}$, and $\nu = \frac{d}{2} - 1$.

Then

$$\zeta_N(\mathbf{x}) = C_N^{(\nu)}(x_1) / K_{N,\nu}, \text{ and } K_{N,\nu} = \left\| C_N^{(\nu)}(x) \right\|_2$$

for which there is an analytical expression (Proposition 11 in Pivato and Zeco, 2003; Abramowitz and Stegun, 1965, ch. 22, p. 773).

The deconvolution $\gamma = \sum_{n=1}^{\infty} \gamma_n$ is valid under the assumption that $\int_{\mathbb{S}^{d-1}} |\gamma(\mathbf{s})|^2 dV(\mathbf{s}) < \infty$, where $V[\mathbf{s}]$ is the standard volume measure, assuming $\int_{\mathbb{S}^{d-1}} dV(\mathbf{s}) = 1$. Notice that the decomposition is orthogonal since the eigenvalues of the Laplacian form an orthonormal system. The essential requirement of the spherical deconvolution is to compute the convolution with Gegenbauer polynomials, which involves an integration over $\mathbb{S}^{d-1}$. The problem has been considered by many authors including Roose and De Doncker (1981) who proposed a trapezoidal rule after a certain transformation. Therefore, such integrals can be evaluated accurately and effortlessly. We also refer to the informative paper of Bazant and Oh (1986).

---

[14] See also Gautier and Kitamura (2011), lemma 2.1 and subsequent discussion. Deconvolution on spheres arises naturally in the context of nonparametric estimation of a random coefficient binary choice model.

[15] The expressions are valid for d≥4.

## 6.6 A more explicit representation

Cheng and Rachev (1995) show that:

$$I_{\mathbf{X}}(\boldsymbol{\tau}) = \log \mathbb{E} \exp(\iota \langle \mathbf{X}, \boldsymbol{\tau} \rangle) = -|\boldsymbol{\tau}|^{\alpha} \int_0^{\pi} \int_0^{\pi} \ldots \int_0^{2\pi} |\cos(\boldsymbol{\tau}, \vartheta)|^{\alpha} \left(1 - \iota \operatorname{sgn} \cos(\boldsymbol{\tau}, \vartheta)\right) \tan \tfrac{\pi\alpha}{2} d\Gamma(\vartheta) + \iota \langle \boldsymbol{\mu}, \boldsymbol{\tau} \rangle, \quad (22) \quad \theta \in \mathbb{R}^d,$$

for $\alpha \neq 1$, where $|\cos(\boldsymbol{\tau}, \vartheta)| = \left(\prod_{i=1}^d \sin \phi_i \sin \vartheta_i\right) + \cos \phi_1 \cos \vartheta_1$, and $\boldsymbol{\tau} = [\rho \sin \phi_1 \ldots \phi_{d-1}, \rho \sin \phi_1 \ldots \cos \phi_{d-1}, \ldots, \rho \cos \phi_1]$.

Denoting $\rho = |\mathbf{X}|$, and $\Theta = [\vartheta_1, \ldots, \vartheta_{d-1}]'$, Cheng and Rachev (1995) show that the normalized spectral measure has density:

$$\gamma_n(\vartheta) = k^{-1} \sum_{t=1}^n \mathbb{I}\left(\Theta_t \leq \vartheta, \rho_t \geq \rho_{n-k+1:n}\right), \qquad (23)$$

where $\rho_{k:n}$ denotes the $k$-th order statistic of $(\rho_1, \ldots, \rho_n)$. Since the value of $k$ is unknown Cheng and Rachev (1995) recommend the value 570 which roughly corresponds to α=1.57. Clearly, it would be best for $k$ to be as large as possible ($1 \leq k \leq n$) since otherwise we discard a lot of information in the data. For almost sure convergence of $\gamma_n(\theta)$, we need $k / \log n \to \infty$ by their Lemma 2.1 (B).

The simple form of $\gamma_n(\vartheta)$ suggests that for a given value of k and given $\rho$ it would be possible to sample directly from the spectral measure as follows: Suppose $\rho_t \geq \rho_{n-k+1:n}$, and $|\Theta|_{\max}$ is the coordinate-wise maximum of $\{\Theta_t, t = 1, \ldots, n\}$. Then $\vartheta$ has a uniform distribution in the set $G(\vartheta) = \left\{\vartheta \in [0, \pi]^{d-1} : \vartheta \geq |\Theta|_{\max}\right\}$, viz. $\vartheta_i$ is uniform in $\left[|\Theta|_{\max}, \pi\right]$, for all $i = 1, \ldots, d-1$. The value of $\vartheta_d$ is obtained in the obvious way.

Suppose a sample $\{\vartheta^{(s)}, s = 1, \ldots, S\}$ is available. Then,

$$I_{\mathbf{X}}(\boldsymbol{\tau}) \approx -|\boldsymbol{\tau}^{(s)}|^{\alpha} \tan \tfrac{\pi\alpha}{2} S^{-1} \sum_{s=1}^S \left|\cos(\boldsymbol{\tau}^{(s)}, \vartheta^{(s)})\right|^{\alpha} \left(1 - \iota \operatorname{sgn} \cos(\boldsymbol{\tau}^{(s)}, \vartheta^{(s)})\right) + \iota \langle \boldsymbol{\mu}, \boldsymbol{\tau}^{(s)} \rangle, \qquad (24)$$

is an estimate of the log characteristic function and can be used to compare directly with the empirical characteristic function, and perform ABC inference. Since the value of $k$ is unknown this procedure has either to be repeated for different values of $k$ or treat it as a parameter. Suppose $\kappa = k/n \in (0,1)$ is the fraction of the sample that we use for tail estimation of the spectral measure.

To examine whether this procedure is reasonable the following computational experiment is used. Fix the dimension, $d$, and suppose the spectral measure, $\Gamma(\vartheta)$, is discrete with point masses 0.25 at 0°, 45°, 180° and 225° degrees[16] as in Byczowski, Nolan, and Rajput (1993, p. 29). Let $\kappa \sim Be\left(\tfrac{1}{2}, \tfrac{3}{2}\right)$. What would be the finite sample properties of the posterior mean of $\alpha$, for different values of the sample size and what is approximately the posterior distribution of $\kappa$? We have considered 1,000 different data sets from the multivariate stable distribution with the specified measure in dimensions $d = 2, 5$ and 10 and values of $\alpha = 1.10, 1.50$ and 1.75. We run the MCMC procedure for 60,000 iterations the first 10,000 of which were discarded, starting from the true values of the parameters. In Table 3 we report the posterior mean of $\alpha$ and the posterior mean of $\kappa$. The procedure is clearly biased towards larger values of $\alpha$. When $\alpha = 1.50$, for example, the results are close to normality and the posterior means of $\alpha$ range from 1.912 to 1.965 for d=10 and across sample sizes. As the sample size increases the procedure seems to perform slightly worse, so we decided not to use it further as it did not seem to be reliable. In addition we did observe that, *approximately*, $k$ seems to scale as $1.3(\log n)^2$. In this sense, Bayesian inference seems to scale $k$ so that the basic relation $k / \log n \to \infty$ is satisfied but other than that estimating correctly the characteristic exponent does not seem possible.

---

[16] We remind that degrees are radians multiplied by $180/\pi$.

Table 3. Simulation means of posterior means of $\alpha$ and $\kappa$.

| $\alpha$ | | n=100 | | n=500 | | n=1,000 | | n=2,000 | |
|---|---|---|---|---|---|---|---|---|---|
| 1.10 | d=2 | 1.365 | 0.30 | 1.316 | 0.12 | 1.362 | 0.08 | 1.377 | 0.05 |
| | d=5 | 1.378 | 0.32 | 1.400 | 0.11 | 1.345 | 0.09 | 1.372 | 0.08 |
| | d=10 | 1.555 | 0.33 | 1.415 | 0.12 | 1.413 | 0.10 | 1.516 | 0.07 |
| 1.50 | d=2 | 1.801 | 0.28 | 1.840 | 0.15 | 1.828 | 0.10 | 1.712 | 0.04 |
| | d=5 | 1.820 | 0.29 | 1.810 | 0.12 | 1.825 | 0.09 | 1.815 | 0.04 |
| | d=10 | 1.912 | 0.21 | 1.904 | 0.12 | 1.911 | 0.04 | 1.965 | 0.04 |
| 1.75 | d=2 | 1.981 | 0.34 | 1.982 | 0.34 | 1.981 | 0.07 | 1.981 | 0.06 |
| | d=5 | 1.980 | 0.32 | 1.990 | 0.34 | 1.993 | 0.06 | 1.992 | 0.05 |
| | d=10 | 1.997 | 0.30 | 1.991 | 0.33 | 1.997 | 0.06 | 1.996 | 0.05 |

## 6.7 Copula approach

Given a random vector $\mathbf{y} \in \mathbb{R}^d$, if all marginal distributions are known to be stable with different scale, location and skewness parameters it is reasonable to use a copula approach to obtain an approximation to the multivariate distribution. A multivariate function $C : [0,1]^d \to [0,1]$ is called a copula if it is a continuous distribution function with uniform marginal, viz. $C(u_1,...,u_d) = \Pr(U_1 \leq u_1,...,U_d \leq u_d)$. The idea is that one can specify the marginal distributions, for example all of them can be general stable with the same parameters $\alpha$ and $\beta$, and then a copula function can be used to "combine" the marginals into a joint multivariate distribution.

While there are many copula functions (Frees and Valdez, 1998)[17], only a few can be used easily in high dimensional problems. The Gaussian (Song, 2000) and Student-$t$ copulae (Demarta and McNeil, 2005) seem to be the most useful in applied work. Pitt, Chan and Kohn (2006) proposed the multivariate normal copula, given by:

$$C(u) = \Phi_d\left(\Phi^{-1}(u_1),...,\Phi^{-1}(u_d)\right),$$

where $\Phi$ denotes the univariate standard normal distribution function and $\Phi_d$ is the distribution function of a d-dimensional normal, viz. $\boldsymbol{\xi} \sim N_d(0,\mathbf{C})$, with density $|\mathbf{C}|^{-1/2} \exp\left[-\frac{1}{2}\boldsymbol{\xi}'\left(\mathbf{C}^{-1} - \mathbf{I}_d\right)\boldsymbol{\xi}\right]$; $\xi_i = F_i(y_i;\theta_i)$, $i = 1...,d$, $\boldsymbol{\xi} = [\xi_1,...,\xi_d]'$, $\mathbf{C}$ is a correlation matrix, and $\theta_i$ is a vector of parameters. For the observations $\{\mathbf{y}_t, t = 1,...,n\}$, $\mathbf{y}_t \in \mathbb{R}^d$, the copula model is

$$y_{ti} = F_i^{-1}\left(\Phi(\xi_{ti});\theta_i\right), \ i = 1,...,d, \qquad (25)$$

where $F_i(\cdot;\theta_i)$ denotes the distribution function of the $i$th component, in our case a member of $\mathscr{S}_{\alpha_i,\beta_i}(\mu_i,\sigma_i)$. The structural parameters are $\theta_i = [\alpha_i,\beta_i,\mu_i,\sigma_i]'$, $i = 1,...,d$, which we denote collectively as $\theta = [\theta_1,...,\theta_d]'$. Suppose also that $f_i(y_{ti};\theta_i)$ denotes densities in $\mathscr{S}_{\alpha_i,\beta_i}(\mu_i,\sigma_i)$, $i = 1,...,d$.

---

[17] For introductions, see Joe (1997) and Nelsen (2006).

The MCMC scheme involves three steps of random drawings: (*i*) From the posterior conditional distribution of $\theta \mid \mathbf{C}, \mathbf{Y}$, (*ii*) from $\boldsymbol{\xi} \mid \theta, \mathbf{C}, \mathbf{Y}$, and (*iii*) from $\mathbf{C} \mid \theta, \boldsymbol{\xi}, \mathbf{Y}$. The prior for $\theta_i$ is uniform in $(0,2] \times (-1,1) \times \mathbb{R} \times \mathbb{R}_+$. For the correlation matrix, $\mathbf{C} = \mathbf{D}'\mathbf{D}$, where $\mathbf{D} = [d_{ij}]$ is a lower triangular matrix we assume that its elements have a prior $p(d_{ij}) \propto \text{const.}$, $i \geq j$.

*Step (i). Random number generation for $\theta_i$.*

Since $u_{ti} = \Phi^{-1}(F_i(y_{ti}))$, the likelihood function is

$$\mathscr{L}(\theta, \mathbf{C}; \mathbf{Y}) = |\mathbf{C}|^{-n/2} \prod_{t=1}^{n} \left\{ \exp\left[-\tfrac{1}{2} \boldsymbol{\xi}_t'(\mathbf{C}^{-1} - \mathbf{I}) \boldsymbol{\xi}_t \right] \prod_{i=1}^{d} f_i(y_{ti}; \theta_i) \right\}, \qquad (26)$$

with $\xi_{ti} = \Phi^{-1}(F_i(y_{ti}; \theta_i))$. Sampling from the posterior conditional distribution $p(\theta_i \mid \mathbf{C}, \mathbf{Y})$ *does not* reduce to sampling from $p(\theta_i \mid \mathbf{C}, \boldsymbol{\xi}, \mathbf{Y}) \propto \prod_{t=1}^{n} f_i(y_{ti}; \theta_i)$, for $i = 1, ..., d$ since $u_{ti} = F_i(y_{ti}; \theta_i)$. The simplification would, indeed, be possible when drawing from $p(\theta_i \mid \mathbf{C}, \boldsymbol{\xi}, \mathbf{Y})$ but as argued convincingly in Pitt, Chan, and Kohn (2006) convergence of MCMC would have been significantly slower.

*Step (ii). Random number generation for the latent variables.*

This step is straightforward since given $\theta_i$, we simply set $\xi_{ti} = \Phi^{-1}(F_i(y_{ti}; \theta_i))$.

*Step (iii). Random number generation for the correlation matrix, $\mathbf{C}$.*

Here, we follow a different approach than Pitt, Chan, and Kohn (2006). Since $\mathbf{C} = \mathbf{D}'\mathbf{D}$, where $\mathbf{D} = [d_{ij}]$ is a lower triangular matrix, we can draw the elements $d_{ij}$, $i \geq j$ using a random walk Metropolis – Hastings algorithm. Using a multivariate normal proposal for the elements $d_{ij}$ is particularly convenient provided $\mathbf{C}$ is a correlation matrix. The required restrictions are $g_1^2 = 1$, $g_2^2 + g_3^2 = 1$, $g_4^2 + g_5^2 + g_6^2 = 1$, *etc.*, where $g_j$ are elements of a vector which is the vectorization of the nonzero $g_{ij}$s. Equivalently, the rows of $\mathbf{D}$ have unit norm, a restriction that can be imposed very easily.

# 7. Results

**Figure 1. Marginal posterior distributions of $\alpha$ and mixture approximations**

The figure presents the marginal posterior distributions of the characteristic exponent, α, resulting from three approximations in the context of an artificial experiment with 1,500 observations. The straight line labeled "Metropolis" represents the "exact" marginal posterior resulting from a Metropolis algorithm. The other two correspond to finite scale mixtures of normal with $M=3$ and $M=5$ components.

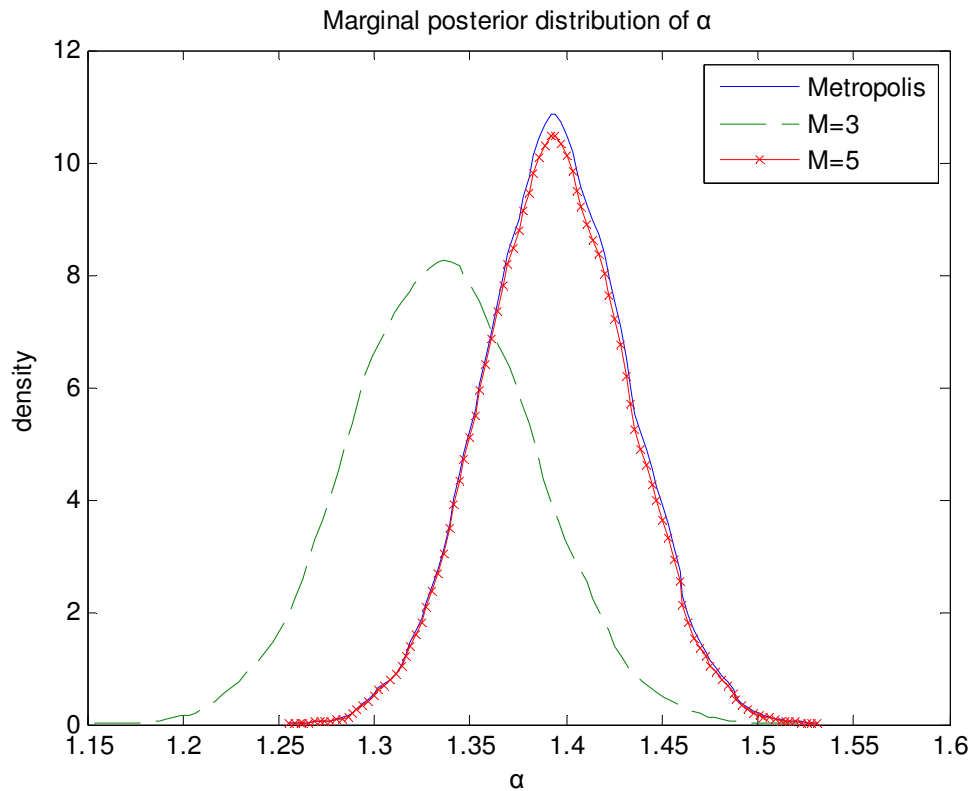

**Figure 2. Scale mixture parameters as functions of $\alpha$, with M=3**

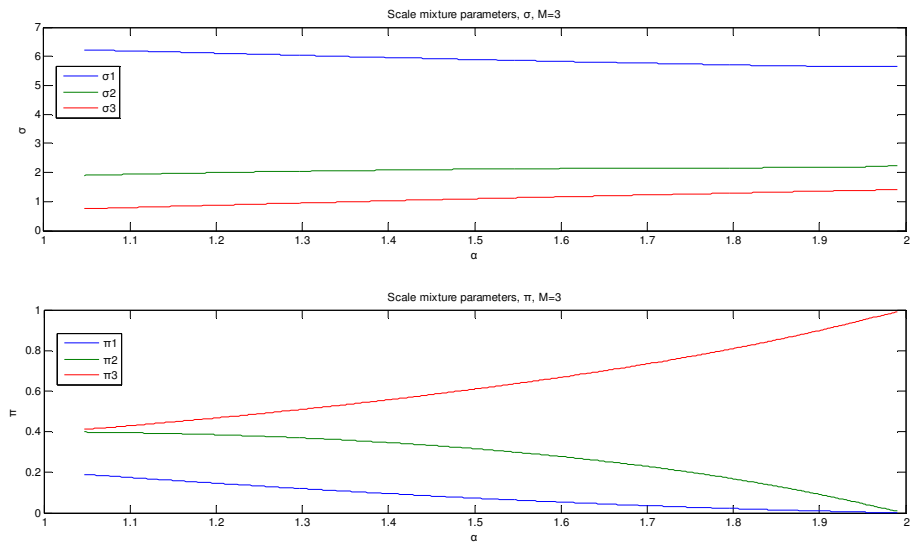

**Figure 3.** Various symmetric stable distributions and their approximations with normal scale mixtures with M=3 components.

For better visualization the range of ordinates is truncated. The original range used for fitting the scale mixtures by the KL criterion was [-16, 16] for $\alpha$ =1.1 and 1.3 and [-10, 10] for $\alpha$ =1.5 and 1.7. The truncation does not affect the tail behavior of the approximation.

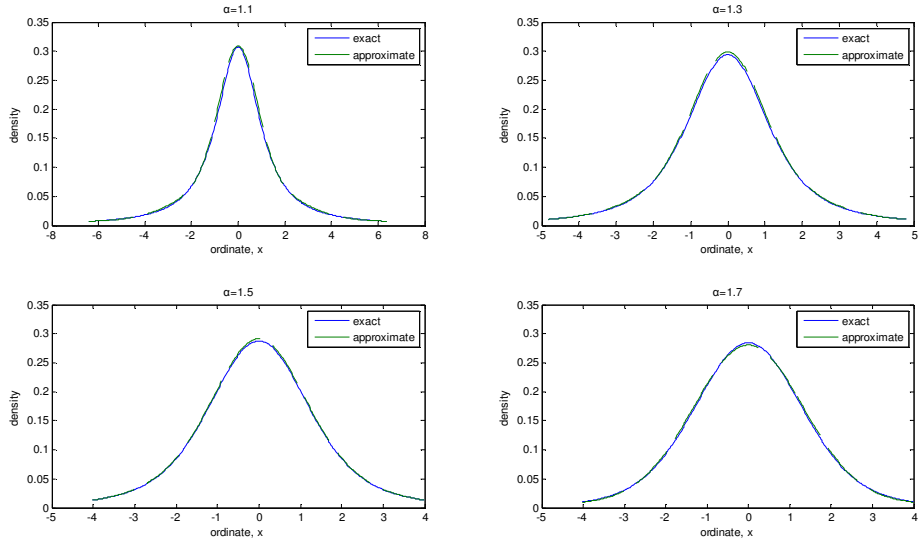

**Figure 4.** Critical values of the maximum absolute difference between the empirical and theoretical characteristic function scaled by $n^{1/2}$.

We consider the statistic $D = \max_{(\tau_i, i=1,\ldots,I)} \left| \varphi(\tau_i) - \hat{\varphi}(\tau_i) \right|$, where $\hat{\varphi}(\tau) = n^{-1} \sum_{t=1}^{n} \exp(\iota \tau Y_t)$, and $\varphi(\tau)$ is the theoretical characteristic function, $\varphi(\tau) = \exp\left(\iota \mu \tau - \sigma |\tau|^{\alpha}\right)$. The figure provides plots of the 90% critical values of the statistic $n^{-1/2} D$ for 20 values of $\alpha$ in the interval 1.1 to 1.9. 10,000 simulations are used to obtain the critical values. The characteristic functions are computed using I=10 equally spaced points in the interval [-0.5, 0.5].

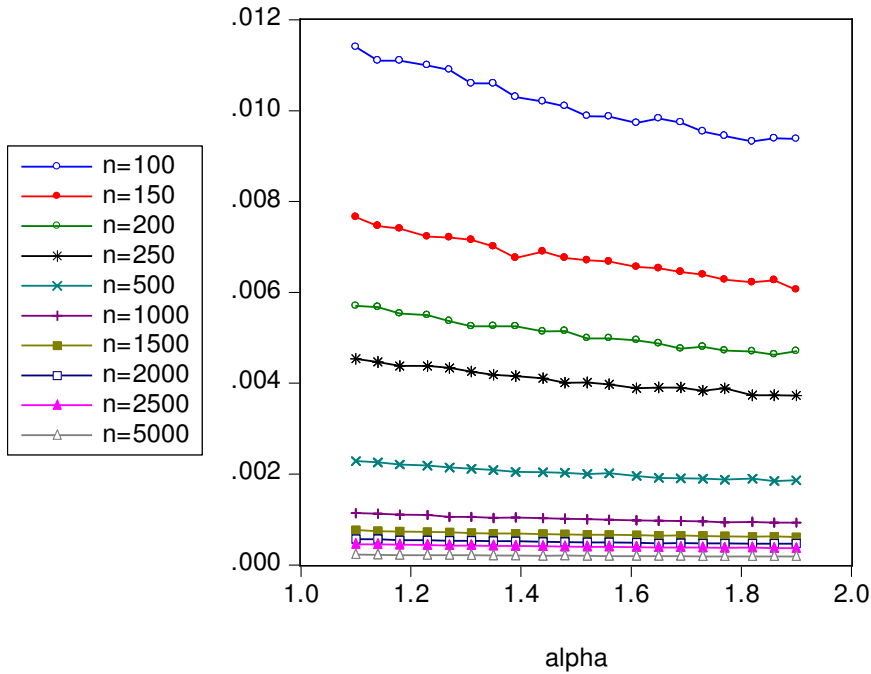

**Table 4a. Critical values of the maximum absolute difference between the empirical and theoretical characteristic function of symmetric stable.**

|              | N=100  | n=1000 | n=5000 |
|--------------|--------|--------|--------|
| $\alpha=1.1$ | 0.114  | 0.0359 | 0.0162 |
| $\alpha=1.5$ | 0.100  | 0.0320 | 0.0142 |
| $\alpha=1.7$ | 0.095  | 0.0304 | 0.0135 |
| $\alpha=1.9$ | 0.094  | 0.0295 | 0.0133 |

**Notes**: The table reports a few (unscaled) 90% critical values of the maximum absolute difference between the empirical and theoretical characteristic function for symmetric stable laws (D statistic). The generation of unscaled critical values is described in the construction of Figure 1.

**Table 4b. Critical values of the maximum absolute difference between the empirical and theoretical characteristic function of general stable.**

|                              | $\beta=-0.9$ | $\beta=-0.5$ | $\beta=0.5$ | $\beta=0.9$ |
|------------------------------|--------------|--------------|-------------|-------------|
| $\alpha=1.1$  n=100          | 0.114        | 0.115        | 0.114       | 0.113       |
| n=500                        | 0.051        | 0.051        | 0.051       | 0.051       |
| n=1000                       | 0.016        | 0.016        | 0.016       | 0.016       |
| $\alpha=1.5$  n=100          | 0.102        | 0.100        | 0.102       | 0.101       |
| n=500                        | 0.045        | 0.045        | 0.045       | 0.045       |
| n=1000                       | 0.014        | 0.014        | 0.015       | 0.014       |
| $\alpha=1.7$  n=100          | 0.096        | 0.097        | 0.096       | 0.097       |
| n=500                        | 0.043        | 0.043        | 0.043       | 0.043       |
| n=1000                       | 0.014        | 0.014        | 0.014       | 0.014       |
| $\alpha=1.9$  n=100          | 0.093        | 0.095        | 0.094       | 0.094       |
| n=500                        | 0.042        | 0.042        | 0.041       | 0.042       |
| n=1000                       | 0.013        | 0.013        | 0.013       | 0.013       |

*Notes*: The table reports a few (unscaled) 90% critical values of the maximum absolute difference between the empirical and theoretical characteristic function for non-symmetric stable laws ($D$ statistic). The generation of unscaled critical values is described in the construction of Figure 1.

**Figure 5a. Marginal posterior distributions of parameters, n=150**

The figure presents marginal posterior distributions of parameters (μ, σ and α) from a random sample of symmetric stable distribution with μ=0, σ=1 and α=1.40. The solid line represents marginal posterior distributions of parameters derived from a Metropolis – Hastings algorithm and is considered to be "exact". The dotted line presents marginal posterior distributions of parameters from ABC. For ABC we have used 50,000 simulations using a rejection Metropolis algorithm (see main text) using as summary statistic the maximum absolute difference D between the empirical and theoretical characteristic function. For the Metropolis – Hastings algorithm we have used 520,000 the first 20,000 are discarded and the remaining are thinned every 10$^{th}$ draw.

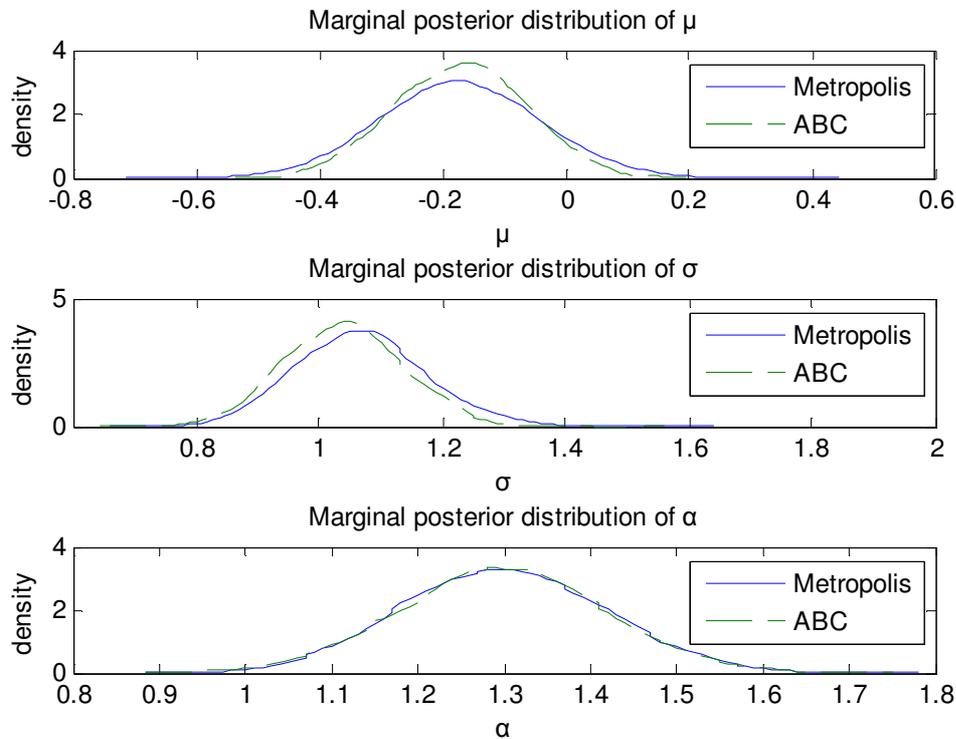

**Figure 5b. Marginal posterior distributions of parameters, n=1500**

For details see description of Figure 4a.

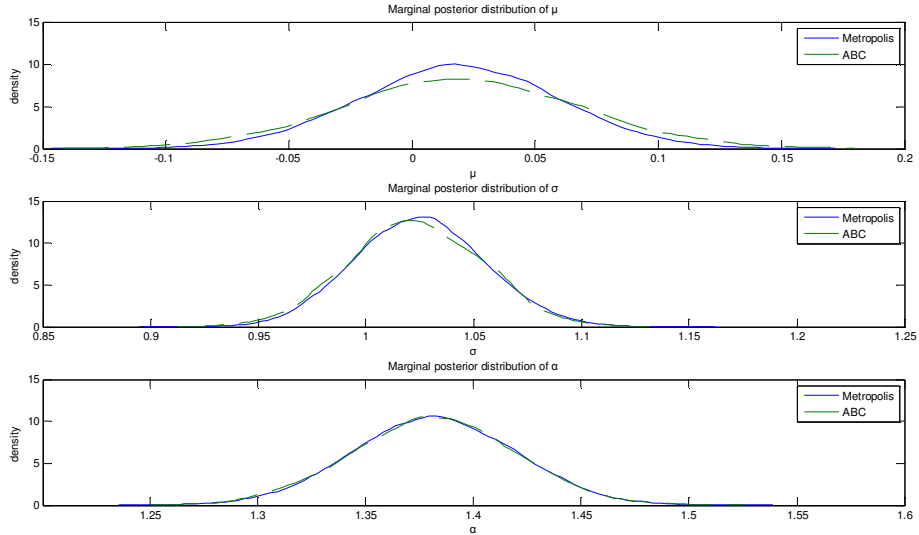

In Figure 6a we present marginal posterior distributions of all four parameters ($\mu,\sigma,\alpha,\beta$) in an artificial experiment. The true values of the parameters are $\mu=0$, $\sigma=1$, $\alpha=1.7$, $\beta=-0.4$, and the sample size is $n=150$. For both MCMC and ABC 250,000 passes have been used, the first 50,000 of which are discarded and thinning every other tenth draw.

**Figure 6a.**

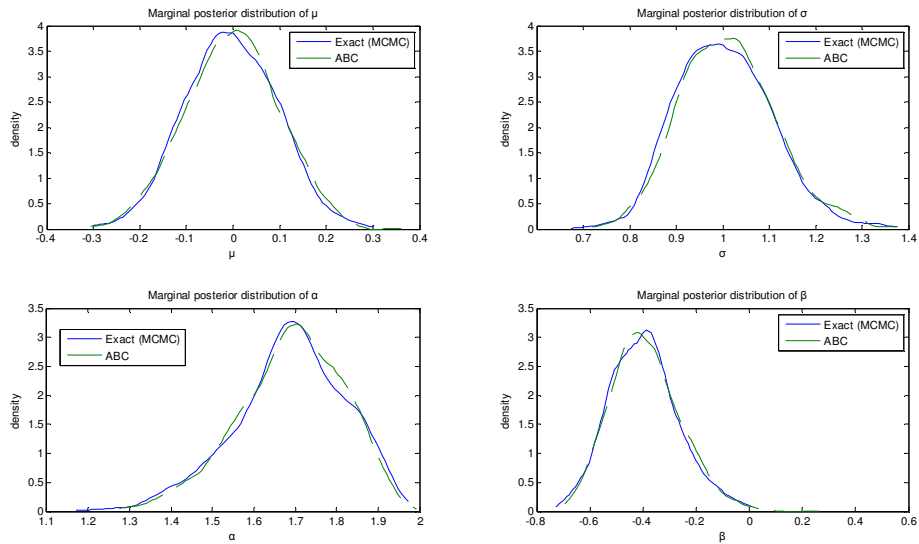

In Figure 6b we examine sensitivity with respect to number of points and endpoints for the grid of the characteristic function, when n=150. Left column has 5 points in $\pm 2$. Center column has 20 points. The right column has 5 points in $\pm 0.5$.

**Figure 6b.**

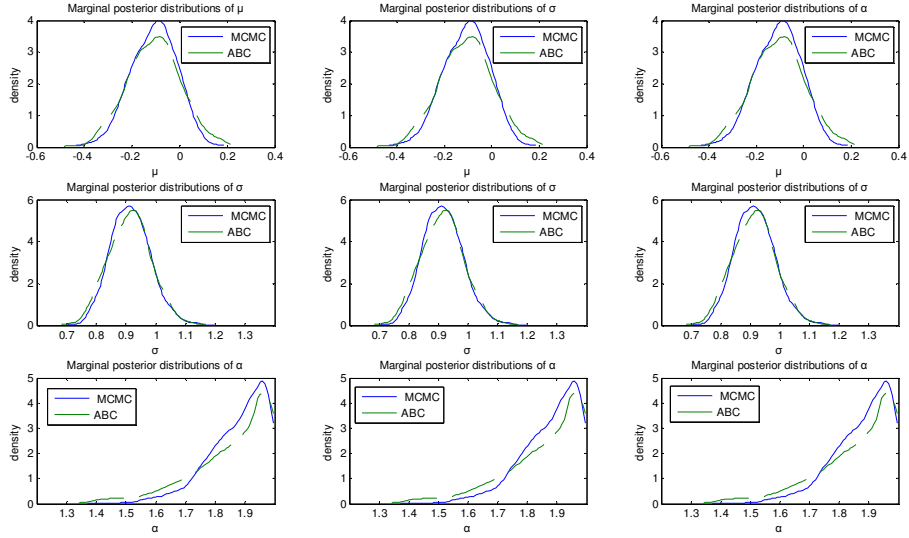

In Figure 6c we present marginal posterior distributions of all three parameters (μ,σ,α) of symmetric stable laws in an artificial experiment. The true values of the parameters are $\mu=0$, $\sigma=1$, $\alpha=1.7$, and the sample sizes are $n=150$ and $n=1500$. For both MCMC and ABC 250,000 passes have been used, the first 50,000 of which are discarded and thinning every other tenth draw.

**Figure 6c.**

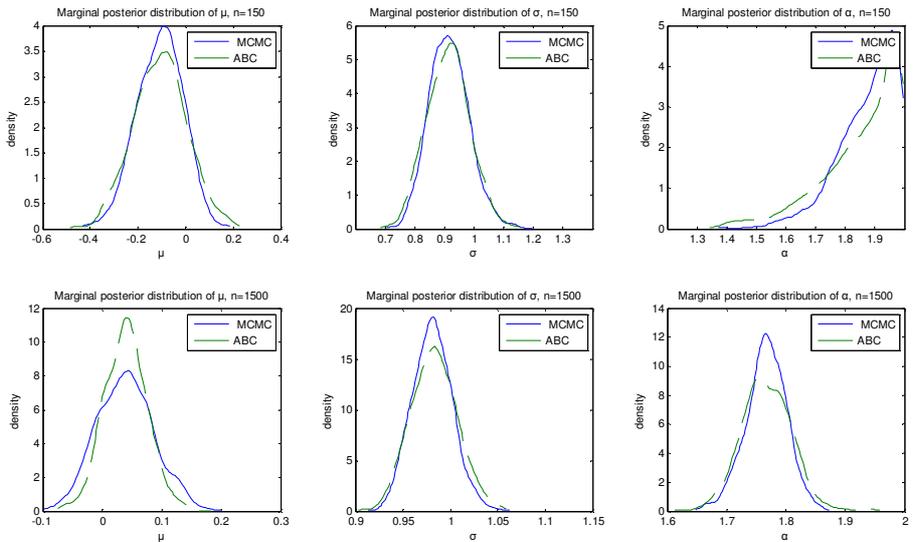

Although some studies examined the choice of grid points, viz. their number and placement, through minimizing the determinant of the covariance matrix in the ANF, they are not extremely relevant when the question is how to choose the configuration in order for the posterior mean to perform well or to facilitate ABC inference using the ANF. In "large samples", these studies provide, of course, useful information but we do not know whether we are indeed in a "large sample" situation for which the configuration of the grid can be determined from asymptotic theory. Another related question is whether the ANF can be used profitably in finite samples.

### Finite sample properties of Bayesian posterior means using the ANF, $\mathscr{S}_{\alpha,0}(\mu,\sigma)$

In Table 5 reported are MSEs for the parameters of a symmetric stable distribution with parameters $\mu,\sigma$ and $\alpha$. The posterior means are obtained through MCMC using the ANF. The sample size is $n$, $G$ is the number of grid points in the interval [-a, a]. The sampling experiment is based on 10,000 replications for the given parameter values $(\mu,\sigma,\alpha)$. MCMC is based on 150,000 draws the first 50,000 of which are discarded and we thin every other $10^{\text{th}}$ draw.

**Table 5.**

|       | A=1.70 (μ=0, σ=1) | | | | |
|---|---|---|---|---|---|
|       | n=50 | n=100 | n=500 | n=1,000 | n=5,000 |
| a=5   G=5  | 8.16 10⁴ 15.9 0.160 | 8.59 10⁴ 3.43 0.165 | 273. 0.894 0.0773 | 2.35 10³ 2.13 0.0759 | 1.39 0.0511 0.0783 |
| a=5   G=20 | 0.558 0.0301 0.0548 | 0.164 0.00976 0.0504 | 0.0266 0.00207 0.0466 | 0.0196 0.00110 0.0491 | 0.0126 0.000214 0.0491 |
| a=0.5 G=5  | 0.224 0.0313 0.0677 | 0.0477 0.0177 0.0545 | 0.0146 0.0104 0.0490 | 0.00949 0.00899 0.0495 | 0.00534 0.00863 0.0508 |
| a=0.5 G=20 | 0.139 0.0323 0.0614 | 0.0310 0.0214 0.0528 | 0.0134 0.0138 0.0498 | 0.00821 0.0121 0.0497 | 0.00447 0.0117 0.0508 |
| a=0.5 G=10 | 0.142 0.0302 0.0606 | 0.0330 0.0201 0.0534 | 0.0131 0.0127 0.0503 | 0.00807 0.0111 0.0506 | 0.00456 0.0107 0.0514 |
| a=0.5 G=15 | 0.143 0.0319 0.0612 | 0.0309 0.0209 0.0529 | 0.0135 0.0132 0.0493 | 0.00818 0.0118 0.0499 | 0.00453 0.0113 0.0509 |
| a=2.5 G=5  | 6.71 0.0674 0.158 | 3.80 0.0158 0.144 | 1.48 0.00425 0.0856 | 0.919 0.00224 0.0728 | 0.252 0.000980 0.0474 |
| a=2.5 G=10 | 0.803 0.0275 0.0623 | 0.196 0.00960 0.0540 | 0.0258 0.00205 0.0469 | 0.0190 0.00109 0.0495 | 0.0120 0.000222 0.0499 |
| a=2.5 G=20 | 0.662 0.0271 0.0613 | 0.202 0.00936 0.0528 | 0.0311 0.00212 0.0440 | 0.0227 0.00108 0.0469 | 0.0149 0.000214 0.0461 |

Since another related question is whether the ANF can be used profitably in finite samples we report below a typical situation when the true values are $\mu=0$, $\sigma=1$, $\alpha=1.40$. Here, Exact (MCMC) is based on the exact density, ANF-ABC is from ABC inference and ANF-MCMC is MCMC based directly on the characteristic function. We have a "small sample" ($n=100$) and a typically "large sample" in economics and finance ($n=2,000$). It seems that in small samples, ANF-MCMC has fat tails and may, as a result, behave erratically. As the sample size increases this phenomenon disappears.

**Figure 7. Typical marginal posterior distributions.**

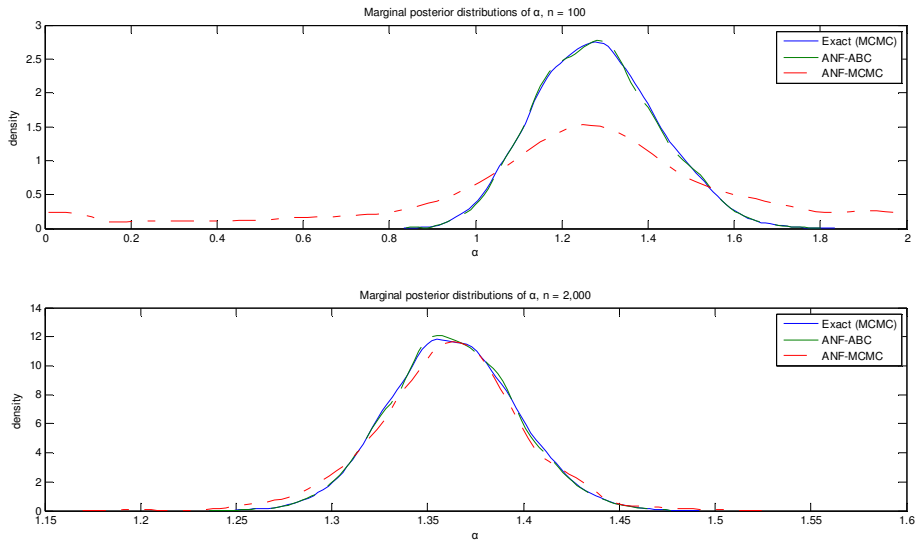

## 7.2 Approximation of general stable distributions

To approximate general standard stable distributions, $\mathscr{S}_{\alpha,\beta}(0,1)$ using finite mixtures of normals we use a $51\times 51$ grid for $\alpha,\beta$. The values of $\alpha$ range from 1.20 to 1.90 with step size 0.014 and the values of $\beta$ are from -0.90 to 0.90 with step size 0.036. For the FFT we use $n=16$ and h=0.0005 yielding $2^n$ points at which the density is computed. We approximate with a location – scale mixture of normals with M=5 components which was found quite adequate, using 200 from the $2^n$ points at which the density was computed. Using more than M components indicated that the extra components have practically zero mixing probabilities, viz. less than about $10^{-5}$ and the KL criterion cannot be improved, as in the case of symmetric stable distributions. We should note that we also tried Student-$t$ location – scale mixtures with 3, 5 and 10 components. It turned out that at the (global) minimum of the KL criterion the approximating mixture had always problems approximating the density at the tails producing multimodal distributions. We have also been unable to find better fit by using a normal mixture with $M-1$ normal components and one component that is Cauchy (stable with $\alpha=1$, $\beta=0$). Therefore, for practical purposes, the use of normal mixtures is recommended.

**Figure 8. KL distance between general standard stable distributions, $\mathscr{S}_{\alpha,\beta}(0,1)$, and approximating location – scale normal mixtures.**

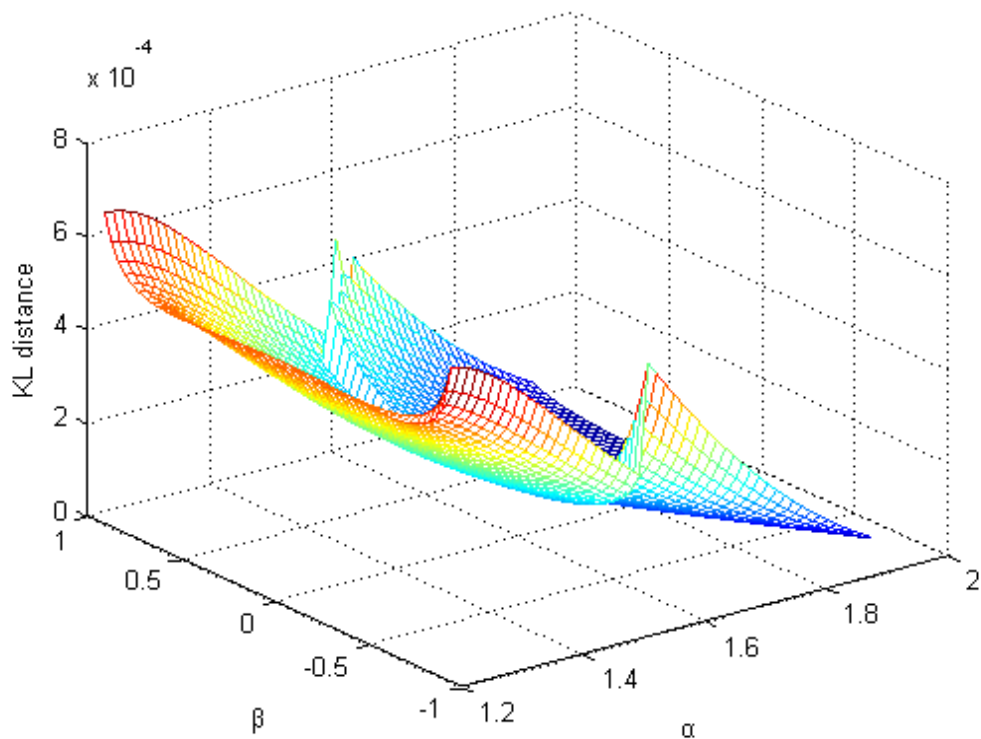

In Figure 9, we plot typical exact and approximate log densities.

Figure 9. Exact and approximate log densities, $\mathscr{S}_{\alpha,\beta}(0,1)$.

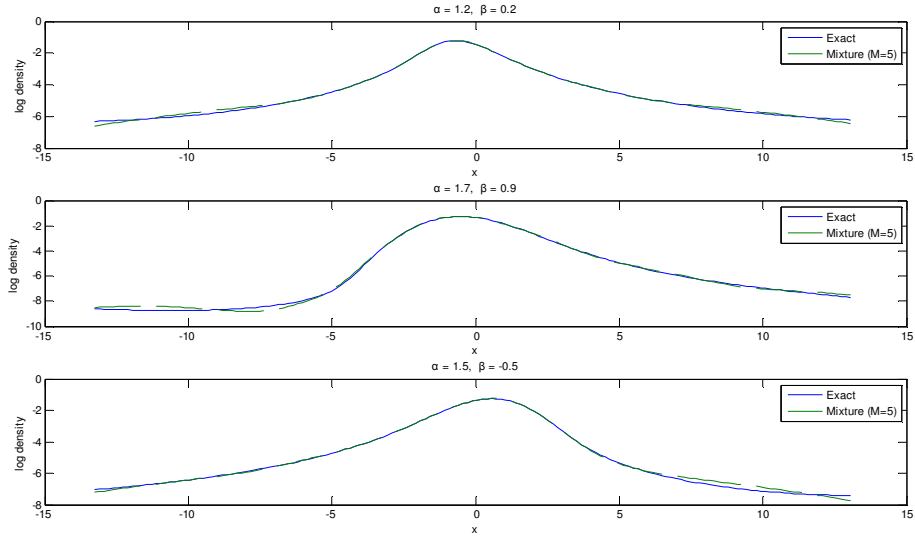

In Figure 10 we show the parameters of the approximating location – scale mixture with M=5 components for $\beta=-0.5$ (left column) and $\beta=0.2$ (right column). The curves as a function of $\alpha$, are reasonably smooth and it has been found that the same is true in the $(\alpha,\beta)$ space. Minor discontinuities shown in the graph are corrected before performing empirical analysis.

Figure 10. Parameters of the approximating normal mixture, $\mathscr{S}_{\alpha,\beta}(0,1)$.

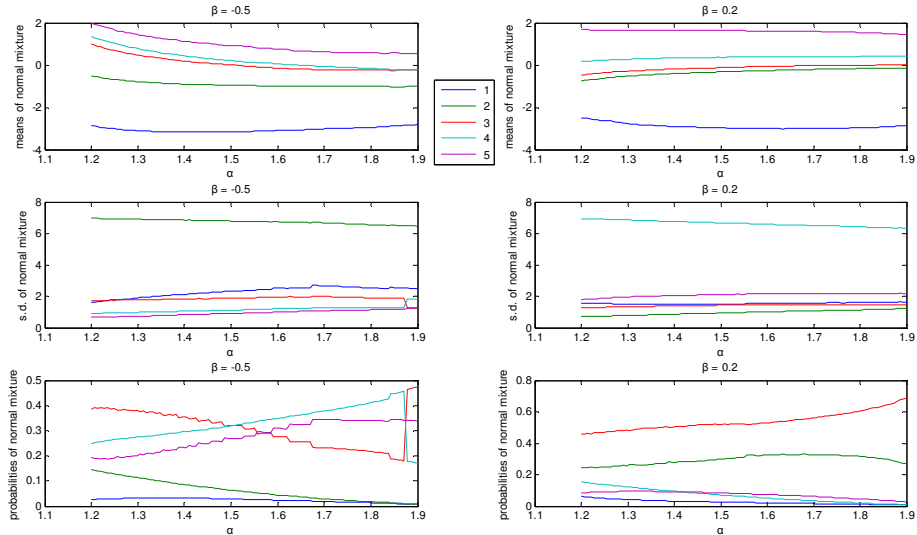

Table 6. Critical values for ABC analysis for general stable distributions, $\mathscr{S}_{\alpha,\beta}(0,1)$.

Reported in the table are 5% and 10% critical values of the absolute difference between the theoretical and empirical characteristic function for representative values of $\alpha$ and $\beta$ for a general stable distribution and sample sizes $n=500$ and $n=1500$. The critical values were obtained using 10,000 simulations. The theoretical and empirical characteristic function were computed on a grid of 20 points in the interval $\pm 2$.

| N | $\alpha$ | $\beta$ | 5% | 10% |
|---|---|---|---|---|
| 500 | 1.10 | -0.90 | 1.842 | 1.870 |
| | | -0.50 | 1.839 | 1.868 |
| | | -0.25 | 1.836 | 1.872 |
| | | 0.25 | 1.837 | 1.860 |
| | | 0.50 | 1.838 | 1.856 |
| | | 0.90 | 1.833 | 1.855 |
| | 1.70 | -0.90 | 1.958 | 1.964 |
| | | -0.50 | 1.959 | 1.965 |
| | | -0.25 | 1.960 | 1.965 |

|   |      |       |             |
|---|------|-------|-------------|
|   |      | 0.25  | 1.961 1.964 |
|   |      | 0.50  | 1.961 1.964 |
|   |      | 0.90  | 1.961 1.963 |
| 1500 | 1.10 | -0.90 | 1.838 1.836 |
|   |      | -0.50 | 1.836 1.834 |
|   |      | -0.25 | 1.833 1.833 |
|   |      | 0.25  | 1.833 1.842 |
|   |      | 0.50  | 1.831 1.838 |
|   |      | 0.90  | 1.832 1.838 |
|   | 1.70 | -0.90 | 1.955 1.955 |
|   |      | -0.50 | 1.958 1.956 |
|   |      | -0.25 | 1.958 1.956 |
|   |      | 0.25  | 1.957 1.957 |
|   |      | 0.50  | 1.958 1.957 |
|   |      | 0.90  | 1.958 1.957 |

## 8. Empirical Application

We apply the general stable distribution to data for the General Electric stock price (January 2 2007 through March 31 2012) for a total of 1,364 observations. We have used MCMC using the density obtained from the FFT (n=16, h=0.001), ABC using the ANF and MCMC using the approximating mixture. Three techniques are considered. *First, "exact" inference using MCMC* based on the density obtained through the FFT. *Second, ABC inference using the ANF* and the empirical characteristic function, and third, inference using the approximating mixture with M=5 components. This is done in two steps. First, a Gibbs sampler has been used to provide inferences for the parameters of the general distribution. Second, the mixture draws are converted to approximate draws from the posterior distribution of the stable model using multivariate spline interpolation from ($\boldsymbol{\mu},\boldsymbol{\sigma},\boldsymbol{\pi}$) to ($\alpha,\beta,\mu,\sigma$).

In practice, it is possible to use the draws of ($\boldsymbol{\mu},\boldsymbol{\sigma},\boldsymbol{\pi}$) and obtain draws for ($\alpha,\beta,\mu,\sigma$) by minimizing the KL distance. This avoids the use of spline interpolation at the cost of performing a huge number of optimizations to solve the KL problem. For all simulations we have used 120,000 passes the first 20,000 of which are discarded to mitigate start up effects and we thin every other $10^{th}$ iteration, so we would have to solve 10,000 optimization problems. We have experimented with solving the problem at a smaller scale (1,000 problems) and it has been found that the spline procedure is extremely accurate. *This procedure, of course, overcomes an important impediment posed by indirect inference in that the parameters of the stable distribution and the approximating indirect model must have a one-to-one correspondence* to allow a homotopy between the parameter spaces of the two models (Lombardi and Calzolari, 2008, 2009, and Lombardi and Veredas, 2009).

**Figure 11. Marginal posterior distributions of $\alpha$ and $\beta$, General Electric returns**

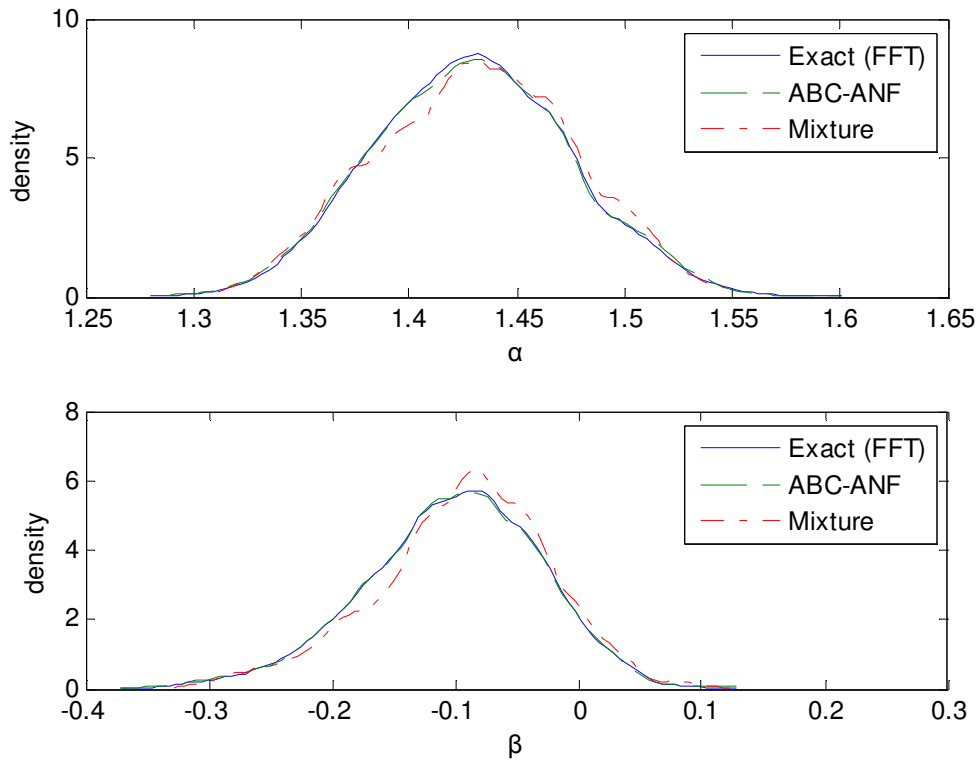

## Table 7. Posterior statistics for the General Electric returns using GARCH and SV models and various approximations to MCMC

The GARCH model is $X_t \mid h_t \sim (\mu, h_t)$, $h_t = \alpha_0 + \alpha_1 (X_{t-1} - \mu)^2 + \alpha_2 h_{t-1}$, where $\mu$ is location parameter (not reported). The distribution of $X_t$ is either normal or stable with parameters $\alpha$ and $\beta$. The SV model is $\log X_t^2 = h_t + \log \varepsilon_t^2$ $h_t = \gamma(1-\rho) + \rho h_{t-1} + v_t$, where $\varepsilon_t$ is either $\mathcal{N}(0,1)$ or standard stable with parameters $\alpha$ and $\beta$, and $v_t \sim iid\,\mathcal{N}(0, \sigma^2)$. All models are fitted using Bayesian methods (exact, ABC or mixtures) using 120,000 iterations the first 20,000 of which are discarded and are thinned every other 10$^{th}$ draw. The GARCH – normal and GARCH – stable[18] (MCMC) models are estimated using a Metropolis / Gibbs sampler whose acceptance rate is targeted at 25%. The SV – normal is estimated using a standard Gibbs sampler using the Kalman filter. The SV – stable (MCMC) model is estimated using the Gibbs sampler.

| Model | $\alpha_1$ | $\alpha_2$ | $\alpha$ | $\beta$ |
|---|---|---|---|---|
| GARCH – normal | 0.0785 (0.010) | 0.914 (0.009) | | |
| GARCH – stable (ABC) | 0.143 (0.025) | 0.857 (0.014) | 1.751 (0.122) | -0.013 (0.141) |
| GARCH – stable (mixture) | 0.140 (0.023) | 0.858 (0.014) | 1.751 (0.120) | -0.011 (0.143) |
| GARCH – stable (MCMC) | 0.143 (0.025) | 0.857 (0.015) | 1.751 (0.122) | -0.013 (0.141) |
| | $\gamma$ | $\rho$ | | |
| SV – normal | -0.117 (0.124) | 0.950 (0.015) | | |
| SV – stable (ABC) | -0.089 (0.085) | 0.832 (0.011) | 1.572 (0.11) | -0.16 (0.015) |
| SV – stable (mixture) | -0.090 (0.083) | 0.830 (0.012) | 1.570 (0.12) | -0.17 (0.016) |
| SV – stable (MCMC) | -0.090 (0.081) | 0.830 (0.011) | 1.550 (0.11) | -0.17 (0.015) |

*Note*: Standard errors appear in parentheses.

Now we provide some details on the results of implementing ABC using the ANF. The posterior results are provided in Figure 12. The top left plot provides the marginal posterior distribution of the optimal number of points, $k$. Given the posterior we select the draws for which $k = 5$, 15 and 25. For each $k$, the posterior means of the optimal grid are reported in the other three plots.

**Figure 12. Posterior distribution of number of points, $k$, in ANF and the optimal placement of the grid points. General Electric stock returns.**

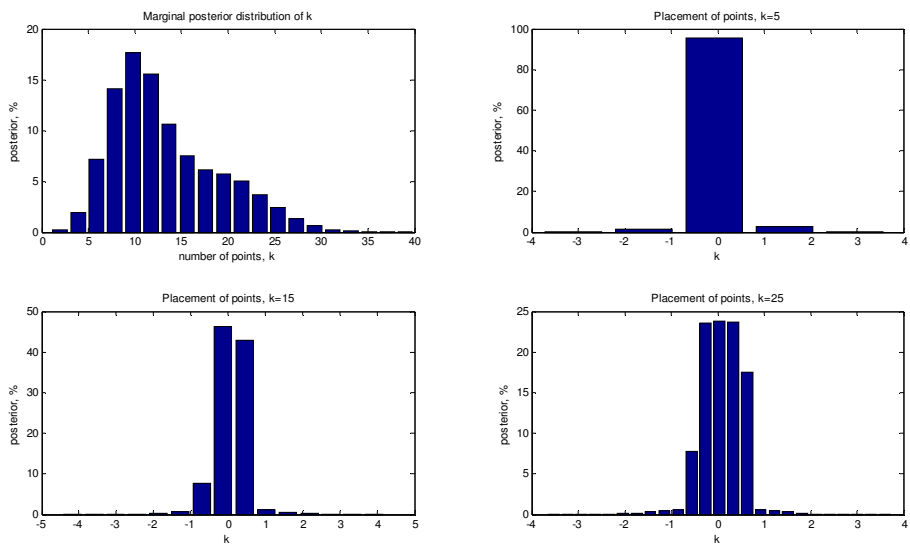

---
[18] The GARCH-stable model seems to have been proposed by Liu and Brorsen (1995).

Figure 13. Marginal posterior distribution of grid points and posterior mean normalized spectral measures.

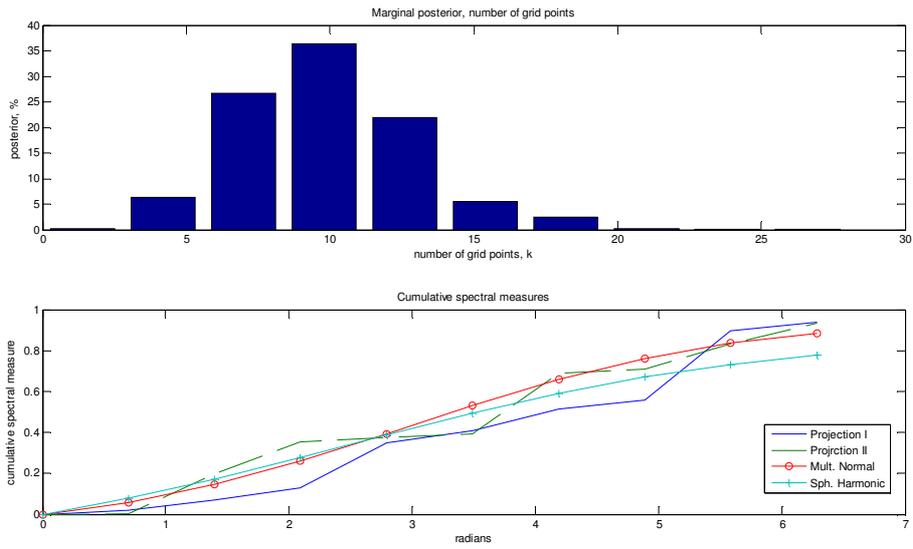

Figure 14. Multivariate Stable distribution: Marginal posterior distributions of $\alpha$ and $\beta$

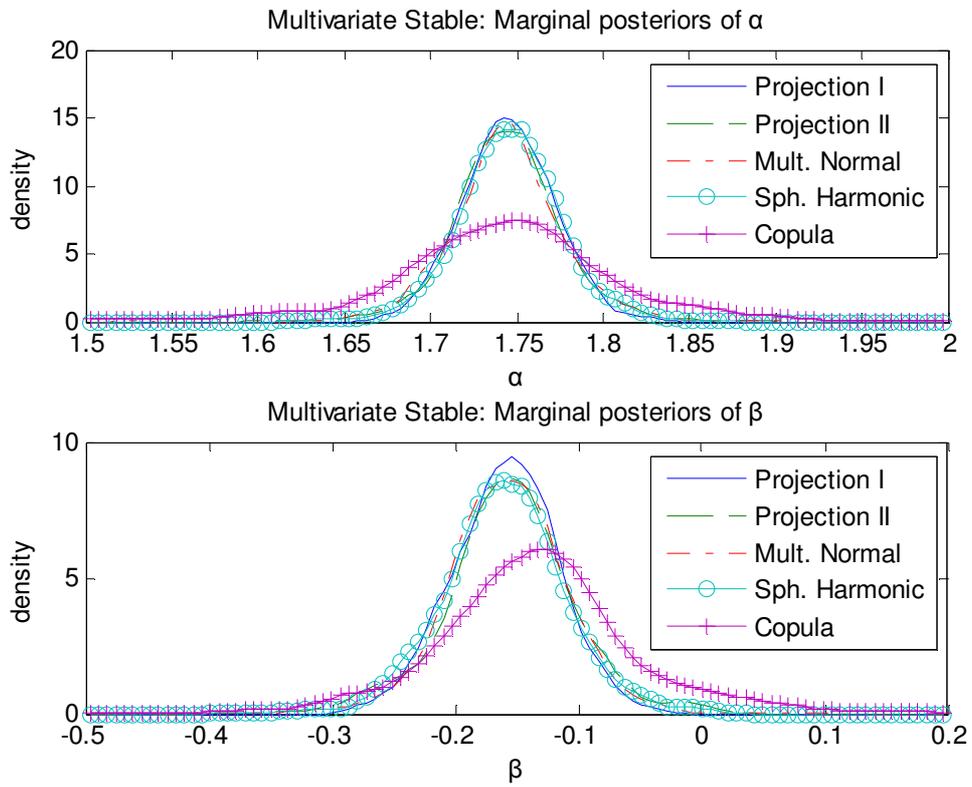

# 9. Stochastic volatility

## 9.1 Introduction

Consider a stochastic volatility model of the form:

$$Y_t = \exp(h_t/2)\varepsilon_t, \qquad (27)$$

where $\varepsilon_t \sim iid\mathcal{S}_{\alpha,\beta}(0,1)$, and $h_t = \delta + \rho h_{t-1} + v_t$, $v_t \sim iidN(0,\omega^2)$. Following Kim, Shephard and Chib (1998)[19] we have:

$$\log Y_t^2 = h_t + \log \varepsilon_t^2. \qquad (28)$$

The distribution of $\log \varepsilon^2$ when $\varepsilon \sim iid\mathcal{S}_{\alpha,\beta}(0,1)$ is not known in closed form. It can be approximated, however, using a finite mixture of normals. We proceed using the characteristic functions. The characteristic function of the normal mixture is:

$$\varphi_N(\tau) = \sum_{m=1}^{M} \pi_m \exp\left(\iota\mu_m\tau - \tfrac{1}{2}\sigma_m^2\tau^2\right).$$

The characteristic function of $\log \varepsilon^2$ is obtained by obtaining a large sample $\varepsilon_{(s)} \sim iid\mathcal{S}_{\alpha,\beta}(0,1)$, set $u_{(s)} = \log \varepsilon_{(s)}^2$, and approximating the characteristic function using $\varphi(\tau) \approx S^{-1}\sum_{s=1}^{S}\exp(\iota\tau u_{(s)})$. If $d(\tau) = \begin{bmatrix}\Re\varphi_N(\tau) - \Re\varphi(\tau)\\ \Im\varphi_N(\tau) - \Im\varphi(\tau)\end{bmatrix}$, we choose the parameters of the approximating normal mixture by minimizing: $\sum_{i=1}^{I} d(\tau_i)^2$. The need for approximating the characteristic function numerically is that the integral:

$$\varphi(\tau) = \int_{-\infty}^{\infty}\exp(i\tau\log\varepsilon^2)f(\varepsilon)d\varepsilon = \int_{-\infty}^{\infty}\exp\left[(\iota\tau+\tfrac{1}{2})\zeta\right]f\left(\exp(\tfrac{\zeta}{2})\right)d\zeta \qquad (29)$$

cannot be obtained in closed form, where $f()$ denotes the density of stable laws, $\mathcal{S}_{\alpha,\beta}(0,1)$. This is unlike the situation in Kim, Shephard, and Chib (1993, section 3) where $\varepsilon_t = \sigma\xi_t$, $\xi_t \sim iid\mathcal{N}(0,1)$ and thus $\xi_t^2 \sim iid\chi^2(1)$. Although the distribution of $\xi_t^2$ is not convenient for MCMC simulation purposes, its density is available in closed form and thus it is possible to approximate by a mixture of normals directly in the space of densities.

In our approximation we use $S = 100,000$ and for $\tau$ we use a grid of 100 equally spaced points in the interval [-0.5, 0.5]. Low order approximations (M=2 or 3) behave quite well and more precision of the order $10^{-6}$ can be obtained using M=10. The approximations are convenient in the sense that most of the mixture probabilities are equal.

Table 8. Approximating normal mixture parameters to the distribution of $\log\varepsilon^2$, M=10 components.

|  | β=-0.9 | β=-0.5 | β=-0.25 | β=0.25 | β=0.5 | β=0.9 |
|---|---|---|---|---|---|---|
| α=1.1 | 2.42 0.780 0.0835<br>2.80 0.512 0.0984<br>3.63 0.0515 0.103<br>3.61 0.177 0.103<br>2.29 3.93 0.102<br>1.59 0.860 0.100<br>3.67 0.167 0.103<br>3.62 0.0954 0.103<br>3.37 0.191 0.103<br>3.68 0.108 0.103 | 1.29 0.869 0.136<br>1.54 0.490 0.0553<br>1.03 0.493 0.0479<br>1.44 0.0519 0.116<br>2.45 0.192 0.0559<br>1.56 4.16 0.136<br>0.625 0.865 0.118<br>3.22 0.189 0.137<br>2.66 0.0976 0.0107<br>1.85 0.192 0.0989<br>3.46 0.114 0.116 | 0.297 0.946 0.214<br>1.03 0.493 0.0481<br>1.44 0.0519 0.116<br>2.10 0.193 0.0119<br>0.708 4.19 0.214<br>-0.406 1.13 0.137<br>3.00 0.202 0.179<br>2.60 0.0976 0.00545<br>1.01 0.193 0.0674<br>3.04 0.115 0.00757 | 0.293 0.949 0.214<br>1.03 0.493 0.0479<br>1.44 0.0519 0.116<br>2.10 0.193 0.0119<br>0.694 4.20 0.214<br>-0.411 1.13 0.137<br>2.99 0.202 0.179<br>2.60 0.0976 0.00545<br>1.01 0.193 0.0671<br>3.04 0.115 0.00756 | 1.30 1.18 0.214<br>1.20 0.515 0.0635<br>2.40 0.0520 0.199<br>2.22 0.193 0.0325<br>1.55 4.16 0.137<br>0.0138 1.23 0.0136<br>3.34 0.189 0.215<br>2.65 0.0976 0.0130<br>1.30 0.195 0.0787<br>3.10 0.115 0.0333 | 2.36 1.26 0.228<br>1.44 0.523 0.0102<br>3.48 0.0519 0.242<br>2.46 0.193 0.0292<br>2.27 4.29 0.0788<br>0.0346 1.27 0.00493<br>3.78 0.183 0.244<br>2.75 0.0976 0.0196<br>1.62 0.195 0.0121<br>3.36 0.114 0.131 |
| α=1.3 | -0.267 4.11 0.151<br>1.18 0.580 0.0153<br>0.484 0.0522 0.121<br>2.28 0.196 0.0145<br>1.90 1.21 0.171<br>-0.490 1.40 0.238<br>1.88 0.199 0.251<br>2.67 0.0977 0.0158<br>1.37 0.198 0.00996<br>2.40 0.114 0.0124 | -0.667 4.03 0.227<br>1.15 0.589 0.0136<br>-0.0631 0.0522 0.155<br>2.27 0.196 0.00828<br>1.46 1.56 0.101<br>-1.19 1.08 0.224<br>1.84 0.207 0.248<br>2.67 0.0978 0.00767<br>1.35 0.199 0.00816<br>2.40 0.115 0.00758 | -0.874 4.08 0.254<br>1.13 0.592 0.0101<br>-0.401 0.0522 0.152<br>2.27 0.196 0.00611<br>1.35 1.87 0.0569<br>-1.63 1.01 0.249<br>1.81 0.210 0.254<br>2.67 0.0978 0.00582<br>1.34 0.199 0.00642<br>2.40 0.115 0.00572 | -0.866 4.08 0.256<br>1.13 0.592 0.0101<br>-0.385 0.0522 0.149<br>2.27 0.196 0.00607<br>1.35 1.88 0.0558<br>-1.64 1.05 0.250<br>1.78 0.210 0.255<br>2.67 0.0978 0.00576<br>1.34 0.199 0.00644<br>2.40 0.115 0.00568 | -0.611 3.92 0.242<br>1.14 0.590 0.0140<br>-0.0705 0.0522 0.161<br>2.27 0.196 0.00785<br>1.37 1.62 0.0921<br>-1.11 1.05 0.218<br>1.87 0.207 0.242<br>2.66 0.0978 0.00725<br>1.35 0.199 0.00827<br>2.40 0.115 0.00722 | -0.187 3.98 0.176<br>1.22 0.590 0.0325<br>0.571 0.0523 0.154<br>2.27 0.195 0.0230<br>1.32 0.653 0.187<br>-0.757 1.20 0.186<br>2.44 0.193 0.187<br>2.62 0.0976 0.0186<br>1.41 0.199 0.0160<br>2.39 0.114 0.0198 |
| α=1.5 | -1.42 3.83 0.195<br>1.12 0.597 0.0212 | -1.54 3.97 0.200<br>1.09 0.600 0.0151 | -1.57 4.01 0.203<br>1.09 0.600 0.0131 | -1.65 4.05 0.203<br>1.09 0.600 0.0131 | -1.55 3.99 0.200<br>1.09 0.599 0.0150 | -1.38 3.82 0.195<br>1.10 0.597 0.0213 |

---

[19] See also Durbin and Koopman (2000).

| | β=-0.9 | β=-0.5 | β=-0.25 | β=0.25 | β=0.5 | β=0.9 |
|---|---|---|---|---|---|---|
| | -0.137 0.0524 0.167 | -0.443 0.0524 0.165 | -0.572 0.0524 0.164 | -0.564 0.0524 0.163 | -0.440 0.0524 0.164 | -0.146 0.0524 0.168 |
| | 2.21 0.195 0.00853 | 2.21 0.196 0.00656 | 2.21 0.196 0.00592 | 2.21 0.196 0.00595 | 2.21 0.196 0.00648 | 2.21 0.195 0.00830 |
| | 0.692 0.737 0.193 | 0.449 0.793 0.195 | 0.342 0.819 0.195 | 0.349 0.817 0.195 | 0.447 0.790 0.195 | 0.694 0.736 0.194 |
| | -1.60 1.09 0.195 | -2.01 1.08 0.200 | -2.20 1.09 0.203 | -2.19 1.10 0.203 | -2.03 1.11 0.200 | -1.62 1.09 0.195 |
| | 2.08 0.194 0.195 | 2.03 0.195 0.199 | 2.01 0.195 0.199 | 2.01 0.195 0.199 | 2.02 0.195 0.199 | 2.07 0.194 0.195 |
| | 2.58 0.0976 0.00663 | 2.58 0.0976 0.00532 | 2.58 0.0976 0.00488 | 2.58 0.0976 0.00489 | 2.58 0.0976 0.00526 | 2.57 0.0976 0.00648 |
| | 1.36 0.199 0.0109 | 1.35 0.199 0.00840 | 1.35 0.199 0.00754 | 1.35 0.199 0.00758 | 1.35 0.199 0.00837 | 1.35 0.199 0.0109 |
| | 2.34 0.114 0.00760 | 2.34 0.114 0.00595 | 2.34 0.114 0.00541 | 2.34 0.114 0.00543 | 2.34 0.114 0.00589 | 2.33 0.114 0.00741 |
| α=1.7 | -2.30 3.89 0.160 | -2.22 3.86 0.165 | -2.33 3.79 0.172 | -2.28 3.78 0.173 | -2.41 3.80 0.170 | -2.30 3.79 0.165 |
| | 1.07 0.537 0.130 | 1.05 0.541 0.125 | 0.981 0.552 0.119 | 0.958 0.546 0.120 | 0.996 0.554 0.122 | 0.986 0.544 0.129 |
| | -0.717 0.0529 0.140 | -0.756 0.0529 0.143 | -0.500 0.0529 0.140 | -0.446 0.0528 0.139 | -0.506 0.0529 0.139 | -0.418 0.0529 0.138 |
| | 2.00 0.193 0.0202 | 2.00 0.193 0.0187 | 2.02 0.193 0.0169 | 2.02 0.193 0.0162 | 2.02 0.193 0.0176 | 2.02 0.193 0.0179 |
| | 1.41 0.786 0.147 | 1.41 0.816 0.146 | 1.49 0.986 0.148 | 1.48 1.00 0.147 | 1.50 0.979 0.149 | 1.47 0.947 0.149 |
| | -2.72 0.135 0.162 | -2.85 0.135 0.166 | -2.80 0.137 0.173 | -2.85 0.137 0.173 | -2.74 0.137 0.171 | -2.74 0.137 0.166 |
| | 0.237 0.143 0.155 | 0.196 0.143 0.158 | 0.163 0.142 0.165 | 0.181 0.142 0.166 | 0.167 0.142 0.163 | 0.228 0.142 0.161 |
| | 2.32 0.0973 0.00648 | 2.32 0.0973 0.00635 | 2.33 0.0973 0.00640 | 2.33 0.0973 0.00623 | 2.34 0.0973 0.00654 | 2.33 0.0973 0.00642 |
| | 1.35 0.197 0.0644 | 1.34 0.197 0.0571 | 1.31 0.198 0.0475 | 1.29 0.198 0.0471 | 1.32 0.198 0.0493 | 1.30 0.197 0.0551 |
| | 2.11 0.114 0.0140 | 2.11 0.114 0.0132 | 2.13 0.114 0.0124 | 2.13 0.114 0.0124 | 2.14 0.114 0.0128 | 2.13 0.114 0.0129 |
| α=1.9 | -2.92 3.66 0.141 | -2.99 3.71 0.138 | -3.02 3.77 0.133 | -2.96 3.70 0.137 | -2.99 3.81 0.126 | -2.86 3.62 0.144 |
| | 1.01 0.521 0.137 | 1.01 0.520 0.137 | 1.01 0.517 0.138 | 1.01 0.521 0.137 | 1.00 0.515 0.139 | 0.996 0.525 0.137 |
| | -0.829 0.0529 0.150 | -0.858 0.0529 0.151 | -0.879 0.0530 0.153 | -0.853 0.0529 0.152 | -0.883 0.0530 0.154 | -0.787 0.0529 0.150 |
| | 1.94 0.192 0.0168 | 1.94 0.192 0.0168 | 1.94 0.192 0.0167 | 1.94 0.192 0.0168 | 1.93 0.192 0.0166 | 1.95 0.192 0.0169 |
| | 1.15 0.605 0.143 | 1.14 0.600 0.143 | 1.14 0.595 0.144 | 1.14 0.603 0.143 | 1.13 0.591 0.145 | 1.14 0.614 0.143 |
| | -2.90 0.134 0.167 | -2.91 0.134 0.168 | -2.95 0.133 0.169 | -2.95 0.134 0.168 | -3.03 0.133 0.171 | -2.94 0.134 0.165 |
| | 0.229 0.143 0.162 | 0.231 0.144 0.162 | 0.237 0.144 0.164 | 0.225 0.144 0.163 | 0.241 0.144 0.165 | 0.208 0.143 0.161 |
| | 2.28 0.0973 0.00540 | 2.28 0.0973 0.00537 | 2.28 0.0973 0.00534 | 2.28 0.0973 0.00539 | 2.28 0.0972 0.00533 | 2.29 0.0973 0.00545 |
| | 1.33 0.196 0.0662 | 1.33 0.196 0.0661 | 1.33 0.196 0.0664 | 1.33 0.196 0.0658 | 1.32 0.196 0.0666 | 1.32 0.196 0.0661 |
| | 2.06 0.114 0.0116 | 2.06 0.114 0.0115 | 2.06 0.114 0.0115 | 2.06 0.114 0.0116 | 2.06 0.114 0.0114 | 2.07 0.114 0.0117 |

*Notes*: A quasi-Newton algorithm was used to fit the normal mixture characteristic function to the characteristic function of stable laws. The objective functions were of the order $10^{-6}$ and the maximum absolute error was of the order $10^{-5}$.

**Table 9. Approximating normal mixture parameters to the distribution of $\log\varepsilon^2$, M=2 components.**

| | β=-0.9 | β=-0.5 | β=-0.25 | β=0.25 | β=0.5 | β=0.9 |
|---|---|---|---|---|---|---|
| α=1.1 | 2.18 3.46 0.146 | 1.45 3.73 0.182 | 0.591 4.03 0.234 | 0.580 4.02 0.237 | 1.42 3.74 0.184 | 2.15 3.51 0.139 |
| | 3.24 0.612 0.854 | 2.22 0.935 0.818 | 1.18 1.37 0.766 | 1.17 1.36 0.763 | 2.22 0.934 0.816 | 3.23 0.647 0.861 |
| α=1.3 | -0.287 3.51 0.243 | -0.753 3.75 0.268 | -1.01 3.86 0.273 | -0.993 3.82 0.289 | -0.714 3.73 0.267 | -0.272 3.49 0.252 |
| | 1.19 1.15 0.757 | 0.627 1.46 0.732 | 0.258 1.73 0.727 | 0.269 1.70 0.711 | 0.612 1.47 0.733 | 1.20 1.13 0.748 |
| α=1.5 | -1.72 3.47 0.221 | -1.89 3.57 0.224 | -1.93 3.58 0.230 | -2.01 3.63 0.229 | -1.86 3.56 0.234 | -1.65 3.44 0.226 |
| | 0.513 1.46 0.779 | 0.257 1.63 0.776 | 0.149 1.70 0.770 | 0.153 1.69 0.771 | 0.265 1.61 0.766 | 0.506 1.45 0.774 |
| α=1.7 | -3.30 3.06 0.159 | -3.29 3.00 0.164 | -3.29 3.05 0.168 | -3.29 2.99 0.171 | -3.29 3.11 0.166 | -3.29 2.99 0.164 |
| | 0.210 1.56 0.841 | 0.145 1.61 0.836 | 0.121 1.62 0.832 | 0.133 1.62 0.829 | 0.142 1.61 0.834 | 0.223 1.55 0.836 |
| α=1.9 | -3.29 2.77 0.192 | -3.39 2.80 0.185 | -3.39 2.82 0.185 | -3.39 2.76 0.187 | -3.31 2.76 0.191 | -3.31 2.71 0.194 |
| | 0.148 1.41 0.808 | 0.130 1.42 0.815 | 0.124 1.43 0.815 | 0.135 1.42 0.813 | 0.139 1.42 0.809 | 0.162 1.40 0.806 |

*Notes*: A quasi-Newton algorithm was used to fit the normal mixture characteristic function to the characteristic function of stable laws. The objective functions and the maximum absolute errors were of the order $10^{-4}$.

**Figure 15. Comparison of exact and approximate (finite normal mixture) densities of $\log \varepsilon^2$, M=2 components**

The "exact" densities were computed using kernel estimation based on the sample of stable random numbers that were used to match the characteristic functions.

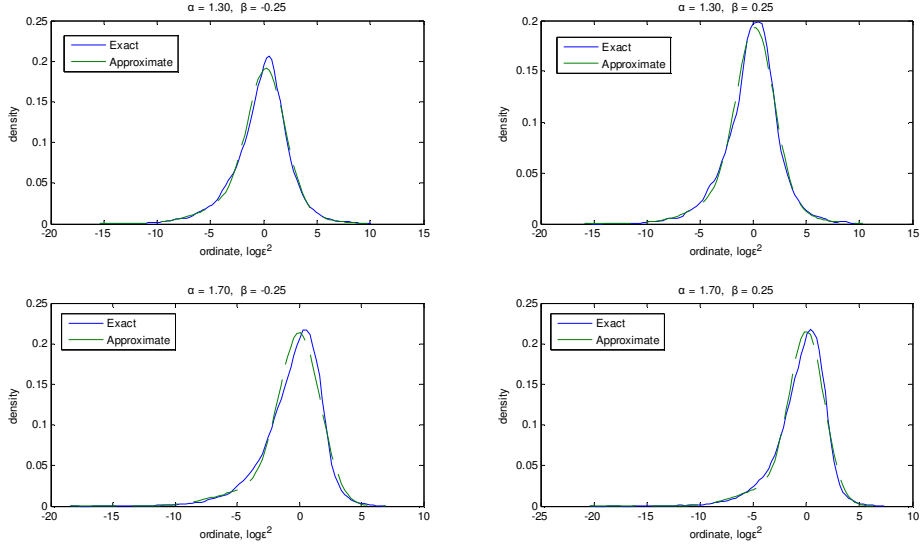

## 9.2 Exact methods

de Vries (1991) noted, for the first time, the relationships between GARCH processes and stable distributions. Meintanis and Taufel (2012) propose a characteristic function – based procedure for conditionally stable – distributed returns whose scale follows an autoregressive scheme with stable innovations. Before proceeding, it is important to mention that despite the fact that the density of $\log \chi_1^2$ is not convenient for ML or MCMC it, nevertheless, has a particularly convenient form of the characteristic function. For example, Knight *et al* (2002), and Yu (2004) show that for the following model with leverage:

$$X_t \triangleq \log Y_t^2 = h_t + \log \varepsilon_t^2, \quad \varepsilon_t \sim iid \mathcal{N}(0,1)$$

$$h_t = \delta + \rho h_{t-1} + v_t, \quad v_t \sim iid \mathcal{N}(0, \omega^2), \text{ and } Cov(\varepsilon_t, v_t) = \psi \omega^2, \tag{30}$$

the joint characteristic function of $X_t, X_{t+1}, ..., X_{t+k-1}$ is given by the following expression (see also Yu, 2004):

$$\varphi(\boldsymbol{\tau}) = \varphi(\tau_1, \tau_2, ..., \tau_k) = \exp\left(\frac{\iota \delta}{1-\rho} A_{k1} - \frac{\omega^2}{2(1-\rho^2)} A_{k2} - \frac{\sigma^2(1-\psi^2)}{2} A_{k3}\right) \cdot$$
$$\frac{\prod_{j=1}^{j} \Gamma\left(\frac{1}{2} + \iota \tau_j\right)}{\Gamma\left(\frac{1}{2}\right)^k} 2^{\iota A_{k1}} \prod_{j=2}^{k} {}_1 F_1\left(\tau_j + \frac{1}{2}, \frac{1}{2}, -\frac{\omega^2 \psi^2}{2} A_{j-1,2}\right), \tag{31}$$

where $A_{k1} = \sum_{j=1}^{k} \tau_j$, $A_{k2} = \left(\sum_{j=1}^{k} \tau_j \rho^{k-j}\right)^2$, $A_{k3} = \sum_{l=2}^{k}\left(\sum_{j=l}^{k} \tau_{k+1-j} \rho^{j-l}\right)^2$, and ${}_1F_1$ denotes the hypergeometric function. In practice, one has to choose the length, $k > 1$, of the moving blocks of data which affects directly the dimensionality of the characteristic function, but otherwise the characteristic function is quite easy to work with. Moreover, the presence of a leverage effect ($\psi \neq 0$) does not complicate the characteristic function.

## 9. Multivariate Stochastic volatility

### 9.1 Basic models

Multivariate Stochastic Volatility (MSV) models are quite difficult to work with and pose a significant impediment for applied work. To introduce the MSV model, suppose $\mathbf{y}_t = [y_{t1}, ..., y_{td}]'$ is a vector of returns in $\mathbb{R}^d$ (Chib, Omori, and Asai, 2009):

$$\mathbf{y}_t = \mathbf{V}_t^{1/2}\mathbf{u}_t, \; t=1,...,n,$$
$$\mathbf{h}_{t+1} = \boldsymbol{\mu} + \boldsymbol{\Phi}(\mathbf{h}_t - \boldsymbol{\mu}) + \boldsymbol{\varepsilon}_t \quad (32)$$
$$\mathbf{V}_t^{1/2} = diag\left[\exp(h_{t1}/2),...,\exp(h_{td}/2)\right], \; \mathbf{h}_t = [h_{t1},...,h_{td}]',$$

where $\boldsymbol{\mu} \in \mathbb{R}^d$ is a vector of parameters, and $\begin{bmatrix}\mathbf{u}_t\\\boldsymbol{\varepsilon}_t\end{bmatrix} | \mathbf{h}_t \sim N_{2d}(\mathbf{0},\boldsymbol{\Sigma})$, where $\boldsymbol{\Sigma} = \begin{bmatrix}\boldsymbol{\Sigma}_{uu} & \boldsymbol{\Psi}\\\boldsymbol{\Psi}' & \boldsymbol{\Sigma}_{\varepsilon\varepsilon}\end{bmatrix}$. For normalization purposes, the diagonal elements of $\boldsymbol{\Sigma}_{uu}$ are set to unity so that it is a correlation matrix. Usually the off-diagonal elements of $\boldsymbol{\Phi}$ are set to zero to simplify the analysis, so $\boldsymbol{\Phi} = diag[\phi_{11},...,\phi_{dd}]$. As explained in the excellent review of Chib, Omori, and Asai (2009) all approaches linearize the model using logs of squared returns and apply different procedures to obtain draws from the conditional distributions of $\mathbf{h}_t$. One prominent procedure is due to Smith and Pitts (2006) who proposed to sample in blocks and then use a Metropolis – Hastings procedure as in Chib and Greenberg (1994, 1995). Wong, Carter and Kohn (2003) propose a reparametrization and a prior for $\boldsymbol{\Sigma}$ in which the Metropolis – Hastings algorithm can be used to obtain random draws for each element in the reparametrization.

To reduce further the curse of dimensionality, factor models have been proposed (Chib, Nardari and Shephard, 2006, and Harvey, Ruiz and Shephard, 1994 among others). To extend the MSV model to the stable distributions, it is possible to assume that the elements of $\mathbf{u}_t$ follow standard $\mathcal{S}_{\alpha,\beta}(0,1)$ distributions. If we linearize the model, we obtain:

$$\mathbf{X}_t = \mathbf{h}_t + \boldsymbol{\xi}_t, \text{ where } \mathbf{X}_t = \left[\log y_{t1}^2,...,\log y_{td}^2\right]',$$

and the elements of $\boldsymbol{\xi}_t$ are independently distributed as $\log u^2$, where $u \sim \mathcal{S}_{\alpha,\beta}(0,1)$. In fact, the linearization is *not* necessary if we adopt the following procedure:

### *Multivariate Asymptotic Normal Form (MANF)*

For a fixed value of $k$, consider a fixed set of $d \times k$ matrices $\mathbb{T} = \left\{\mathbf{T}^{(1)},...,\mathbf{T}^{(G)}\right\}$.

1. Given the data $\mathbb{Y} = \{\mathbf{y}_t, t=1,...,n\}$ compute the empirical characteristic function:
$$\hat{\varphi}_{(g)}(\mathbf{T}) = (n-k+1)^{-1}\sum_{t=1}^{n-k+1}\exp\left(\iota \mathbf{Y}_{(t,k)}\mathbf{T}^{(g)}\right), \text{ where } \mathbf{Y}_{(t,k)} = [\mathbf{y}_t, \mathbf{y}_{t+1},...,\mathbf{y}_{t+k-1}], \text{ Let } \hat{\mathbf{z}}_{(g)}(\mathbf{T}) = \begin{bmatrix}\mathcal{R}\hat{\varphi}_{(g)}(\mathbf{T}^{(g)})\\\mathcal{I}\hat{\varphi}_{(g)}(\mathbf{T}^{(g)})\end{bmatrix}.$$

2. Denote the parameter vector by $\boldsymbol{\theta} = \{\boldsymbol{\mu},\boldsymbol{\Phi},\boldsymbol{\Sigma},\alpha,\beta\}$. Draw a parameter vector $\boldsymbol{\theta}$.

3. Simulate artificial data $\left\{\tilde{\mathbf{y}}_{t,s} \equiv \tilde{\mathbf{y}}_{t,s}(\boldsymbol{\theta}), t=1,...,n; s=1,...,S\right\}$, and compute the characteristic function $\tilde{\varphi}_{(s)}(\mathbf{T}^{(g)}) = (n-k+1)^{-1}\sum_{t=1}^{n-k+1}\exp\left(\iota\tilde{\mathbf{Y}}_{(t,k),s}\mathbf{T}^{(g)}\right)$, where $\tilde{\mathbf{Y}}_{(t,k),s} = [\tilde{\mathbf{y}}_{t,s},\tilde{\mathbf{y}}_{t+1,s},...,\tilde{\mathbf{y}}_{t+k-1,s}]$, and $\mathbf{T}$ is $d \times k$ a matrix.

4. Define $\tilde{\mathbf{z}}(\mathbf{T}^{(g)}) = S^{-1}\sum_{s=1}^{S}\begin{bmatrix}\mathcal{R}\tilde{\varphi}_{(s)}(\mathbf{T}^{(g)})\\\mathcal{I}\tilde{\varphi}_{(s)}(\mathbf{T}^{(g)})\end{bmatrix} \triangleq S^{-1}\sum_{s=1}^{S}\tilde{\mathbf{z}}_{(s)}(\mathbf{T}^{(g)})$, $g=1,...,G$.

5. Compute $\hat{\boldsymbol{\Omega}}(\mathbb{T},\boldsymbol{\theta}) = \text{cov}\left(\tilde{\mathbf{z}}(\mathbf{T}^{(g)}),\tilde{\mathbf{z}}(\mathbf{T}^{(g')})\right)$, $g,g' = 1,...,G$, and cov denotes the empirical covariance of $\tilde{\mathbf{z}}_{(s)}(\mathbf{T}^{(g)})$, an approximation to the optimal covariance matrix of the MANF.

6. Define the likelihood function of the MANF:
$$L_{MANF}(\boldsymbol{\theta}) \propto \left|\hat{\boldsymbol{\Omega}}(\mathbb{T},\boldsymbol{\theta})\right|^{-1/2}\exp\left[-\tfrac{n}{2}\left(\tilde{\mathbf{z}}_{(s)}(\mathbb{T},\boldsymbol{\theta}) - \hat{\mathbf{z}}(\mathbb{T})\right)'\hat{\boldsymbol{\Omega}}(\mathbb{T},\boldsymbol{\theta})^{-1}\left(\tilde{\mathbf{z}}_{(s)}(\mathbb{T},\boldsymbol{\theta}) - \hat{\mathbf{z}}(\mathbb{T})\right)\right]$$

Given a prior $p(\mathbf{\theta})$, one can define the posterior $p_{MANF}(\mathbf{\theta}\mid\mathbb{Y},\mathbb{T})\propto L_{MANF}(\mathbf{\theta};\mathbb{Y},\mathbb{T})p(\mathbf{\theta})$ and apply any MCMC procedure to draw. An alternative is to consider ABC inference based on the comparison of the empirical and simulated – theoretical characteristic function. This can be accomplished either using critical values[20] or selecting the parameter $\varepsilon$ so that approximately 50% of the proposed draws are accepted. The proposal distribution for ABC inference was a simplified form of the MANF using $\hat{\mathbf{\Omega}}(\mathbb{T},\mathbf{\theta})=\mathbf{I}$.

The major problem in the implementation of MANF is the choice of the set of $d\times k$ matrices $\mathbb{T}=\left\{\mathbf{T}^{(1)},...,\mathbf{T}^{(G)}\right\}$. Each row of the matrix corresponds to $k$ elements of the moving blocks of the data, and each column corresponds to each one of the $d$ variables. In addition we need G such matrices exactly as we need G grid points in the univariate characteristic function. For empirical purposes we prefer to work with a fixed set $\mathbb{T}$, where each element $\mathbf{T}^{(g)}$ is a random draw from a standard multivariate normal distribution, and the matrix is normalized to unity in the $L_2$ norm, viz. $\mathbf{T}^{(g)}=\left[\tau_{ij}\right]$, $\tau_{ij}\overset{IID}{\sim}N(0,1)$, subject to $\sum_{i=1}^{d}\sum_{j=1}^{k}\tau_{ij}=1$.

The other choices we have to make are: G (the number of matrices or the size of the grid in $\mathbb{R}^d\times\mathbb{R}^k$, the size of the moving blocks, $k$, the number of simulations, $S$, to obtain the simulated – theoretical characteristic function and the covariance $\hat{\mathbf{\Omega}}(\mathbb{T},\mathbf{\theta})$ of the MANF. In artificial experiments we have found that $S\approx 500$ is acceptable for $\alpha=1.50$ and $\beta=-0.50$ and in their vicinity ($\pm 0.2$). For the other parameters it is preferable to conduct sensitivity analysis with actual data to understand what values are plausible.

It is important to mention that we leave the parameters of $\mathbf{\Phi}$ unrestricted so unlike previous studies we do not assume that this matrix is diagonal. Moreover, we allow for a general $\mathbf{\Psi}$ matrix, allowing for general patterns of leverage. The priors of the parameters are as follows. For the parameters of $\mathbf{\Phi}$ we have:

$$p(\mathbf{\Phi})=\prod_{i,j=1}^{d}p(\phi_{ij}),\ \phi_{ii}\sim N(0.50,\ 0.2^2),\ i=1,...,d,\ \phi_{ij}\sim N(0,\ 0.1^2),\ i,j=1,...,d,\ i\neq j. \quad (33)$$

Matrix $\mathbf{\Sigma}$ is reparametrized as $\mathbf{\Sigma}=\mathbf{C}'\mathbf{C}$, where $\mathbf{C}$ is a lower triangular matrix whose elements $c_{ij}\overset{IID}{\sim}N(0,1)$, subject to the restriction that the diagonal elements of $\mathbf{\Sigma}_{uu}$ are unity.

## 9.2 Multivariate Stochastic Volatility and the Spectral Measure

For stable distributions it is somewhat unnatural to proceed as in the previous section because the formal definition of multivariate stable distributions involves the spectral measure. This does not, of course, preclude the empirical validity of the model we have proposed but we feel it is more natural to proceed through the formal spectral measure in the case of multivariate stable distributions. There is an additional motivation to do so, in that *through the spectral measure the MSV problem can be reduced to a univariate stochastic volatility problem.*

Given $\mathbf{X}_t\in\mathbb{R}^d$, the negative log characteristic function is $\mathcal{M}_{\mathbf{X}_t}(\mathbf{\tau})\triangleq-\log\varphi_{\mathbf{X}_t}(\mathbf{\tau})=I_{\mathbf{X}_t}(\mathbf{\tau})-\iota\langle\mathbf{\mu},\mathbf{\tau}\rangle$, where

$$I_{\mathbf{X}_t}(\mathbf{\tau})=\int_{\mathbb{S}^{d-1}}\psi_\alpha(\langle\mathbf{\tau},\mathbf{s}\rangle)\Gamma_t(d\mathbf{s}). \quad (34)$$

Suppose we use a discrete approximation to the spectral measure, so that

$$\Gamma_t(d\mathbf{s})=\sum_{i=1}^{N}\gamma_{i,(t)}\delta_{\{\mathbf{s}_i\}}(\mathbf{s}), \quad (35)$$

where $\gamma_{i,(t)}$ are time-varying weights, for all $t=1,...,n$, and fixed $\mathbf{s}_i\in\mathbb{S}^{d-1}$, $i=1,...,N$. If we make use of the normal approximation then

$$I_{\mathbf{X}_t}(\mathbf{\tau})\approx\mathbb{E}_{\Gamma(d\mathbf{s})}\psi_\alpha(\langle\mathbf{\tau},\mathbf{s}^{(t)}\rangle), \quad (36)$$

where $\mathbb{E}_{\Gamma(d\mathbf{s})}$ denotes expectation taken with respect to $\mathbf{s}^{(t)}\sim\mathcal{N}_d(\mathbf{0},\omega_t^2\mathbf{I})\mid\mathbf{s}^{(t)}\in\mathbb{S}^d$, given the time-varying parameters $\omega_t^2$, $t=1,...,n$. In the discrete case, we assume

$$\log\mathbf{\gamma}_{(t)}=\mathbf{\delta}+\mathbf{\Lambda}\log\mathbf{\gamma}_{(t-1)}+\mathbf{\varepsilon}_{(t)}, \quad (37)$$

---

[20] 90% critical values have been computed for different sample sizes ranging from n=100 to 5,000, values of α and β in [1.10, 1.90]x[-0.90, 0.90] and various combinations of the $\mathbf{T}$ matrices. There is only slight dependence of the critical values on $\mathbf{T}$. We have considered dimensions d=2, 5, 10 and 50. The results are available on request but are not reported to save space.

where $\boldsymbol{\gamma}_{(t)} = \left[\gamma_{1,(t)},...,\gamma_{N,(t)}\right]'$, $\boldsymbol{\delta}$ and $\boldsymbol{\Delta}$ are $N \times 1$ and $N \times N$ parameters, and $\boldsymbol{\varepsilon}_{(t)} \sim \mathcal{N}_N(\mathbf{0}, \boldsymbol{\Phi})$. In the normal approximation, we assume $\log \omega_t^2 = \delta + \Delta \log \omega_{t-1}^2 + \varepsilon_t$, where $\varepsilon_t \sim \mathcal{N}(0, \Phi)$, $\Phi > 0$. We use the following ABC procedure:

- Propose draws for the parameters $\alpha, \beta, \mu, \sigma$ and $(\boldsymbol{\delta}, \boldsymbol{\Delta}, \boldsymbol{\Phi})$ in the discrete case or $(\delta, \Delta, \Phi)$ in the normal approximation.
- Simulate artificial data for $\boldsymbol{\gamma}_{(t)}$ or $\omega_t^2$ using parameters $(\boldsymbol{\delta}, \boldsymbol{\Delta}, \boldsymbol{\Phi})$ or $(\delta, \Delta, \Phi)$.
- Compute the spectral measure $\Gamma_t(d\mathbf{s}) = \sum_{i=1}^{N} \gamma_{i,(t)} \delta_{\{\mathbf{s}_i\}}(\mathbf{s})$ and $I_{\mathbf{X}_t}(\boldsymbol{\tau}) = \sum_{i=1}^{N} \psi_\alpha(\langle \boldsymbol{\tau}, \mathbf{s}_i \rangle) \gamma_{i,(t)}$ in the discrete case, and $I_{\mathbf{X}_t}(\boldsymbol{\tau}) = \mathbb{E}_{\Gamma(d\mathbf{s})} \psi_\alpha(\langle \boldsymbol{\tau}, \mathbf{s}^{(t)} \rangle)$, where $\mathbb{E}_{\Gamma(d\mathbf{s})}$ denotes expectation taken with respect to $\mathbf{s}^{(t)} \sim \mathcal{N}_d(\mathbf{0}, \omega_t^2 \mathbf{I}) \mid \mathbf{s}^{(t)} \in \mathbb{S}^d$, given the time-varying parameters $\omega_t^2$ in the normal case.
- Compute $\mathcal{M}_{\mathbf{X}_t}(\boldsymbol{\tau}) \triangleq -\log \varphi_{\mathbf{X}_t}(\boldsymbol{\tau}) = I_{\mathbf{X}_t}(\boldsymbol{\tau}) - \iota \langle \boldsymbol{\mu}, \boldsymbol{\tau} \rangle$ and the empirical equivalent $\hat{\mathcal{M}}(\boldsymbol{\tau}) = -\log\left[n^{-1} \sum_{t=1}^{n} \exp(\iota \langle \boldsymbol{\tau}, \mathbf{X}_t \rangle)\right]$ or the moving-blocks approximation.
- Accept the draw if $\left| \hat{\mathcal{M}}(\boldsymbol{\tau}) - \sum_{t=1}^{n} \mathcal{M}_{\mathbf{X}_t} \right| \leq \varepsilon$, for some constant $\varepsilon > 0$.

There are two problems to address. First, how to propose draws for the parameters and second, how to determine $\varepsilon$. The second problem can always be handled using adaptation so that the overall acceptance rate of the ABC procedure is close to about 50%. To address the first problem we use a well-crafted proposal distribution. The construction of the proposal is as follows.

- We partition the sample into $P$ subsamples of approximately equal size, say $n_o$. Denote each partition by $\{\mathbf{X}_t^{(p)}, t = 1,...,n_o\}$, $p = 1,...,P$.
- For each subsample obtain $\Gamma^{(p)}(d\mathbf{s})$ using either the normal or the discrete approximation. As a result we have estimates of the stable distribution parameters $\hat{\theta}^{(p)} = \left[\hat{\alpha}^{(p)}, \hat{\beta}^{(p)}, \hat{\mu}^{(p)}, \hat{\sigma}^{(p)}\right]'$, along with $\boldsymbol{\gamma}^{(p)}$ in the discrete case or $\omega^{(p)}$ in the normal approximation case. The estimates are obtained using ABC-ANF for each subsample when $\boldsymbol{\gamma}$ or $\omega$ are fixed.
- Use least squares to fit $\log \omega^{2,(p)} = \hat{\delta} + \hat{\Delta} \log \omega^{2,(p-1)} + \varepsilon^{(p)}$ or $\log \boldsymbol{\gamma}^{(p)} = \hat{\boldsymbol{\delta}} + \hat{\boldsymbol{\Delta}} \log \boldsymbol{\gamma}^{(p-1)} + \boldsymbol{\varepsilon}^{(p)}$, $p = 2,...,P$, where $\omega^{2,(o)}$ and $\boldsymbol{\gamma}^{(o)}$ are obtained from ABC-ANF for the entire sample, and $\mathbb{V}\varepsilon^{(p)} = \hat{\Phi}$, $\mathbb{V}\boldsymbol{\varepsilon}^{(p)} = \hat{\boldsymbol{\Phi}}$, where $\mathbb{V}$ denotes the empirical (co)variance, $\mathbb{V}\mathbf{x} = n^{-1} \sum_{i=1}^{n}(\mathbf{x}_i - \bar{\mathbf{x}})(\mathbf{x}_i - \bar{\mathbf{x}})'$.
- Set $\hat{\boldsymbol{\theta}} = P^{-1} \sum_{p=1}^{P} \hat{\theta}^{(p)}$, $\hat{\mathbf{V}}_\theta = P^{-1} \sum_{p=1}^{P} (\hat{\theta}^{(p)} - \hat{\theta})$, and for $\hat{\zeta} = (\hat{\delta}, \hat{\Delta})$ or $\hat{\boldsymbol{\zeta}} = (\hat{\boldsymbol{\delta}}, \hat{\boldsymbol{\Delta}})$ denote by $\hat{\mathbf{V}}_\zeta$ the least squares quantities.
- The proposal is $\mathcal{N}_s\left(\begin{bmatrix}\hat{\boldsymbol{\theta}} \\ \hat{\boldsymbol{\zeta}}\end{bmatrix}, \begin{bmatrix}\hat{\mathbf{V}}_\theta & \mathbf{O} \\ \mathbf{O} & \hat{\mathbf{V}}_\zeta\end{bmatrix}\right)$, where $s = 4 + N + N^2$ for the discrete measure, and $s = 6$ for the normal approximation.

The proposals for the scale parameters are $\frac{\mathbb{V}\varepsilon^{(p)}}{\hat{\Phi}} \sim \chi^2(P)$, and $\hat{\boldsymbol{\Phi}} \sim \mathcal{W}\left(P, N(N+1), \mathbb{V}\boldsymbol{\varepsilon}^{(p)}\right)$, a Wishart distribution. Apparently, unless we have $P > N(N+1)$ this procedure cannot work. If we assume that $\boldsymbol{\Delta}$ is diagonal we need $P > 2N$ so with 20 partitions we cannot have more than 10 points in the support of the spectral measure. The normal approximation, on the other hand, is not subject to this "curse of dimensionality" and can be applied easily. We call these procedures "spectral" because they rely explicitly on the spectral measure. We have respectively the **Spectral-Discrete** (**SD**) and the **Spectral-Normal** procedures (**SN**).

## 10.2 An alternative approach: Principal Directions

In a relatively unnoticed but very important paper, Meerschaert and Scheffler (1999) showed that the uncentered sample moment matrix, the familiar $\mathbf{X}'\mathbf{X}$, contains useful information about tail behaviour as well as dependence. As they noted "*[t]he eigenvectors indicate a set of marginals which completely determine the moment behavior of the data, and the eigenvalues can be used to estimate the tail thickness of each marginal*". This, of course, stands in sharp contrast to methods that estimate the tail behavior or the characteristic exponent for each time series individually (Hill, 1975, McCulloch, 1997).

Meerschaert and Scheffler (1999) showed that if $\mathbf{X}$ is a random variable in $\mathbb{R}^k$ which belongs to the domain of attraction of stable laws, then there exist scalars $0 < \alpha_j \leq 2$, $j = 1,...,k$, such that $\mathbb{E}|X_j|^\alpha < \infty$ for $\alpha > \alpha_j$, $\mathbb{E}|X_j|^\alpha = \infty$ for $0 < \alpha < \alpha_j$, and moreover: For any unit vector $\theta$, $\mathbb{E}|\langle \mathbf{X}, \theta \rangle| < \infty$, for $0 < \alpha < \alpha(\theta)$, $\mathbb{E}|\langle \mathbf{X}, \theta \rangle| = \infty$ if $\alpha > \alpha(\theta)$, $\alpha(\theta) \triangleq \min\{\alpha_j : \theta^{(j)} \equiv \langle \theta, \theta_i \rangle \neq 0\}$. Since we allow for "time-changing" tail indices the following definitions are necessary: Let $X, X_1, X_2,...$ be iid random variables in $\mathbb{R}^k$. Then $X$ belongs to the *domain of attraction* of the k-dimensional random variable $Y$ if there exist $k \times k$ linear operators $A_n$ and constant vectors $a_n \in \mathbb{R}^k$, such that $A_n \sum_{t=1}^n X_t + a_n \overset{D}{\to} Y$ (Meerschaert and Scheffler, 2001). Notice that $Y$ is operator-stable with matrix exponent B, which means that if $Y_1, Y_2,...$ are iid with $Y$, for every $n$ there exists $b_n \in \mathbb{R}^d$ such that $n^{-B} \sum_{t=1}^n Y_t - b_n \overset{D}{=} Y$, and $n^{-B} = \exp(-B \log n)$. For more details see the excellent survey paper by Meerschaert and Scheffler (2003).

Let $\mathbb{X} = [\mathbf{x}'_t, t = 1,...,n]$ be the $n \times k$ matrix of observations on a k-variate process, and $\mathbb{X}'\mathbb{X}$ is the $k \times k$ uncentered sample moment matrix. Suppose now $\lambda_{n1},...,\lambda_{nk}$ denote the eigenvalues of $\mathbf{M}_n = \mathbb{X}'\mathbb{X}$. Then $2 \log n / \log \lambda_{nj} \to \alpha_j$, in probability for all $j = 1,...,k$. Convergence is almost sure if $\mathbb{E}\|X\|^2 < \infty$ (so that conditions of the classical central limit theorem apply) and all eigenvalues are distinct. Estimation can be also based on $n^{-1}\mathbb{X}'\mathbb{X}$ or the centered matrix, see Meerschaert and Scheffler (1999). The procedure yields a coordinate system in which the marginal distributions determine completely the tail behaviour, as well as a tail thickness estimate for each marginal, that is for each time series. Tail behaviour in any direction is determined by the heaviest tail marginal which has a non-vanishing component in this direction. The coordinate vectors are the eigenvectors of $\mathbb{X}'\mathbb{X}$. Eigenvector, say $\mathbf{p}_{\min}$ and $\mathbf{p}_{\max}$ corresponding to the minimum and maximum eigenvalue, provide directions for the multivariate distribution in which the tails are lightest and heaviest, respectively. Considering $\mathbf{p}'_i \mathbf{x}_t$ ($i = \min, \max$) is important in order to understand the temporal movements in the tail behaviour of the multivariate distribution. In connection with multivariate stable distributions this approach is important because it does not only yield useful estimates of the tail indices but it also provides useful estimates of the principal directions. Preliminary work in the context of exchange rates (Tsionas, 2012) has shown that only few currencies "load" in the principal directions (two or three) and the coefficients have an easy interpretation. The following procedure can be used to craft a proposal distribution using this approach.

- Use the method of Meerschaert and Scheffler (2003) to estimate principal directions in the direction of heaviest tails, $\mathbf{p}_{\max}$. As in Tsionas (2012) this can be applied[21] in $P$ subsamples to obtain an estimate $\hat{\boldsymbol{\tau}}$.
- For each subsample the method of Meerschaert and Scheffler (2003) also yields estimates $\hat{\alpha}^{(p)}$, $p = 1,...,P$, from which the empirical covariance $\mathbb{V}(\hat{\alpha}, \hat{\mu}, \hat{\sigma}, \hat{\boldsymbol{\tau}})$ can be obtained.
- In the case of *general* (non-symmetric) multivariate stable distributions the samples $\boldsymbol{\tau}' \mathbf{X}_t \sim \mathscr{S}_{\alpha, \beta}(0,1)$, so estimates $\hat{\alpha}^{(p)}, \hat{\beta}^{(p)}$, $p = 1,...,P$ can be obtained along with $\hat{\zeta} \triangleq (\hat{\alpha}, \hat{\beta}, \hat{\mu}, \hat{\sigma}, \hat{\boldsymbol{\tau}})$ and $\mathbb{V}\hat{\zeta}$.

From this point onwards, we can follow two routes. *First*, since we have accurate estimates of the stable parameters and the principal directions, it is straightforward to use ABC or ABC-ANF using as proposal a multivariate normal $\mathcal{N}_{4+d}\left(\hat{\zeta}, \mathbb{V}\left(\hat{\alpha},\hat{\beta},\hat{\mu},\hat{\sigma},\hat{\boldsymbol{\tau}}\right)\right)$. *Second*, this procedure is not explicitly based on the spectral measure so it cannot provide estimates of it.

---

[21] Since both τ and –τ are directions we impose the restriction that diagonal elements are positive, for identification purposes. The restriction is standard in factor analysis (Geweke and Zhou, 1997).

However, it is simple, in this case, to invert $I_{\mathbf{X}_t}(\boldsymbol{\tau}) = \int_{\mathbb{S}^{d-1}} \psi_\alpha\left(\langle\boldsymbol{\tau},\mathbf{s}\rangle\right)\Gamma_t(d\mathbf{s})$ when $\Gamma_t(d\mathbf{s}) = \sum_{i=1}^{N}\gamma_{i,(t)}\delta_{\{\mathbf{s}_i\}}(\mathbf{s})$ and obtain estimates of the spectral measure since the principal directions are known. This can be applied in each of the $P$ subsamples to obtain $\hat{\zeta} \triangleq \left(\hat{\alpha},\hat{\beta},\hat{\mu},\hat{\sigma},\hat{\boldsymbol{\tau}},\hat{\boldsymbol{\gamma}}\right)$ along with the empirical covariance $\mathbb{V}\hat{\zeta}$. *Notice that in this case, the estimates include not only the principal directions but also the weights of the spectral measure.* Again, we can use as proposal a multivariate normal $\mathcal{N}_{4+d}\left(\hat{\zeta}, \mathbb{V}\hat{\zeta}\right)$ in connection with ABC or ABC-ANF procedures.

To craft a proposal for $(\boldsymbol{\delta},\boldsymbol{\Delta},\boldsymbol{\Phi})$ in the discrete case or $(\delta,\Delta,\Phi)$ in the normal approximation we can follow the same least squares procedure as before. We call these procedures "***Principal-Directions-Based***" (***PD***) depending on whether simple ABC or ABC in the context of the ANF is used. Moreover, the procedures differ in terms of whether a discrete or a normal approximation is used for the spectral measure.

Before proceeding, it should be mentioned that the discrete spectral measure procedures can, in fact, be used to estimate the parameters $(\delta,\Delta,\Phi)$ in the normal approximation. This is useful in its own right but also because we need a common "benchmark" to compare the different procedures in the *same* model below, where we consider a Monte Carlo experiment. Since parameters $(\boldsymbol{\delta},\boldsymbol{\Delta},\boldsymbol{\Phi})$ are not comparable to $(\delta,\Delta,\Phi)$ because they describe the time-varying process of spectral weights but not the time-varying process of the volatility in the normal case, we proceed as follows.

- Given parameters $(\boldsymbol{\delta},\boldsymbol{\Delta},\boldsymbol{\Phi})$ compute the discrete approximation to the spectral measure for all subsamples through $\left(\mathbf{s},\boldsymbol{\gamma}^{(p)}\right)$, $p=1,...,P$.
- Compute $\tilde{\omega}^{2,(p)} = \sum_{i=1}^{N}(s_i - \bar{s})^2 \gamma_i^{(p)}$, $\bar{s} = \sum_{i=1}^{N} s_i \gamma_i^{(p)}$, for each $p=1,...,P$.
- Use LS fit to obtain $\log\tilde{\omega}^{2,(p)} = \hat{\delta} + \hat{\Delta}\log\tilde{\omega}^{2,(p-1)} + \varepsilon^{(p)}$, $\mathbb{V}\hat{\varepsilon}^{(p)} = \hat{\Phi}$.
- Use $\left(\hat{\delta},\hat{\Delta},\hat{\Phi}\right)$ and $\mathbb{V}\left(\hat{\delta},\hat{\Delta},\hat{\Phi}\right)$ to formulate a multivariate normal proposal for ABC or ANF.

Part of the attraction of the procedure of Meerschaert and Scheffler (1999) is that the principal directions are easy to compute so efficient MCMC proposals can be employed to provide full Bayesian inference for these parameters. The question is how this approach compares with the "explicit" approach based on the spectral measure. To examine the issues we use the following Monte Carlo experiment. Given the dimensionality $d$ we assume that the normal approximation to the spectral measure is, in, fact, exact and $\log\omega_t^2 = \delta + \Delta\log\omega_{t-1}^2 + \Phi^{1/2}\xi_t$, where $\xi_t \overset{IID}{\sim} \mathcal{N}(0,1)$, $\log\omega_0^2 = -1$, $t=1,...,n$, $\delta = -0.1$, $\Delta = 0.9$, $\Phi = 0.01$ where $n$ is the sample size. Moreover, the points of the support of the spectral measure $\mathbf{s}^{(t)} \sim \mathcal{N}_d\left(\mathbf{0},\omega_t^2\mathbf{I}\right) \mid \mathbf{s}^{(t)} \in \mathbb{S}^d$, given the time-varying parameters $\omega_t^2$, $t=1,...,n$. We fix the sample size to $n=1500$ which is typical for most applications of the univariate stochastic volatility model. For the Monte Carlo experiment a different set of $\mathbf{s}^{(t)} \sim \mathcal{N}_d\left(\mathbf{0},\omega_t^2\mathbf{I}\right) \mid \mathbf{s}^{(t)} \in \mathbb{S}^d$ has been used. The experiment is based on 1,000 replications to minimize computational costs. MCMC is based on 120,000 draws the first 10,000 of which are discarded and we thin every other 10[th] draw. This finally produces 10,000 draws per replication.

Table 10. Results of Monte Carlo experiment, symmetric case, $\alpha = 1.75$, $\beta = 0$

|  | $\delta$ | $\Delta$ | $\Phi^{1/2}$ | $\alpha$ | $\beta$ |
|---|---|---|---|---|---|
| dimensionality, $d=2$ | | | | | |
| Spectral-Discrete | -0.11 (0.016) | 0.91 (0.020) | 0.12 (0.010) | 1.76 (0.012) | -0.015 (0.070) |
| Spectral-Normal | -0.11 (0.014) | 0.93 (0.015) | 0.12 (0.016) | 1.75 (0.013) | -0.013 (0.061) |
| PD-ABC-discrete | -0.13 (0.019) | 0.94 (0.022) | 0.11 (0.015) | 1.77 (0.013) | -0.018 (0.072) |
| PD-ABC-normal | -0.11 (0.011) | 0.94 (0.019) | 0.14 (0.013) | 1.74 (0.013) | -0.015 (0.072) |
| PD-ANF-discrete | -0.11 (0.015) | 0.93 (0.020) | 0.12 (0.015) | 1.76 (0.011) | -0.017 (0.071) |
| PD-ANF-normal | -0.12 (0.015) | 0.92 (0.011) | 0.13 (0.013) | 1.76 (0.023) | -0.012 (0.085) |
| PDS-ANF-discrete | -0.14 (0.021) | 0.85 (0.032) | 0.16 (0.022) | 1.81 (0.043) | --- |
| PDS-ANF-normal | -0.14 (0.015) | 0.92 (0.021) | 0.11 (0.012) | 1.77 (0.013) | --- |
| dimensionality, $d = 5$ | | | | | |
| Spectral-Discrete | -0.15 (0.032) | 0.95 (0.044) | 0.15 (0.035) | 1.79 (0.041) | 0.12 (0.091) |
| Spectral-Normal | -0.15 (0.011) | 0.95 (0.012) | 0.15 (0.011) | 1.79 (0.012) | -0.07 (0.031) |
| PD-ABC-discrete | -0.13 (0.041) | 0.92 (0.043) | 0.12 (0.053) | 1.72 (0.025) | 0.03 (0.028) |
| PD-ABC-normal | -0.11 (0.013) | 0.90 (0.011) | 0.11 (0.010) | 1.74 (0.011) | 0.02 (0.024) |
| PD-ANF-discrete | -0.12 (0.043) | 0.91 (0.031) | 0.12 (0.030) | 1.75 (0.021) | 0.02 (0.022) |
| PD-ANF-normal | -0.11 (0.011) | 0.90 (0.012) | 0.10 (0.012) | 1.76 (0.010) | 0.015 (0.022) |
| PDS-ANF-discrete | -0.12 (0.041) | 0.91 (0.035) | 0.12 (0.022) | 1.76 (0.032) | --- |
| PDS-ANF-normal | -0.11 (0.012) | 0.92 (0.014) | 0.11 (0.007) | 1.75 (0.008) | --- |
| dimensionality, $d = 10$ | | | | | |
| Spectral-Discrete | -0.24 (0.065) | 0.77 (0.076) | 0.22 (0.067) | 1.63 (0.077) | 0.24 (0.034) |
| Spectral-Normal | -0.11 (0.007) | 0.93 (0.008) | 0.10 (0.003) | 1.75 (0.006) | -0.012 (0.011) |
| PD-ABC-discrete | -0.11 (0.066) | 0.93 (0.078) | 0.10 (0.082) | 1.75 (0.078) | -0.012 (0.044) |
| PD-ABC-normal | -0.12 (0.004) | 0.91 (0.003) | 0.11 (0.001) | 1.73 (0.002) | 0.012 (0.011) |
| PD-ANF-discrete | -0.10 (0.071) | 0.90 (0.066) | 0.12 (0.035) | 1.73 (0.067) | 0.012 (0.046) |
| PD-ANF-normal | -0.10 (0.003) | 0.90 (0.005) | 0.11 (0.001) | 1.75 (0.002) | 0.012 (0.011) |
| PDS-ANF-discrete | -0.11 (0.077) | 0.90 (0.067) | 0.12 (0.082) | 1.74 (0.071) | --- |
| PDS-ANF-normal | -0.11 (0.004) | 0.90 (0.003) | 0.10 (0.002) | 1.75 (0.002) | --- |

*Notes*: PDS stands for the Principal Directions Method when a symmetry assumption is explicitly made. The table reports sampling averages of posterior means and sampling standard deviations in parentheses. All discrete procedures use $N = 10$ points in the support of the spectral measure. In all approaches we keep the same number of subsamples ($P = 10$) of equal size ($n_o = 150$) to have a common ground for comparison of estimates and standard deviations.

From the Monte Carlo experiment, the main message is that the performance of the Spectral-Discrete procedure deteriorates rapidly as the dimensionality of the problem increases from 2 to 10. The performance of the Spectral-Normal procedure remains robust and compares favorably with the much simpler computationally PD procedures. We have failed to document any significant differences between the ABC and ANF in this setup and it seems that they behave similarly although ANF is slightly better. The discrete approximations to the spectral measure provide *relatively* accurate parameter as the dimensionality increases but from the reported standard errors it seems that their quality deteriorates fast. Of course the fact that they remain unbiased is of little use when the standard errors increase rapidly *relative* to the other approximations.

Since the spectral measures are of independent interest, in Figure 16 we report sampling expectations (across all 10,000 replications) of the spectral measure as estimated by different procedures in four different time periods ($10^{th}$, $100^{th}$, $500^{th}$, and $1000^{th}$). For visual purposes we report the true measure and two approximations: The first is based on the Spectral-Discrete approach and the second on the PD-ANF procedure.

**Figure 16. True and estimated spectral measures, $d = 10$,
Multivariate Stochastic Volatility Stable Model.**

*Notes*: PD-ANF denotes the ANF procedure in conjunction with the Principal Directions technique. SD denotes the Spectral-Discrete approach. The SD approach uses 10 points in the support of the spectral measure.

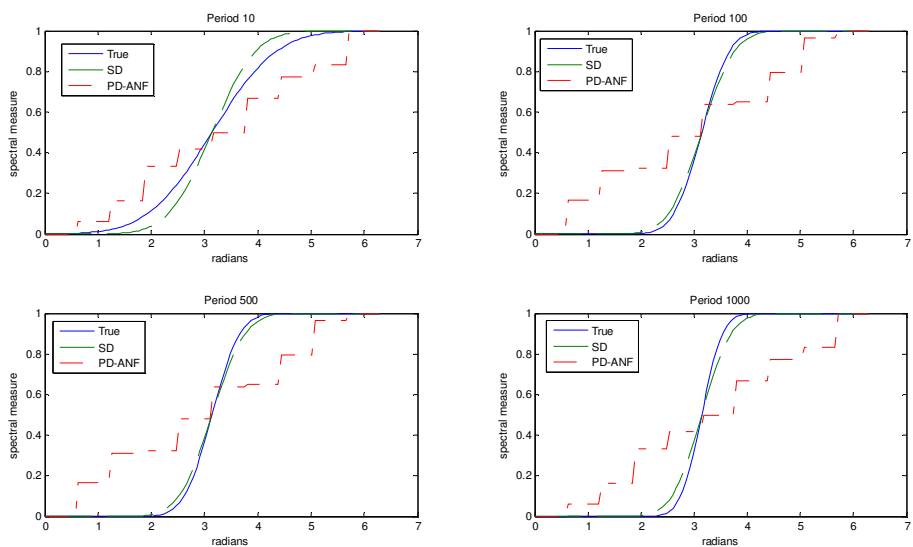

Apparently, the PD approach works much better and provides a close approximation to the true measure (which was computed using the exact parameter values for the multivariate volatility process) while the SD procedure seems to overestimate heavily the spectral measure at low "frequencies" and underestimate them heavily at high "frequencies". Results (not reported here to save space) show clearly that the SD approach performs much better when $d = 2$ but begins to deteriorate as shown in the Figure above in higher dimensions.

## 9.3 Direct Multivariate Stochastic Volatility Stable Models

In the context of multivariate stable distributions we described above what we believe to be a reasonable approach to stochastic volatility, viz. either stable models whose spectral weights evolve over time or normal spectral measures whose scale parameter follows a univariate stochastic volatility process. We have described numerical MCMC procedures which were shown to perform well and appear quite reasonable in the context of actual data.

Next, we consider an alternative or direct model for multivariate stable distributions with stochastic volatility. The idea is that we can keep the spectral measure time-invariant (either discrete or normal) and model directly the scale parameters of the stable distributions. As before, we can use the following model:

$$\mathbf{y}_t = \mathbf{V}_t^{1/2} \mathbf{u}_t + \boldsymbol{\xi}_t, \ t = 1, ..., n,$$

$$\mathbf{h}_{t+1} = \boldsymbol{\mu} + \boldsymbol{\Phi}(\mathbf{h}_t - \boldsymbol{\mu}) + \boldsymbol{\varepsilon}_t,$$

$$\mathbf{V}_t^{1/2} = diag\left[\exp(h_{t1}/2), ..., \exp(h_{td}/2)\right], \ \mathbf{h}_t = \left[h_{t1}, ..., h_{td}\right]',$$

where $\boldsymbol{\mu} \in \mathbb{R}^d$ is a vector of parameters, and $\begin{bmatrix}\boldsymbol{\varepsilon}_t \\ \boldsymbol{\xi}_t\end{bmatrix} \sim \mathcal{N}_{2d}(\mathbf{0}, \boldsymbol{\Sigma})$. Usually the off-diagonal elements of $\boldsymbol{\Phi}$ are set to zero to simplify the analysis, so $\boldsymbol{\Phi} = diag[\phi_{11}, ..., \phi_{dd}]$ but here we leave this matrix unrestricted. Moreover $\mathbf{u}_t$ follows a *standard* multivariate stable distribution with spectral measure $\Gamma(d\mathbf{s})$ which implies that the negative log characteristic function is $\mathcal{M}_{\mathbf{u}_t}(\boldsymbol{\tau}) \triangleq -\log \varphi_{\mathbf{u}_t}(\boldsymbol{\tau}) = I_{\mathbf{u}_t}(\boldsymbol{\tau})$, where $I_{\mathbf{u}_t}(\boldsymbol{\tau}) = \int_{\mathbb{S}^{d-1}} \psi_\alpha(\langle\boldsymbol{\tau},\mathbf{s}\rangle)\Gamma_t(d\mathbf{s})$. Notice that in this model the dependence between errors in the mean and volatility is captured by the correlation between the additional normal error term $\boldsymbol{\xi}_t$ and the volatility error, $\boldsymbol{\varepsilon}_t$.

As we mentioned, it can be shown that $\mathbf{u}_t = \sum_{j=1}^{D_N} \gamma_j^{1/\alpha} \mathcal{Z}_j \mathbf{s}_j$, where $\mathcal{Z}_j \sim iid\mathcal{S}_{\alpha,1}(\mu^*, 1)$, $j = 1, ..., N$, see Modarres and Nolan (1994). The interpretation is that a multivariate α-stable random vector can be represented as a finite mixture of univariate α-stable variates which are totally skewed to the right (that is, they have skewness coefficients $\beta = 1$). For $\alpha = 1$ we have $\mathbf{u}_t = \sum_{j=1}^{D_N} \gamma_j^{1/\alpha} \left(\mathcal{Z}_j + \frac{2}{\pi}\log\gamma_j\right)\mathbf{s}_j$.

Suppose we use a discrete approximation to the spectral measure, so that $\Gamma(d\mathbf{s}) = \sum_{i=1}^{N} \gamma_i \delta_{\{\mathbf{s}_i\}}(\mathbf{s})$, where $\gamma_i$ are weights, $i=1,...,n$, and fixed $\mathbf{s}_i \in \mathbb{S}^{d-1}$, $i=1,...,N$. The parameters of the model are $\zeta = (\boldsymbol{\mu}, \boldsymbol{\Phi}, \boldsymbol{\Sigma}, \boldsymbol{\gamma}, \alpha)$. Unlike the case with the stochastic volatility stable models in the previous section there is no curse of dimensionality in terms of $\boldsymbol{\gamma}$ although there is one in terms of $\boldsymbol{\Phi}$. The major impediment in existing MCMC analysis of the MSV model is drawing the latent states $\{\mathbf{h}_t, t=1,...,n\}$. Once a well-crafted proposal can be constructed for $\zeta$ this is no longer a problem for ABC-type inference.

- Draw parameters $\zeta = (\boldsymbol{\mu}, \boldsymbol{\Phi}, \boldsymbol{\Sigma}, \boldsymbol{\gamma}, \alpha)$ from a proposal distribution.
- Simulate artificial data $\{\tilde{\mathbf{y}}_t(\zeta), t=1,...,n\}$, fix the length, $B$, of moving blocks, and let $\tilde{\mathbf{Y}}_i(\zeta) = [\tilde{\mathbf{y}}_{(i-1)B+1 \,:\, iB}]$, $i=1,...,n/B$.
- Compute the simulated log characteristic function $\log \tilde{\varphi}_{\tilde{\mathbf{Y}}_i(\zeta)}(\boldsymbol{\tau})$ and the empirical characteristic function $\log \hat{\varphi}_{\mathbf{Y}_i}(\boldsymbol{\tau})$, where $\mathbf{Y}_i = [\mathbf{y}_{(i-1)B+1 \,:\, iB}]$, $i=1,...,n/B$, is the moving blocks in the data.
- Accept the draw using the ABC or ANF criteria.

The $\boldsymbol{\tau}$s can either be fixed in advance or can be made parameters to draw in the context of MCMC. They can be fixed at the Principal Direction estimates which as we showed produce very accurate estimates. If they are treated as parameters, to draw them we can still use the Principal Direction estimates and their empirical covariance to craft a reasonable proposal distribution, say $Q(\boldsymbol{\tau})$. There remains the problem to craft a proposal for $\zeta = (\boldsymbol{\mu}, \boldsymbol{\Phi}, \boldsymbol{\Sigma}, \boldsymbol{\gamma}, \alpha)$.

A proposal for $(\boldsymbol{\mu}, diag\boldsymbol{\Phi}, diag\boldsymbol{\Sigma})$ can be obtained from the univariate log-squared-return processes which can be estimated using a normal mixture approximation for their error terms. Unfortunately this procedure is not capable of providing a reasonable approximation to a scale matrix that can be used to construct a relatively accurate proposal. For this reason, it is applied in $P$ subsamples of the original data set from which their empirical covariance can be obtained and used as a scale matrix. To the same subsamples we fit multivariate α-stable distributions from which estimates and the empirical covariance of $(\boldsymbol{\gamma}, \alpha)$ can be constructed. Given this construction the product measure $Q(\boldsymbol{\tau}) \times Q(\boldsymbol{\mu}, \boldsymbol{\Phi}, \boldsymbol{\Sigma}) \times Q(\boldsymbol{\gamma}, \alpha)$ is used as a proposal distribution to implement an efficient Metropolis-Hastings algorithm for ABC or ANF. For the non-diagonal elements of $\boldsymbol{\Phi}, \boldsymbol{\Sigma}$ (for the entire matrices, to be more precise) the required proposal is based on the estimates and the empirical covariance from the subsamples where the univariate log-squared-return processes are estimated.

An alternative proposal can be constructed if we proceed in a somewhat different way. *The log-squared-return processes are estimated in their multivariate form* using a normal approximation for the error terms. Using MCMC we obtain posterior means and the posterior covariance of parameters $(\boldsymbol{\mu}, \boldsymbol{\Phi}, \boldsymbol{\Sigma})$. The joint proposal for $(\boldsymbol{\tau}, \boldsymbol{\gamma}, \alpha)$ is obtained from PD analysis in the entire sample and their covariance is obtained from the $P$ subsamples.

Due to the dimensionality of the problem and the large number of parameters it is not possible to provide critical values for the distance between the simulated and the empirical characteristic function that can be useful for further research. Therefore, we decided to implement the ABC and ANF procedures by tuning the various constants so that the acceptance rate is not too high or too low (90% and 10% respectively). Without any adjustments we were able to obtain acceptance rates between 60% and 75% both in artificial and real data.

## 9.4 Empirical results

We use two data sets, one for 100 stocks of the Standard & Poor's index (minute data, 23-29/10/2009)[22] and ten major currencies. The stock data have been used in Plataniotis and Dellaportas (2012). The exchange rate data is daily, against the US dollar over the period July 3 1996 to May 21 2012. The currencies are Canadian dollar, Euro, Japanese yen, British pound, Swiss franc, Australian dollar, Hong-Kong dollar, New Zealand dollar, South Korean won and Mexican peso.

In Figure 17 we report histograms of posterior means of parameters $(\delta, \Delta, \Phi^{1/2}, \alpha)$ from the approximating univariate stable – stochastic volatility models.

---

[22] I wish to thank P. Dellaportas and A. Plataniotis for providing the data of their study.

**Figure 17. Histograms of posterior means of parameters from the approximating univariate stable – stochastic volatility models**

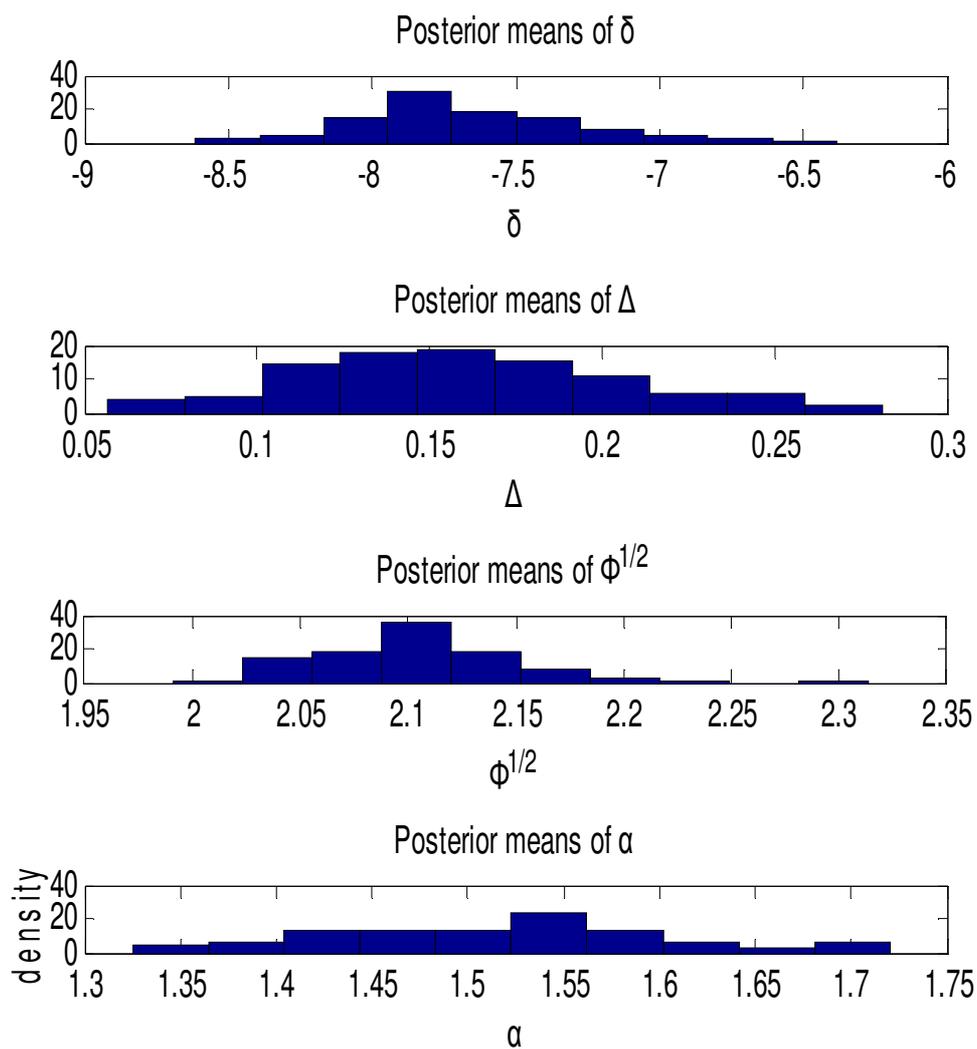

Typical marginal posterior distributions of the autoregressive volatility parameter are reported in Figure 18, for 20 stocks of the SP100.

# Figure 18. Marginal posterior distributions of Δ

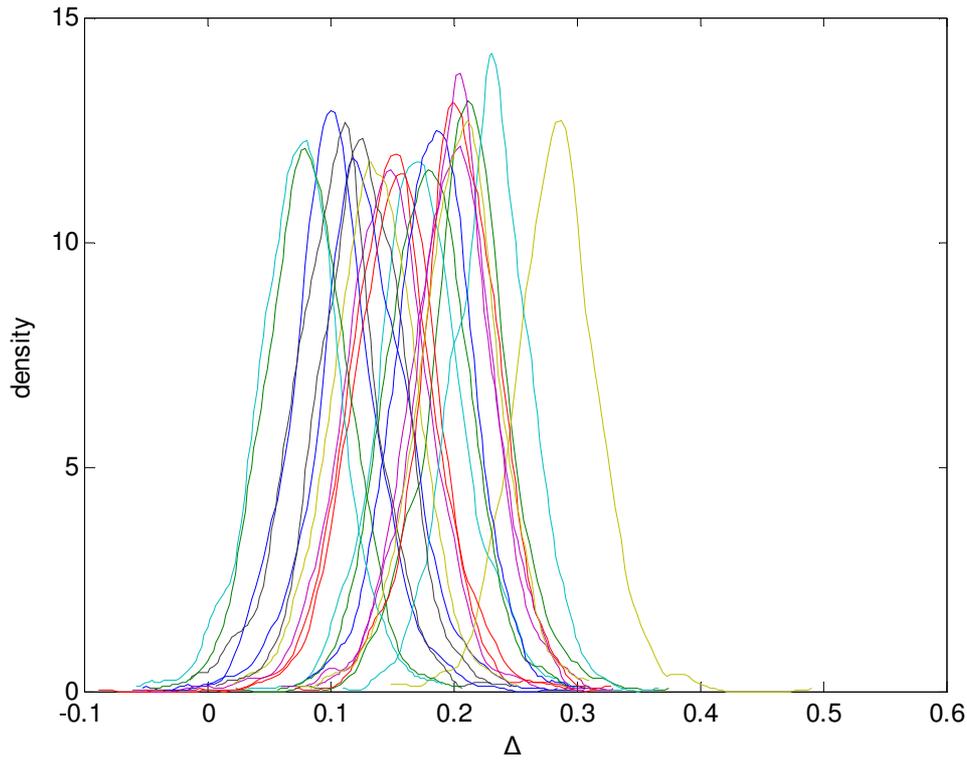

In Figure 19a we present plots of the posterior means of directions resulting from the ANF and the proposal based on PD for both the SP100 data (left) and the ten major exchange rates (right). We should note that for the SP100 data, 20 stocks account for almost 50% of the variation in all 100 stocks (based on the eigenvalues of the cross-product matrix of the data scaled by the grand median). For the exchange rates, 30% of the variation is explained by nine exchange rates. For the SP100 posterior directions show that at most five are needed. Dependence is significant as can be seen from Figure 19b. We have also found that this dependence is not due to specific stocks, at least for the most part. From posterior directions (not reported as they form a 100×100 matrix) only 22 stocks "load" on others (by more than two s.d.) and only 9 times out of 100 a stock "loads" on all others (this is stock number 100). Stocks 32 and 94 load on 8 other while stocks 59,60,76,78,88 on 7 others.

Figure 19a. Posterior means of directions

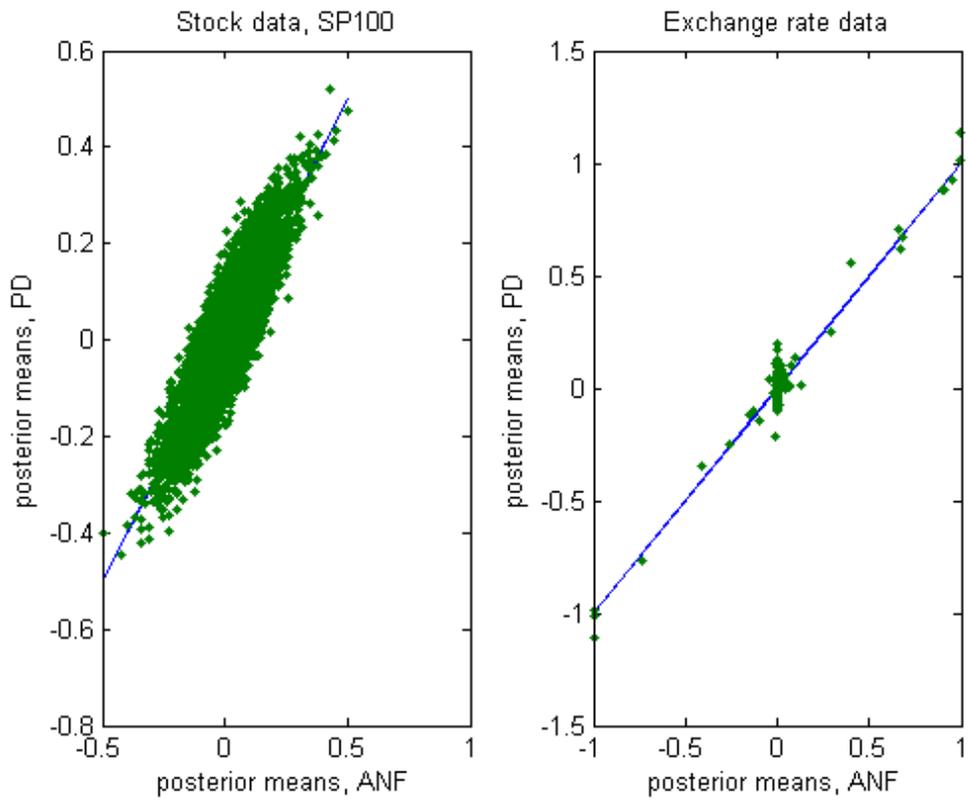

Figure 19b. Posterior means of $\alpha$ from ANF and the PD proposal, SP100

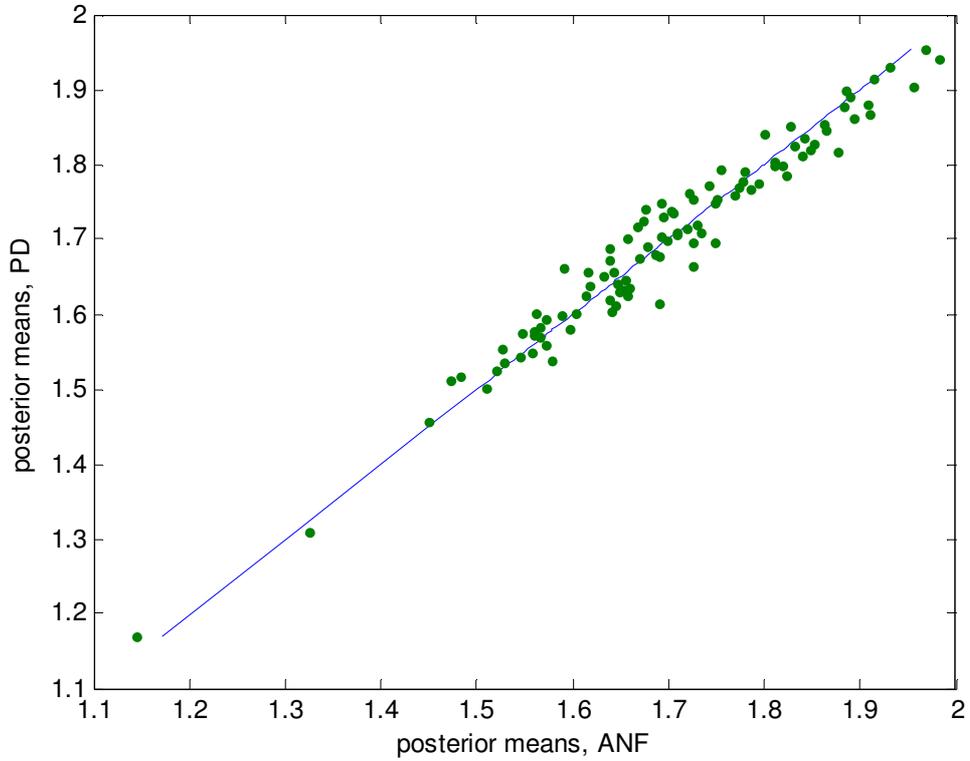

Figure 20. Marginal posterior distributions of α, Exchange Rates

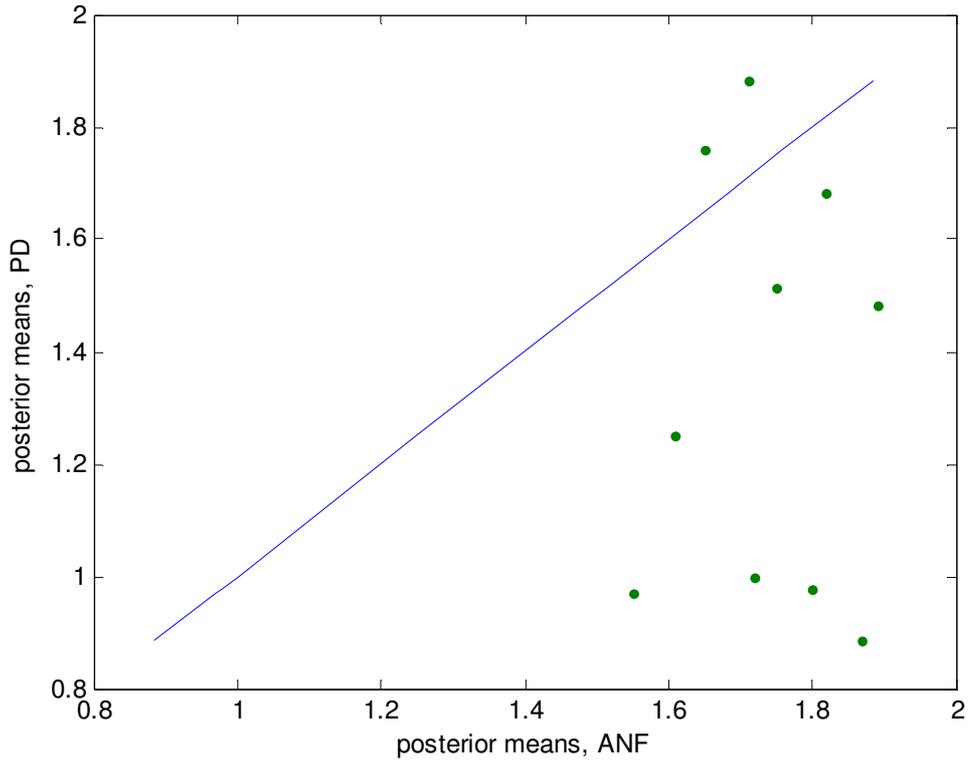

The marginal posterior distributions of $\alpha$ and $\beta$ are reported in Figure 21.

Figure 21. Marginal posterior distributions of $\alpha$ and $\beta$, SP100

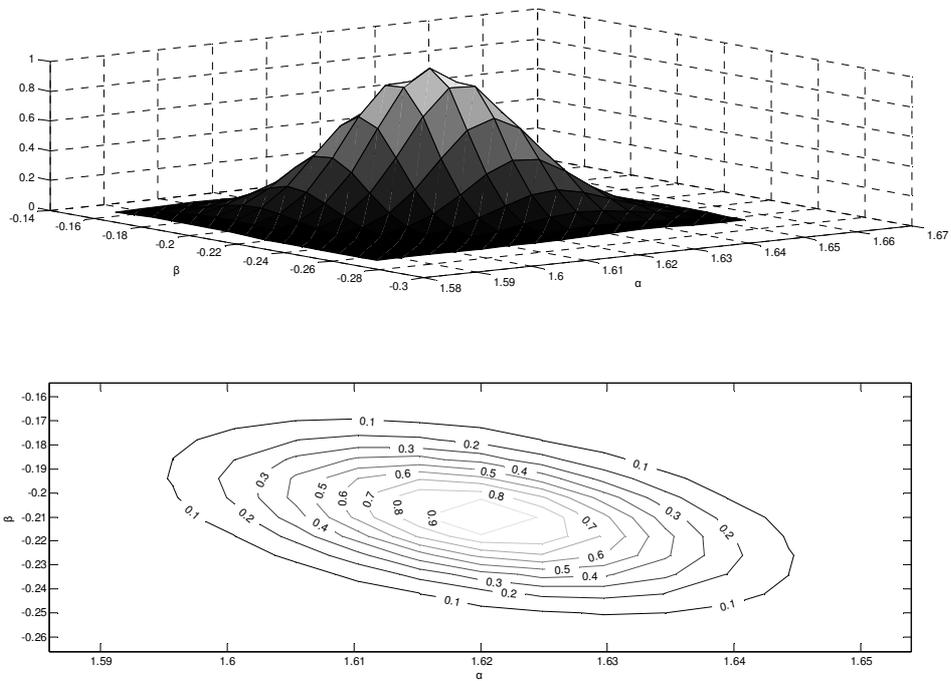

Table 11. Computational efficiency and convergence, SP100 data

|  | RNE | | CD | | Max autocorrelation at lag 10 | |
|---|---|---|---|---|---|---|
|  | PD | ANF | PD | ANF | PD | ANF |
| $\alpha$ | 0.33 | 0.12 | 0.43 | 1.32 | 0.51 | 0.12 |
| $\beta$ | 0.21 | 0.35 | 1.12 | 1.20 | 0.35 | 0.17 |
| $\tau$ | 0.41-1.35 | 0.45-0.72 | 0.86-1.71 | 0.61-1.28 | -0.12-0.61 | -0.17-0.62 |
| $\delta$ | 0.37 | 0.40 | 1.33 | 0.93 | 0.40 | 0.21 |
| $\Delta$ | 0.45 | 0.45 | 0.16 | 1.15 | 0.32 | 0.11 |
| $\Phi$ | 0.32 | 0.55 | 1.11 | 0.32 | 0.45 | 0.21 |

*Notes*: RNE is relative numerical efficiency. CD is the absolute value of the convergence diagnostic. For the directions, $\tau$, the statistics reported are minimum and maximum. PD stands for "Principal Directions" and ANF for "Asymptotic Normal Form". The CD is a *t*-statistic computed for the means of the first 50% and last 25% of the final draws.

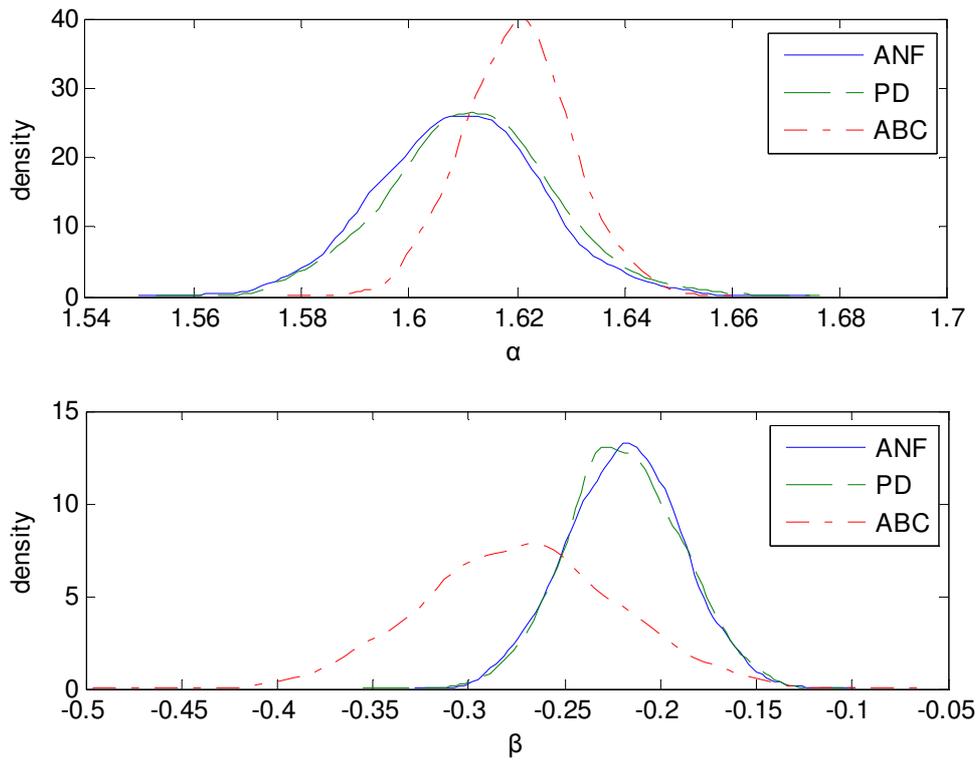

Figure 22. Marginal posterior distributions of $\alpha$ and $\beta$.

Figure 23. Median absolute autocorrelation functions of τ

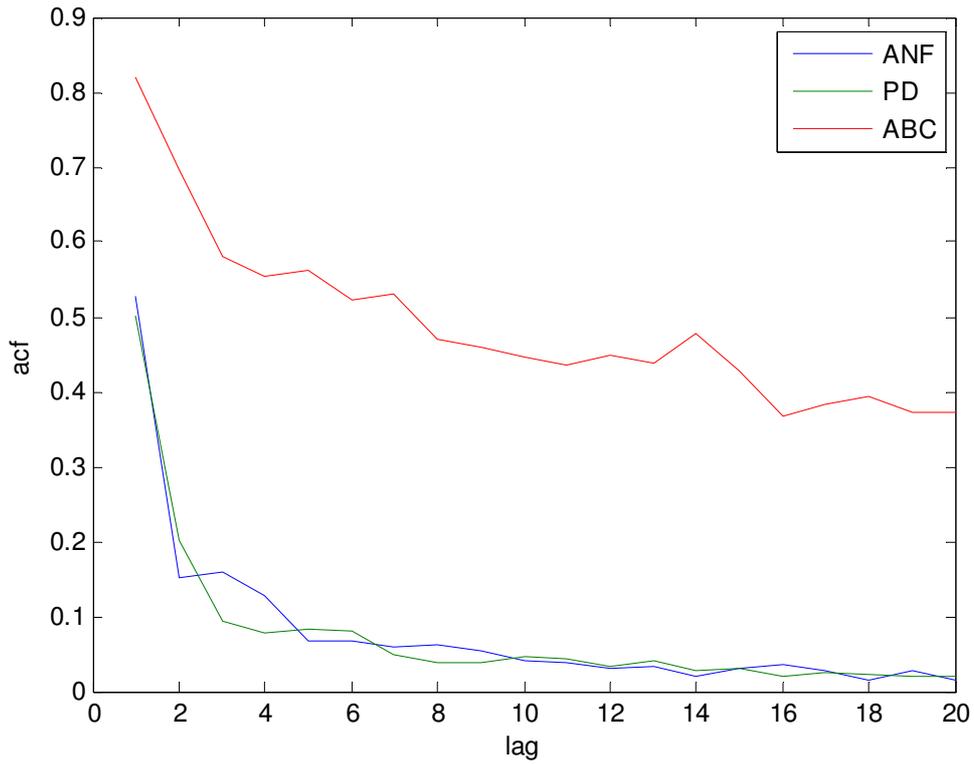

Figure 24. Rank correlations from the Copula model and median absolute acf of draws, Exchange rate data

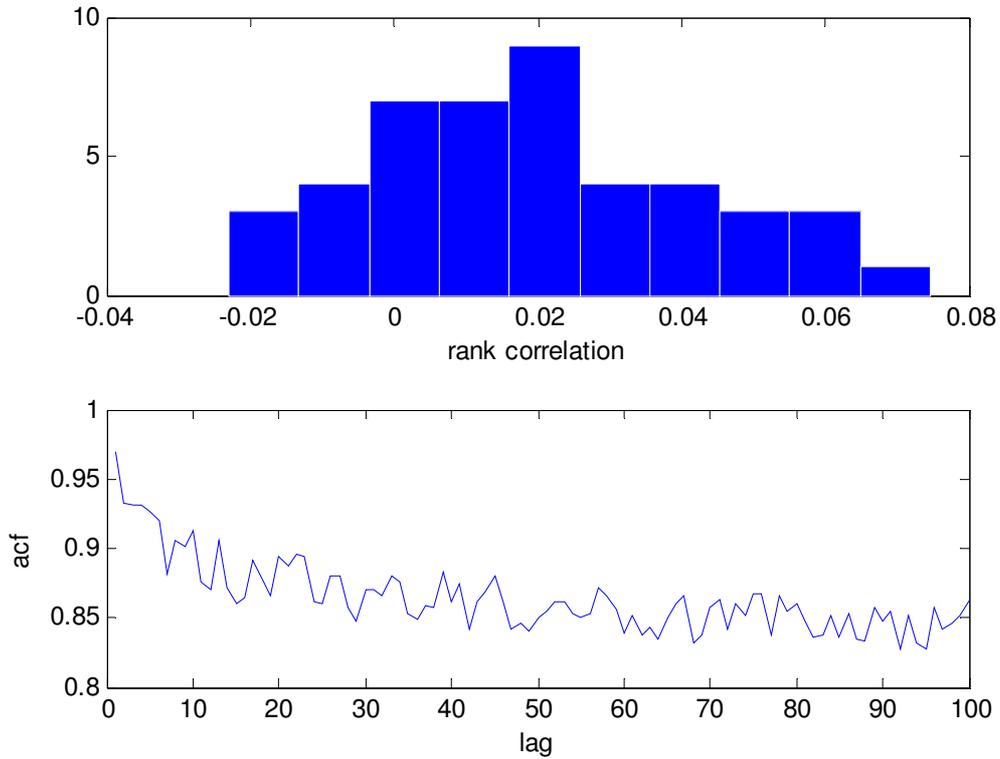

**Figure 25. Marginal posterior distributions of $\alpha$ and $\beta$, Copula approach, Exchange Rate data.**
Rows represent exchange rates, columns are for $\alpha$ (left) and $\beta$ (right). Straight lines represent ANF posteriors. Dotted lines represent posteriors from the copula approach.

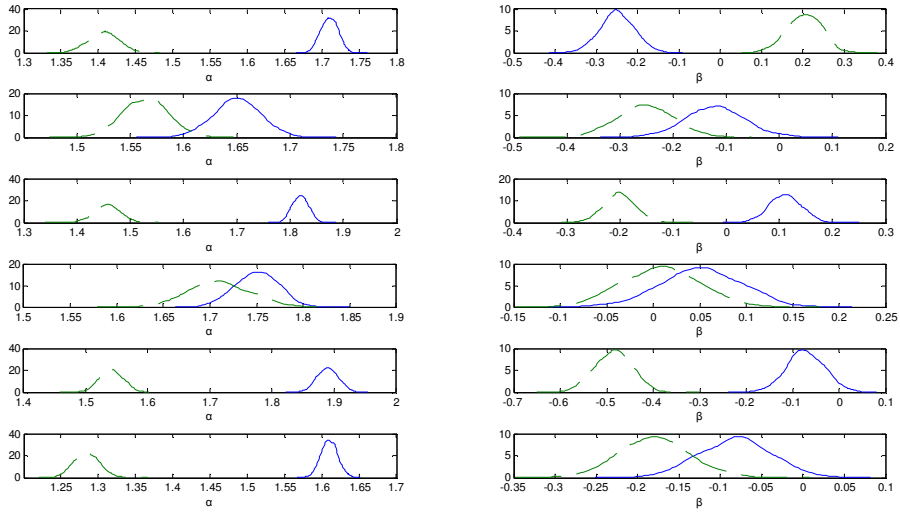

# 10. Stable Factor Models

## 10.1 Static factor models

Given that many financial time series are asymmetric and heavy-tailed, in this section we take up the problem of Bayesian inference in factor models from the stable family of distributions.

Consider the factor model

$$\mathbf{y}_t = \boldsymbol{\mu} + \boldsymbol{\Lambda}\mathbf{f}_t + \mathbf{u}_t, \ t=1,...,n, \tag{38}$$

where $\mathbf{y}_t$ is the $d \times 1$ vector of observed data, $\mathbf{f}_t = [f_{t1},...,f_{tk}]'$ is the $k \times 1$ vector of factors, with $k \leq d$, $\boldsymbol{\Lambda}$ is a $d \times k$ vector of loadings, $\boldsymbol{\mu}$ is a $d \times 1$ vector of location parameters, and $\mathbf{u}_t$ is a $d \times 1$ vector of error terms. In classical factor analysis one assumes that $\mathbf{f}_t \sim iid\mathcal{N}_k(\mathbf{0},\mathbf{I}_d)$, and $\mathbf{u}_t \sim iid\mathcal{N}_d(\mathbf{0},\boldsymbol{\Sigma})$ independently of $\mathbf{f}_t$, so that $Cov(\mathbf{y}_t) = \boldsymbol{\Lambda}\boldsymbol{\Lambda}' + \boldsymbol{\Sigma}$, where $\boldsymbol{\Sigma} = diag(\sigma_1^2,...,\sigma_d^2)$.

There are various ways to generalize the factor model in the stable family of distributions. The most general is to assume that $\mathbf{u}_t \sim \mathcal{S}_{\alpha,\beta}^d(\mathbf{0},\boldsymbol{\Gamma})$, and independently the factors are $iid\mathcal{S}_{\alpha_i,\beta_i}(0,1)$, for $i=1,...,k$. In this setup, the factors are distributed as standard multivariate stable, with different shape parameters $\alpha_i, \beta_i$ and the errors follow a multivariate stable distribution with spectral measure $\boldsymbol{\Gamma}$. In classical factor analysis one assumes that the $\mathbf{u}_t$s are independent so that the correlation of the observed data arises solely from the common factors. In that case it would be reasonable to extend this assumption as: $\mathbf{u}_t = [u_{t1},...,u_{td}]'$, and $u_{ti} \sim iid\mathcal{S}_{\alpha',\beta'}(0,\sigma_i^2)$, where $\sigma_i^2$ denotes the scale parameter, for all $i=1,...,d$, and $\alpha', \beta'$ denote the shape parameters of the stable distributions for the factors. Let us denote this distribution by $\mathbf{u}_t \sim \mathcal{S}_{\alpha',\beta',d}(\mathbf{0},\boldsymbol{\Sigma})$.

In classical factor analysis there are various identification problems. For any $k \times k$ orthonormal matrix $\mathbf{P}$ if we define $\boldsymbol{\Lambda}^* = \boldsymbol{\Lambda}\mathbf{P}'$ and $\mathbf{f}_t^* = \mathbf{P}\mathbf{f}_t$ then the model is not identified by the covariance matrix. With $(\alpha',\beta') \neq (\alpha,\beta)$ this condition can no longer be used for identification and it cannot be used even when the equality holds since the covariance does not exist and a non-Gaussian distribution is involved[23]. However, it is quite unlikely that the distributional assumptions can aid in significantly mitigating the identification problem. We use the traditional zero upper-triangular parametrization of $\boldsymbol{\Lambda}$ to define identifiable models, the parametrization in which the first k variable have "distinguished status" (Geweke and Zhou, 1996, Aguilar and West, 2000, Lopes and West, 2003). See also Geweke and Singleton (1980) for rank deficiency problems with this matrix. It is clear that

$$\mathbf{y}_t \mid \mathbf{f}_t, \boldsymbol{\mu}, \boldsymbol{\Lambda}, \boldsymbol{\Omega}, \alpha', \beta' \sim \mathcal{S}_{\alpha',\beta',d}(\boldsymbol{\mu} + \boldsymbol{\Lambda}\mathbf{f}_t, \boldsymbol{\Sigma}), \tag{39}$$

$$\mathbf{f}_t \mid \boldsymbol{\Sigma}, \alpha, \beta, k \sim \mathcal{S}_{\alpha,\beta,k}(\mathbf{0}, \mathbf{I}_k), \ t=1,...,n. \tag{40}$$

The kernel posterior distribution is

$$p(\boldsymbol{\zeta},\mathbf{f},k \mid \mathbf{Y}) \propto \left(\prod_{i=1}^d \sigma_i^{-n/2}\right)\left[\prod_{t=1}^n \prod_{i=1}^d f_{\alpha',\beta'}\left(\frac{y_{ti} - \mu_i - \boldsymbol{\lambda}_i' \mathbf{f}_t}{\sigma_i}\right) f_{\alpha,\beta}(f_{ti})\right] p(\boldsymbol{\zeta},k), \tag{41}$$

where $\boldsymbol{\zeta} = (\boldsymbol{\mu},\boldsymbol{\Lambda},\boldsymbol{\Sigma})$ is the vector of parameters, and $\boldsymbol{\Lambda} = [\boldsymbol{\lambda}_1',...,\boldsymbol{\lambda}_d']'$, where $\boldsymbol{\lambda}_i$ is the $k \times 1$ vector of elements in the $i$th row of $\boldsymbol{\Lambda}$. As in the main text, $f_{\alpha,\beta}(\cdot)$ denotes the density of distributions $\mathcal{S}_{\alpha',\beta'}(0,1)$. We denote $p(\boldsymbol{\zeta},\mathbf{f},k \mid \mathbf{Y}) \propto \mathcal{L}(\boldsymbol{\zeta},\mathbf{f},k \mid \mathbf{Y}) p(\boldsymbol{\zeta})$ by Bayes' theorem. Following Geweke and Zhou (1996, pp. 565-566) we assume for identification purposes that is a lower triangular matrix whose diagonal elements are strictly positive $\Lambda_{ij} \sim \mathcal{N}(0,C_0^2)$, for $i \neq j$, and $\Lambda_{ii} \sim \mathcal{N}(0,C_0^2) \mid \Lambda_{ii} \geq 0$, where[24] $C_0$ is a large positive constant which here we take equal to 10. For

---

[23] The point is also made in Liu, Xiu and Chu (2004) and Viroli (2009) who used a mixture of normal distributions for the factors.
[24] We keep the positivity restriction despite the fact that the model does not suffer from non-identification of the signs of the factors when at least one factor or error term is strictly non-symmetric stable. The reason is that the signs could be poorly identified when all distributions are close to symmetry. Moreover the variances are ordered in

the scale parameters, $\sigma_i^2, \omega_i^2 \sim IG(\bar{\nu}/2, \overline{\nu q}/2)$, where the hyperparameters $\bar{\nu} = 2.2$ and $\bar{q} = 0.1$. For $\alpha, \beta$ and $\alpha', \beta'$ we use a uniform prior in the allowable range $(0,2] \times [-1,1]$. For the number of factors we assume the prior $p(k) \propto \exp(-k)$, $k = 1, 2, \ldots$, a Poisson restricted to positive integers. Integrating out the matrix of factors, $\mathbf{f}$, from (A.4) is impossible unlike the case of normal factor analysis. Analytical integration is also impossible when at least one of $\mathbf{f}_t$ or $\mathbf{u}_t$ is stable non-Gaussian.

We consider the following ABC scheme. To start with, compute the empirical characteristic function $\hat{\varphi}(\boldsymbol{\tau}) = n^{-1} \sum_{t=1}^{n} \exp(\iota \langle \boldsymbol{\tau}, \mathbf{y}_t \rangle)$.

- Fit the normal k-factor model using MCMC[25] to obtain posterior draws for $\boldsymbol{\zeta} = (\boldsymbol{\mu}, \boldsymbol{\Lambda}, \{\mathbf{f}_t, t=1,\ldots,n\}, \boldsymbol{\Sigma})$, for $k = 1, 2, \ldots, \bar{k}$, where $\bar{k}$ is an a priori known bound on $k$. For $\boldsymbol{\theta} = (\alpha', \beta', \boldsymbol{\mu}, \boldsymbol{\Sigma})$ estimate univariate stable $\mathscr{S}_{\alpha',\beta'}(\mu_i, \sigma_i^2)$ distributions for each time series $i = 1, \ldots, d$. Take as proposal a multivariate Student-t (see below) based on maximum likelihood estimation[26] with mean $\bar{\boldsymbol{\theta}} = d^{-1} \sum_{i=1}^{d} \hat{\boldsymbol{\theta}}_i$, $\hat{\boldsymbol{\theta}}_i \triangleq (\hat{\alpha}'_i, \hat{\beta}'_i, \hat{\mu}_i, \hat{\sigma}_i^2)$, and covariance[27] $\bar{\mathbf{V}} \equiv \bar{\mathbf{V}}_k = \frac{1}{2} d^{-1} \sum_{i=1}^{d} \left[ \hat{\mathbf{V}}_i + (\hat{\boldsymbol{\theta}}_i - \bar{\boldsymbol{\theta}})(\hat{\boldsymbol{\theta}}_i - \bar{\boldsymbol{\theta}})' \right]$, where $\hat{\boldsymbol{\theta}}_i, \hat{\mathbf{V}}_i$ denote the ML quantities. For $(\alpha, \beta)$ take the same proposal independently of $(\alpha', \beta', \boldsymbol{\mu}, \boldsymbol{\Sigma})$.
- Use a multivariate Student-t proposal for $\boldsymbol{\zeta}$, with the stated parameters. The degrees of freedom, $\nu = n - p$, where $p$ is the number of parameters.
- Draw a candidate parameter vector $\boldsymbol{\zeta}$ from the proposal and simulate model (39)-(40) to obtain artificial data $\tilde{\mathbf{Y}}(\boldsymbol{\zeta}) = (\tilde{\mathbf{y}}_t(\boldsymbol{\zeta}), t = 1, \ldots, n)$. Compute the simulated characteristic function $\tilde{\varphi}(\boldsymbol{\tau}; \boldsymbol{\zeta}) = n^{-1} \sum_{t=1}^{n} \exp(\iota \langle \boldsymbol{\tau}, \tilde{\mathbf{y}}_t(\boldsymbol{\zeta}) \rangle)$.
- If $|\tilde{\varphi}(\boldsymbol{\tau}; \boldsymbol{\zeta}) - \hat{\varphi}(\boldsymbol{\tau})| \leq \varepsilon$, for some positive constant $\varepsilon$, accept the draw with probability.
- Use (41) and the Laplace approximation to obtain the log-marginal likelihood for $k = 1, \ldots, \bar{k}$, and draw a value for $k$. If the log-marginal likelihood is denoted by $l_k$, the probability of model $k$ is[28] $p_k = \exp(l_k - k) / \sum_{m=1}^{\bar{k}} \exp(l_m - m)$, $k = 1, \ldots, \bar{k}$.

It is well known[29] that

$$l_k \simeq \log\left[ S^{-1} \sum_{s=1}^{S} \mathscr{L}\left(\bar{\boldsymbol{\zeta}}, \mathbf{f}^{(s)}, k \mid \mathbf{Y}\right) \right] + \frac{p}{2} \log |\bar{\mathbf{V}}_k| + \frac{1}{2} (\boldsymbol{\zeta} - \bar{\boldsymbol{\zeta}})' \bar{\mathbf{V}}_k^{-1} (\boldsymbol{\zeta} - \bar{\boldsymbol{\zeta}}), \tag{42}$$

where $\{\mathbf{f}^{(s)}, s = 1, \ldots, S\}$ denotes draws for the common factors from distribution $\mathscr{S}_{\alpha,\beta,k}(\mathbf{0}, \mathbf{I}_k)$ as in (39). The first term accounts for approximate integration of the posterior kernel in (41) with respect to $\mathbf{f}$.

---

increasing order to avoid a "labelling" problem. Notice that the lower triangularity with strictly positive diagonal elements guarantees full rank of $\boldsymbol{\Lambda}$ and avoids the problems discussed in Geweke and Singleton (1980).
[25] See Geweke and Zhou (1996), equations (15)-(17) and their Appendix.
[26] Maximum likelihood estimation is implemented using the FFT transform.
[27] This estimate is an equally weighted average of the ML estimates for time series i and the cross-sectional covariance matrix of individual estimates.
[28] It is important to note that this step must be actually performed first, preceding every other MCMC computation to guarantee that this "collapsed" Gibbs step guarantees convergence to the correct posterior distribution.
[29] See DiCiccio, Kass, Raftery and Wasserman (1997). As discussed in this paper extreme draws have to be avoided to ensure numerical stability: One way to make it sure is to use the draws for which the quadratic form in the second term of (A.5) is less than 0.05 or 0.10. Here, we have used 5%. The likelihood is computed using the FFT via interpolation from $2^{16}$ base points.

In the following Table A1, we report 90% critical values of the distribution of $D = |\tilde{\varphi}(\boldsymbol{\tau};\boldsymbol{\zeta}) - \hat{\varphi}(\boldsymbol{\tau})|$, for different sample sizes $n$ as well as dimensionality $d$, the number of factors, $k$, and parameters $\alpha, \beta$. First, for each value of $\boldsymbol{\mu}, \boldsymbol{\Sigma}, \{\mathbf{f}_t\}$ we consider S=1,000 Monte Carlo simulations where different data sets are generated according to (39) and (40) and $D$ is computed and recorded to find the 90% critical value of $D$. To make these critical values useful for practitioners it is necessary to take into account their dependence on $\boldsymbol{\mu}, \boldsymbol{\Sigma}, \{\mathbf{f}_t\}$. Therefore, to provide a rough idea about this dependence we consider for all S=1,000 simulations, 1,000 different values of the parameters generated as $\boldsymbol{\mu} \sim \mathcal{N}_d(\mathbf{0}, \mathbf{I}_d)$, $\sigma_i \sim \mathcal{U}(0,1)$, the standard uniform distribution, and $\mathbf{f}_t \mid \boldsymbol{\Sigma}, \alpha, \beta, k \sim \mathcal{S}_{\alpha,\beta,k}(\mathbf{0}, \mathbf{I}_k)$.

In Table A.1 reported are (*i*) the median of the 90% critical values of $D$, and (*ii*) the 90% confidence interval of the 90% critical values of $D$. The values of $\mu_i$ and $\sigma_i$ should be appropriate for most financial time series, at least after the usual transformations employed in practice. These values should be useful in choosing a reasonable value of $\varepsilon$ or adjusting this constant around these values to provide reasonable acceptance rates, say close to 50%[30].

Table A1. Critical values of D

|  | I | II | III | IV |
|---|---|---|---|---|
|  | $\alpha = 1.70$, $\beta = -0.20$ | $\alpha = 1.20$, $\beta = -0.80$ | $\alpha = 1.20$, $\beta = 0.80$ | $\alpha = 1.70$, $\beta = 0.20$ |
| **d=5, k=2** | | | | |
| n=100 | 0.372 | 0.345 | 0.343 | 0.373 |
|  | 0.246 − 0.617 | 0.217 − 0.610 | 0.220 − 0.606 | 0.240 − 0.613 |
| n=500 | 0.359 | 0.309 | 0.306 | 0.358 |
|  | 0.209 − 0.581 | 0.184 − 0.557 | 0.180 -0.558 | 0.210 − 0.584 |
| n=1,000 | 0.357 | 0.322 | 0.319 | 0.357 |
|  | 0.213 − 0.601 | 0.178 − 0.593 | 0.174 − 0.606 | 0.211 − 0.601 |
| n=2,000 | 0.352 | 0.303 | 0.303 | 0.354 |
|  | 0.207 − 0.579 | 0.171 − 0.556 | 0.171 − 0.556 | 0.206 − 0.578 |
| **d=10, k=4** | | | | |
| n=100 | 0.224 | 0.194 | 0.195 | 0.227 |
|  | 0.167 − 0.352 | 0.151 − 0.291 | 0.147 − 0.283 | 0.166 − 0.353 |
| n=500 | 0.164 | 0.112 | 0.111 | 0.165 |
|  | 0.099 − 0.310 | 0.079 − 0.189 | 0.080 − 0.193 | 0.098 − 0.308 |
| n=1,000 | 0.150 | 0.095 | 0.095 | 0.154 |
|  | 0.083 − 0.307 | 0.063 − 0.179 | 0.064 − 0.181 | 0.084 − 0.305 |
| n=2,000 | 0.154 | 0.087 | 0.087 | 0.154 |
|  | 0.080 − 0.322 | 0.053 − 0.177 | 0.054 − 0.176 | 0.079 − 0.321 |
| **d=20, k=8** | | | | |
| n=100 | 0.180 | 0.176 | 0.172 | 0.181 |
|  | 0.136 − 0.265 | 0.137 − 0.252 | 0.135 − 0.252 | 0.138 − 0.262 |
| n=500 | 0.081 | 0.077 | 0.080 | 0.082 |
|  | 0.064 − 0.120 | 0.060 − 0.113 | 0.062 − 0.114 | 0.063 − 0.118 |
| n=1,000 | 0.059 | 0.055 | 0.056 | 0.059 |
|  | 0.046 − 0.085 | 0.043 − 0.079 | 0.043 − 0.081 | 0.046 − 0.088 |
| n=2,000 | 0.044 | 0.040 | 0.041 | 0.044 |
|  | 0.033 − 0.065 | 0.031 − 0.058 | 0.031 − 0.058 | 0.033 − 0.067 |
| **d=100, k=2** | | | | |
| n=1,000 | 0.311 | 0.086 | 0.083 | 0.309 |
|  | 0.201 − 0.440 | 0.063 − 0.124 | 0.062 − 0.123 | 0.201 − 0.439 |
| n=3,000 | 0.309 | 0.069 | 0.070 | 0.308 |
|  | 0.203 − 0.437 | 0.050 − 0.111 | 0.050 − 0.110 | 0.206 − 0.436 |

*Notes*: The table reports medians of the 90% critical values of the maximum absolute difference between the empirical and simulated characteristic function ($D$). The median is computed across 1,000 different parameter sets. The interval below the reported median is the 90% interval of the distribution of $D$ across the 1,000 different parameter sets.

Stable Factor Analysis has been implemented for the SP100 data set using the MCMC procedure described previously using 120,000 iterations the first 20,000 of which are discarded and the remaining are thinned every other 10[th] draw. Rough values of the constant $\varepsilon$ were obtained by extrapolation from Table A1 when $d = 100$ and $k = 1$.

---

[30] Another tuning constant may be introduced which multiplies the covariance matrix of the proposal distributions. After adjusting ε this is, usually, not necessary as we have found in experiments with artificial data to validate and debug the numerical procedures.

The constant was adapted during the "burn in" phase to achieve a target acceptance rate of 50% and the final acceptance rate was 30%. Initially we set $\bar{k}=20$ but after the "burn in" phase the results indicated that $\bar{k}=7$ was enough so we set $\bar{k}=10$ to be on the safe side and minimize somewhat the computational cost. Posterior results are reported in Figures A1 through A3. The posterior mean model probabilities reported in Figure A1 indicate clearly the presence of a single common factor. Implementation of a normal factor model using MCMC is quite easy and the posterior means of common factors are reported in Figure A2. Evidently, the results from Stable Factor Analysis are quite different and slightly negatively correlated from those obtained from Factor Analysis under the assumption of normality. Finally, marginal posterior distributions of the shape parameters are reported in Figure A3. Three models are allowed: *First*, a general Stable Factor Analysis allowing for different shape parameters in the factors and the data. *Second*, a model with common shape parameters and *third*, a symmetric stable model with different tail index for the factors and the data.

**Figure 26. Posterior probability of number of factors**

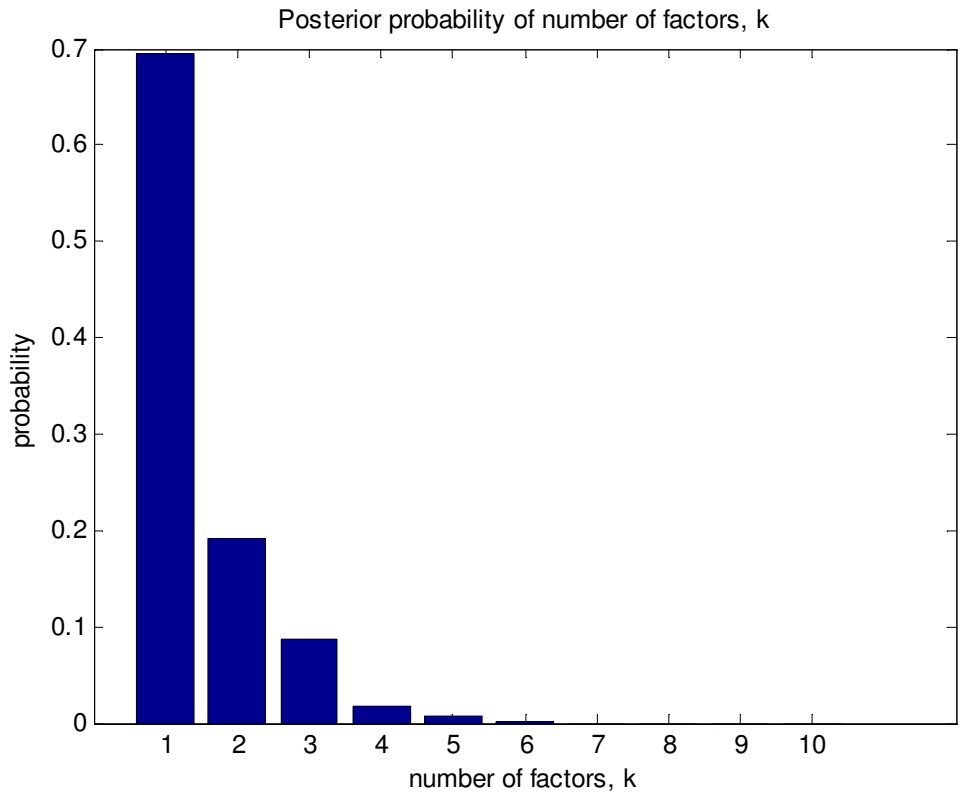

**Figure 27. Comparison of posterior means of the first common factor from normal and general stable factor analysis**

The straight line denotes the 45º line.

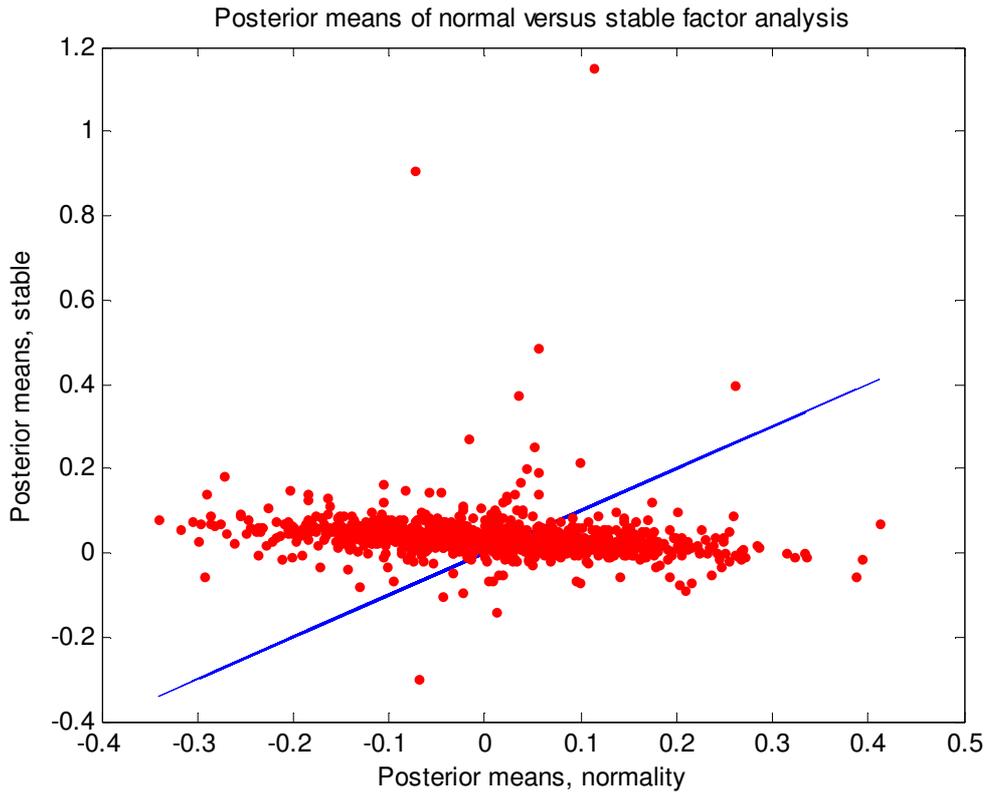

**Figure 28. Marginal posterior distributions of $\alpha$ and $\beta$ in Stable Factor Analysis**

The curves labelled $\alpha$ and $\alpha'$ denote marginal posterior distributions of $\alpha$ and $\alpha'$ respectively. The curves labelled $\beta$ and $\beta'$ denote marginal posterior distributions of $\beta$ and $\beta'$ respectively. "Common $\alpha, \beta$" denotes the marginal posterior of $\alpha$ (upper panel) or $\beta$ (bottom panel) when $\alpha = \alpha'$ and $\beta = \beta'$. "Symmetric stable" denotes the marginal posterior of $\beta$ when $\beta = 0$.

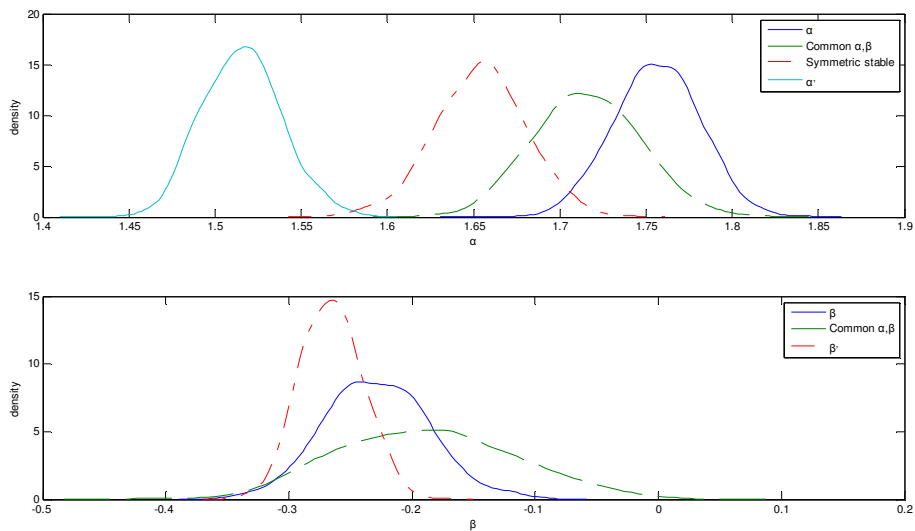

## 10.2 A Markov Stable Factor model

Modelling financial time series as realizations from factor models may be too simplistic in practice when regime-switching is possible. Here we consider an extension of the Stable Factor Model to allow for regime changes when the regime can be described by the state $\mathscr{F}_t \in \{0,1\}$ of the random variable $(\alpha,\beta)$ whose state space is $(\alpha_0,\alpha_1,\beta_0,\beta_1)$ in $(0,2]^2 \times (-1,1)^2$ for the common factors. The state evolves according to a Markov chain with transition probabilities

$$\mathbb{P}(\mathscr{F}_{t+1}=j \mid \mathscr{F}_t=i) = \pi_{ij}, \quad i,j \in \{0,1\}. \tag{43}$$

The random variable $\boldsymbol{\theta} = (\alpha',\beta',\{\mu_i,\sigma_i, i=1,...,d\})$ for the data or the disturbances evolves according to a Markov chain with transition probabilities

$$\mathbb{P}(\mathscr{D}_{t+1}=j \mid \mathscr{D}_t=i, \mathscr{F}_t=m) = \pi_{ij}^{(m)}, \quad i,j,m \in \{0,1\}. \tag{44}$$

The state space is $\{\boldsymbol{\theta}_0,\boldsymbol{\theta}_1\}$. In this model, a Markov chain determines first the state of the stable shape parameters of the factors and, conditional on that state, the state of the shape, location and scale parameters of the data is determined. It is more flexible than a two-state Markov chain that would attempt to exhaust the description of time-varying behaviour in the entire state space $(\mathscr{F}_t,\mathscr{D}_t)$.

The parameters of the Markov chain are $\boldsymbol{\pi} = \left[\pi_{00},\pi_{11},\pi_{00}^{(0)},\pi_{11}^{(0)},\pi_{00}^{(1)},\pi_{11}^{(1)}\right]'$. Denote $\gamma = (\alpha,\beta)$ and $\gamma' = (\alpha',\beta')$, the shape parameters for the factors and the disturbances respectively. Their states are $\{\gamma_0,\gamma_1\}$ and $\{\gamma_0',\gamma_1'\}$ respectively. The model is as follows.

$$\mathbf{y}_t \mid \mathbf{f}_t,\boldsymbol{\mu},\boldsymbol{\Lambda},\boldsymbol{\Omega},\gamma', \mathscr{F}_t=m, \mathscr{D}_t=i \sim \mathscr{S}_{\gamma_m',d}(\boldsymbol{\mu}_{im} + \boldsymbol{\Lambda}_m \mathbf{f}_t, \boldsymbol{\Sigma}_{im}), \quad i,m \in \{0,1\}, \tag{45}$$

$$\mathbf{f}_t \mid \boldsymbol{\Sigma},\gamma,k, \mathscr{F}_t=m \sim \mathscr{S}_{\gamma_m,k}(\mathbf{0},\mathbf{I}_k), \quad t=1,...,n, \quad m \in \{0,1\}. \tag{46}$$

and (43), (44). Depending on the state of the factors ($m$) and the state of the disturbance ($i$) the data, from (43), have different location, scale and shape parameters. For simplicity it is assumed that $\boldsymbol{\Lambda}_{0m} = \boldsymbol{\Lambda}_{1m} = \boldsymbol{\Lambda}_m$, $m \in \{0,1\}$.

The kernel posterior in (41) has to be modified as follows.

$$p(\boldsymbol{\zeta},\mathbf{f},k \mid \mathbf{Y}) \propto \left[\prod_{i=1}^d \sigma_i^{-n/2}\right] \left[\prod_{t=1}^n \prod_{i=1}^d \sum_{i,m\in\{0,1\}} \left\{\pi_{i0}^{(m)} f_{\gamma_0'}\left(\frac{y_{ti}-\mu_{i,0}-\boldsymbol{\lambda}_{i0}'\mathbf{f}_t}{\sigma_{i,0}}\right) + \pi_{i1}^{(m)} f_{\gamma_1'}\left(\frac{y_{ti}-\mu_{i,1}-\boldsymbol{\lambda}_{i1}'\mathbf{f}_t}{\sigma_{i,1}}\right)\right\} \sum_{i\in\{0,1\}} \left\{\pi_{i0} f_{\gamma_0}(f_{ti}) + \pi_{i1} f_{\gamma_1}(f_{ti})\right\}\right] p(\boldsymbol{\zeta},k) \tag{47}$$

where the parameter vector $\boldsymbol{\zeta} = (\boldsymbol{\mu}_{im},\boldsymbol{\Sigma}_{im},\boldsymbol{\Lambda}_m,\boldsymbol{\pi},\gamma_m,\gamma_m')_{i,m\in\{0,1\}}$. We assume that the scale parameters are ordered $\sigma_{0,0} \leq \sigma_{1,0}$, $\sigma_{0,1} \leq \sigma_{1,1}$, the initial state $\mathscr{I}_0 = (\mathscr{F}_0,\mathscr{D}_0)$, is unknown and their point probability is $\mathbb{P}(\mathscr{I}_0=s)=q_s$, $s \in \{00,01,10,11\}$ which introduces three additional parameters $q_{00},q_{01},q_{11}$.

Table A2. Posterior statistics for the Markov Stable model

|  | Posterior mean (posterior s.d.) | RNE | CD | NSE |
|---|---|---|---|---|
| $\mu_0$, state factor 0 | 0.0012 (0.015) | 0.353 | -1.212 | 0.0027 |
| $\mu_0$, state factor 1 | -0.0027 (0.022) | 0.451 | -1.555 | 0.0041 |
| $\mu_1$, state factor 0 | -0.0021 (0.017) | 0.810 | 1.356 | 0.0045 |
| $\mu_1$, state factor 0 | -0.0034 (0.027) | 0.766 | 1.212 | 0.0013 |
| $\sigma_0$, state factor 0 | 0.015 (0.0017) | 0.561 | 0.897 | 0.00045 |
| $\sigma_0$, state factor 1 | 0.171 (0.055) | 0.212 | 1.245 | 0.0061 |
| $\sigma_1$, state factor 0 | 0.027 (0.016) | 0.353 | -1.451 | 0.0034 |
| $\sigma_1$, state factor 0 | 0.188 (0.044) | 0.477 | 0.561 | 0.0054 |
| $\alpha,\beta$, factor state 0 | 1.971    0.012 | 0.671 | 0.446 | 0.0034  0.0011 |

| | | | | | |
|---|---|---|---|---|---|
| | (0.022) (0.033) | | | | |
| $\alpha, \beta$, factor state 1 | 1.712 -0.012 (0.015) (0.041) | 0.566 | -1.391 | 0.0034 0.0037 | |
| $\alpha', \beta'$, factor state 0 | 1.989 0.0034 (0.011) (0.0025) | 0.345 | 0.556 | 0.0045 0.0037 | |
| $\alpha', \beta'$, factor state 1 | 1.560 -0.21 (0.014) (0.022) | 0.444 | 0.812 | 0.0022 0.0039 | |
| $\pi_{00}$ | 0.273 (0.015) | 0.819 | -1.210 | 0.0012 | |
| $\pi_{11}$ | 0.612 (0.017) | 0.710 | 1.337 | 0.0025 | |
| $\pi_{00}^{(m)}, m=0$ | 0.115 0.055 (0.035) (0.012) | 0.650 | 1.215 1.122 | 0.0031 0.001 | |
| $\pi_{11}^{(m)}, m=0$ | 0.122 0.031 (0.021) (0.011) | 0.710 0.717 | 1.341 -1.22 | 0.0027 0.002 | |
| $q_{00}, q_{01}, q_{11}$ | 0.035 0.712 0.021 (0.017) (0.022) (0.026) | 0.610 | -1.320 | 0.0021 | |

*Notes*: RNE is relative numerical efficiency, CD is Geweke's (1994) convergence diagnostic, and NSE is the numerical standard error.

## 10.3  General Dynamic Stable Factor model

Important generalization of the model can be introduced along two dimensions: First, generalizing the disturbance structure so that the errors are not necessarily independent and second, by introducing dynamics. In the normal factor model one can assume $\mathbf{u}_t \sim iid \mathcal{N}_d(\mathbf{0}, \mathbf{\Sigma})$ where matrix $\mathbf{\Sigma}$ is not necessarily diagonal. This is called the generalized factor model (Forni and Lippi, 2001, Forni, Hallin, Lippi and Reichlin, 2000, 2005). The model is as follows.

$$\mathbf{y}_t \mid \mathbf{f}_t, \boldsymbol{\mu}, \mathbf{\Lambda}, \mathbf{\Omega}, \alpha', \beta' \sim \mathscr{S}_{\alpha', \beta', d}(\boldsymbol{\mu} + \mathbf{\Lambda}\mathbf{f}_t, \mathbf{\Sigma}, \Gamma), \tag{48}$$

$$\mathbf{f}_t = (\mathbf{I}_k - \mathbf{\Delta})\boldsymbol{\delta} + \mathbf{\Delta}\mathbf{f}_{t-1} + \boldsymbol{\varepsilon}_t, \ \boldsymbol{\varepsilon}_t \mid \mathbf{\Sigma}, \alpha, \beta, k \sim \mathscr{S}_{\alpha,\beta,k}(\mathbf{0}, \mathbf{I}_k), \ t=1,...,n, \tag{49}$$

where $\Gamma = \Gamma(d\mathbf{s})$ denotes the spectral measure defined over the boundary of the unit hyper-ball in $\mathbb{R}^{d-1}$, and $\mathbf{\Delta}$ is a $k \times k$ matrix. We maintain the assumption that $\mathbf{\Sigma}$ is diagonal so $\mathbf{\Sigma} = diag(\sigma_1^2, ..., \sigma_d^2)$ and the assumption that $\mathbf{\Lambda}$ is lower triangular with strictly positive diagonal elements and $\mathbf{\Delta} = diag(\delta_1, ..., \delta_k)$. A similar model, under normality, has been proposed by Geweke (1977), Sargent and Sims (1977) and Engle and Watson (1981). A variation of the model for possibly non-stationary times would be: $\mathbf{f}_t = \boldsymbol{\delta} + \mathbf{\Delta}\mathbf{f}_{t-1} + \boldsymbol{\varepsilon}_t$. A model encompassing (48) − (49) is the so called factor augmenting vector autoregression where $\mathbf{y}_t, \mathbf{f}_t$ are allowed to interact, see Stock and Watson (2005) and Bernanke, Boivin, and Eliasz (2005). The methods developed here can be easily adapted to handle such models so further analysis will not be undertaken here.

The main complication in (48) − (49) is the fact that we have to use the spectral measure and estimate it along with the other parameters of the model. Conditional on the latent dynamic factors, (48) is a multivariate regression model whose disturbances belong to the most general multivariate stable distribution. As with other multivariate stable distributions, the spectral measure can be approximated using either a discrete measure or the normal distribution over the unit hyper-sphere. Relative to other models based on the multivariate stable distribution, the complication is that the number of factors is unknown and a "collapsed" Gibbs step has to be used based on approximations of the log-marginal likelihood. A third impediment is that sampling the latent factors is not easy. Under normality and static factors ($\boldsymbol{\delta} = \mathbf{0}_k, \mathbf{\Delta} = \mathbf{O}_{k \times k}$) it can be shown that the conditional posterior distribution of latent factors is normal with moments

$$\mathbb{E}(\mathbf{f}_t \mid \boldsymbol{\mu}, \mathbf{\Lambda}, \mathbf{\Sigma}, k, \mathbf{Y}) = \mathbf{\Lambda}'(\mathbf{\Lambda}\mathbf{\Lambda}' + \mathbf{\Sigma})^{-1}(\mathbf{y}_t - \boldsymbol{\mu}), \tag{50}$$

$$Cov(\mathbf{f}_t \mid \boldsymbol{\mu}, \mathbf{\Lambda}, \mathbf{\Sigma}, k, \mathbf{Y}) = \mathbf{I}_k - \mathbf{\Lambda}'(\mathbf{\Lambda}\mathbf{\Lambda}' + \mathbf{\Sigma})^{-1}\mathbf{\Lambda}. \tag{51}$$

If there does not exist "too much" persistence we can use these moments to formulate a multivariate Student-$t$ proposal distribution with degrees of freedom $\nu$ (a parameter to be determined) or use our previous Principal-Directions (PD) based sampling scheme. A third alternative is to use a proposal based on simulating (A.12) as we did before in the case of static stable factor models. For a normal *dynamic* factor model the latent factors can be sampled independently as:

$$\mathbf{f}_t \mid \boldsymbol{\mu}, \mathbf{\Lambda}, \mathbf{\Sigma}, \mathbf{Y}, k \sim \mathcal{N}_k(\mathbf{A}(\mathbf{y}_t - \boldsymbol{\mu}), \mathbf{I}_k - \mathbf{A}\mathbf{Q}\mathbf{A}'), \ t=1,...,n. \tag{52}$$

where $\mathbf{Q} = \mathbf{\Lambda}\mathbf{\Lambda}' + \mathbf{\Sigma}$, and $\mathbf{A} = \mathbf{\Lambda}'\mathbf{Q}^{-1}$, see Aguilar and West (2000, Appendix B). With heavy-tailed distributions it is not clear how useful these expressions can be although they are, certainly, extremely convenient. One approach that has proved useful in connection with artificial data is to use a *mixture of PD and (A.15)* –specifically the proposal for the factors is set initially to a 50:50 mixture of PD and (A.15) and it is adapted during the "burn in" phase so that the acceptance rate is close to 50%. The final proportion was, approximately, 73:27, showing once more the usefulness of the PD construction.

Aspects of posterior analysis for the factor models are reported in Figures A4 through A6. In Figure A4 we report marginal posterior probabilities for the number of factors in the generalized static factor model (left panel) and the generalized dynamic factor model (right panel) for the SP100 data. In the static model two factors are favored whereas in the dynamic factor model, the posterior evidence in favor of a single factor are overwhelming. In Figure A5, reported are posterior mean estimates of the normalized spectral measures for the generalized static factor model (left panel) and the generalized dynamic factor model (right panel). Two approximations to the spectral measure are used, the discrete approximation (continuous line, which in fact represents a step function) and the normal distribution approximation over the boundary of the d-dimensional unit hyper-ball. The normal approximation is quite close to the discrete counterpart suggesting its usefulness in connection with multivariate general stable distributions. In Figure A6, reported are marginal posterior distributions of the tail index and the skewness parameters of the multivariate general stable distributions for the dynamic factor model. We remind that $\alpha', \beta'$ denote the parameters of the factors and $\alpha, \beta$ denote the parameters of the disturbances or the data. From the results it is evident that the factor is stable distributed with shape close to 1.75 and skewness -0.1 while the disturbances are likely to be symmetric and there is considerable evidence that they are, in fact, normally distributed. In that sense the non-Gaussianity of stock returns comes from the common factors and conditional on them, the data are likely to be normal.

**Figure 29. Posterior probability of number of factors in generalized static and dynamic stable factor models, SP100 data**

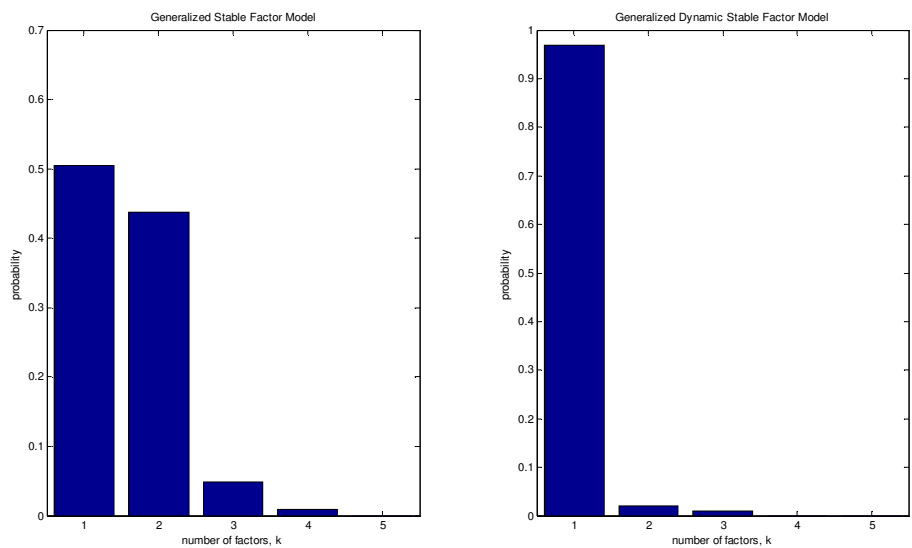

Figure 30. Posterior mean normalized spectral measures, SP100 data

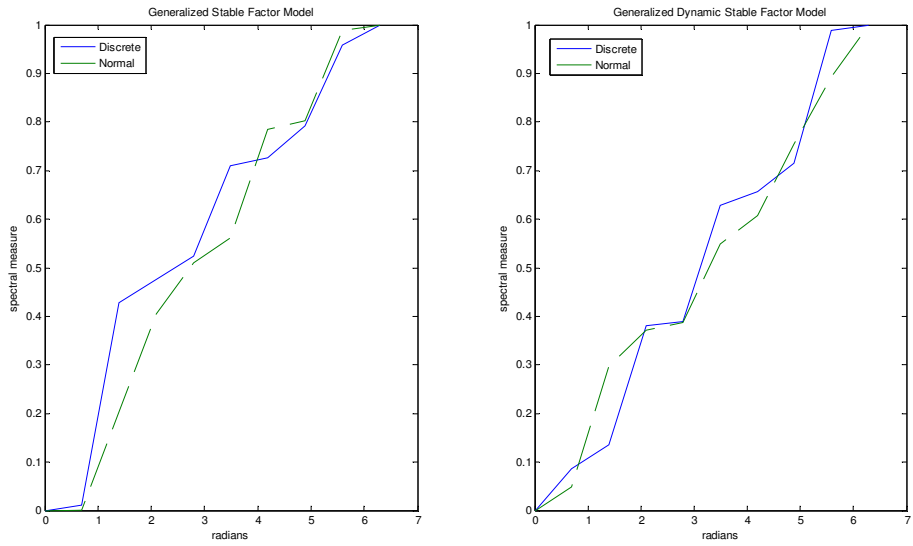

Figure 31. Marginal posterior densities of stable shape parameters, Generalized Dynamic Stable Factor Model, SP100 data

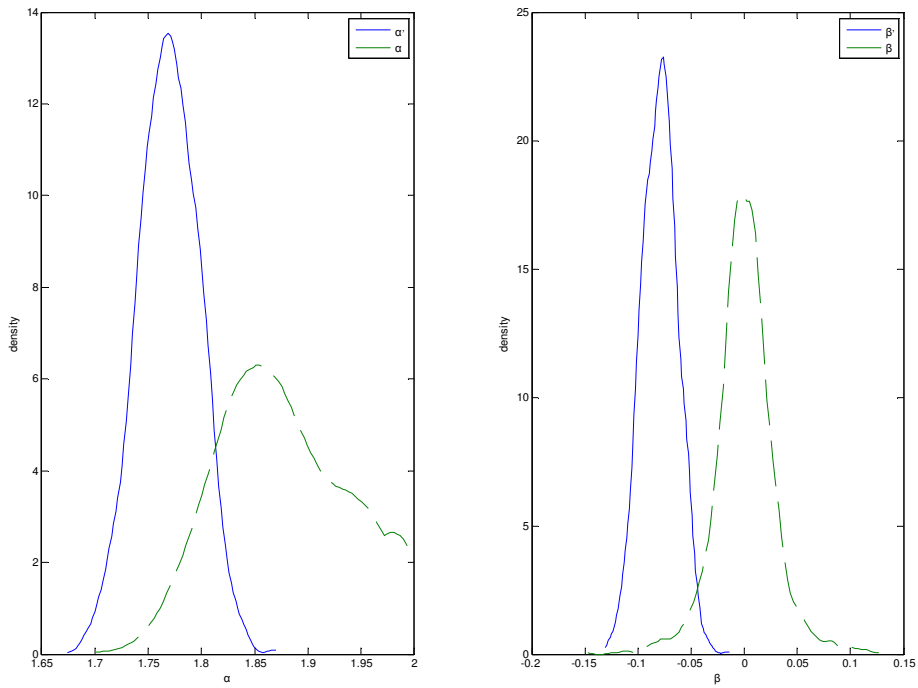

In Figure A7, we report the histogram of the posterior means of the loadings in $\underset{(d\times 1)}{\mathbf{\Lambda}}$. It turns out that most loadings are close to zero and only a few stand out.

Figure 32. Factor loadings in the Generalized Dynamic Stable Factor Model, SP100

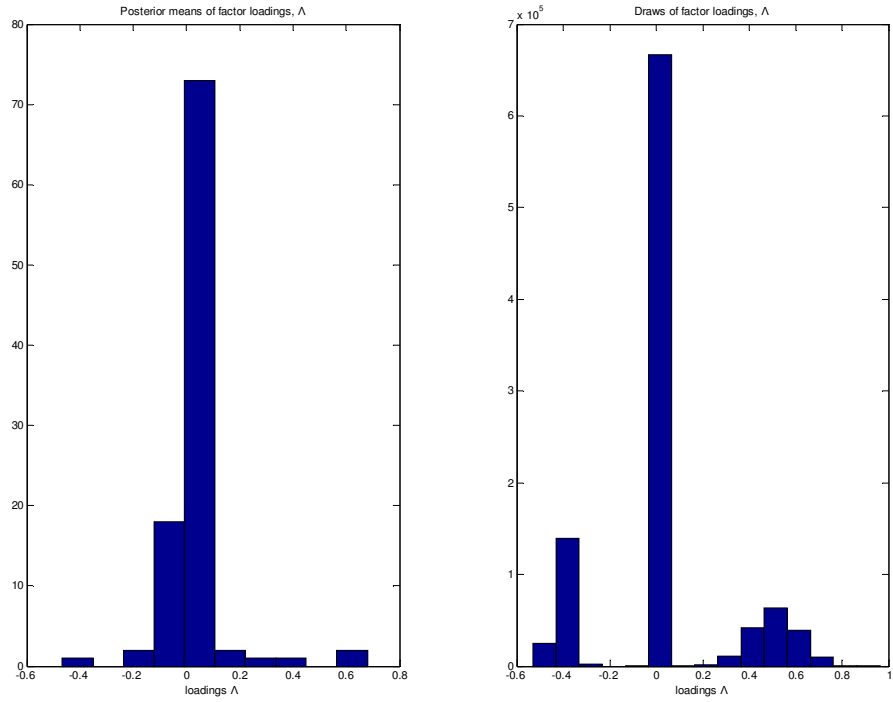

Figure 33. Maximum autocorrelations for the Static and Dynamic factor models

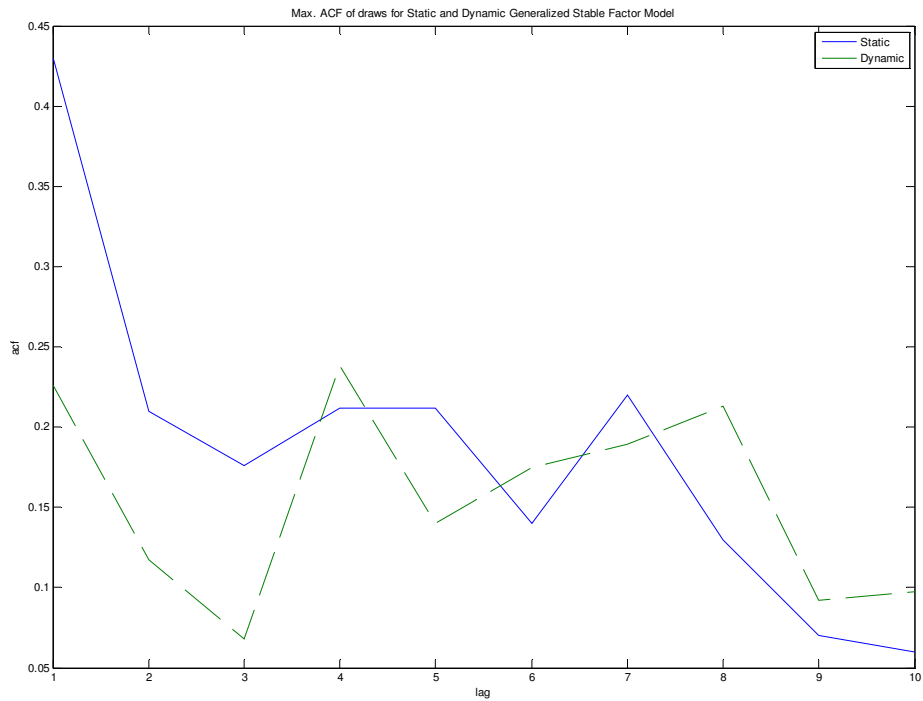

# Concluding remarks

In this paper we have presented the application of several tailored applications of ABC inference in univariate and multivariate stable Paretian models, along with their extension to multivariate stochastic volatility, as well as static and dynamic factor models. In connection with univariate stable distributions ABC along with the asymptotic normal form likelihood is an extremely good competitor to exact inference using the fast Fourier transform. Approximation of general stable distributions by scale mixtures of normal distributions perform equally well, and the optimal number of components is quite small. In the case of multivariate stable distributions we have proposed and explored the performance of several methods to perform statistical inferences for the associated spectral measure of the distributions which has been, so far, the major impediment in the development of empirical models for stable distributions. Statistical inferences are also made for the grid points of the characteristic function, thus removing a major obstacle for the econometric implementation of stable distributions. In that way full likelihood inference is possible, *unconditional* on the number or configuration of the grid.

A particular form of ABC along with the method of principal directions or a multivariate normal approximation for the spectral measure were found to perform well in applications to exchange rates (ten major currencies and stocks of SP100). The techniques are well suited to handle multivariate stable with stochastically − varying spectral measure and stochastic volatility thus extending stable Paretian distributions in an empirically important way that it will, hopefully, find other applications in econometrics. Moreover, multivariate stable Paretian models have been generalized in the context of static and dynamic factor models, whose disturbances and factors are distributed according to stable distributions, thus removing the assumption of normality from such models. The proposed methods were found, again, to perform well in high-dimensional data sets.

We have also provided critical values to guide practitioners in implementation of efficient ABC inference in connection with stable distributions, and conducted Monte Carlo experiments to examine the performance of Bayesian techniques from the sampling-theory viewpoint. In multivariate distributions we feel that the method of principal directions holds great potential for implementation of likelihood inference in stable distributions and opens the way for routine estimation of complicated models in stable, and similar, distributions. In particular, the method seems very capable of providing inferences for the tail and skewness parameters that are close to inferences from ABC and related "exact" Monte Carlo models.